%% file: trotter-omelyan.tex
\documentclass[a4paper]{article}
\usepackage{a4wide}
\usepackage{graphicx}
\usepackage[T1]{fontenc}
\usepackage[utf8]{inputenc}
\usepackage[UKenglish]{babel}
\usepackage{placeins}
\usepackage[shortlabels]{enumitem}
\usepackage[binary-units=true]{siunitx}
\usepackage{color}
\usepackage{amsmath}
\usepackage{amssymb}
\usepackage{amsthm}
\usepackage{mathtools}
\usepackage{bbold}
\usepackage{braket}
\usepackage[backend=bibtex,style=ieee,citestyle=numeric-comp]{biblatex}
\usepackage{hyperref}
\usepackage{cleveref}
\usepackage[final]{changes}

\graphicspath{{figures/},{inkscape/},{simulation/benchmark/}}

\crefformat{footnote}{#2\footnotemark[#1]#3}

\DeclareMathOperator{\im}{i}
\DeclareMathOperator{\real}{Re}

\DeclareMathOperator{\tr}{Tr}

\DeclareMathOperator{\ord}{\mathcal{O}}

\newcommand{\eto}[1]{\ensuremath{\mathrm{e}^{#1}}}

\newcommand{\ordnung}[1]{\ensuremath{\ord\left(#1\right)}}

\newcommand{\liverpool}{
	\\\textit{\footnotesize Department of Mathematical Sciences,
		University of Liverpool, United Kingdom}
}

\begin{document}
	
	\title{\vspace*{-2\baselineskip}Optimised Trotter Decompositions\\ for Classical and Quantum Computing}
	
	\author{Johann Ostmeyer\footnote{\href{mailto:ostmeyer@liverpool.ac.uk}{ostmeyer@liverpool.ac.uk}}\liverpool}
	\date{\today}
	\maketitle
	
	
	\begin{abstract}
		Suzuki-Trotter decompositions of exponential operators like $\exp(Ht)$ are required in almost every branch of numerical physics. Often the exponent under consideration has to be split into more than two operators $H=\sum_k A_k$, for instance as local gates on quantum computers.
		We demonstrate how highly optimised schemes originally derived for exactly two operators $A_{1,2}$ can be applied to such generic Suzuki-Trotter decompositions, providing a formal proof of correctness as well as numerical evidence of efficiency.
		A comprehensive review of existing symmetric decomposition schemes up to order $n\le4$ is presented and complemented by a number of novel schemes, including both real and complex coefficients. We derive the theoretically most efficient unitary and non-unitary 4th order decompositions. The list is augmented by several exceptionally efficient schemes of higher order $n\le8$.
		Furthermore we show how Taylor expansions can be used on classical devices to reach machine precision at a computational effort at which state of the art Trotterization schemes do not surpass a relative precision of~$10^{-4}$.
		Finally, a short and easily understandable summary explains how to choose the optimal decomposition in any given scenario.
	\end{abstract}


	
	\allowdisplaybreaks[1]
	\unitlength = 1em
	
	\section{Introduction}
	
	\textbf{(Suzuki-)Trotter decomposition schemes}, or \textbf{Trotterizations}, as they are called in most physics related contexts, also referred to as \textbf{splitting methods}, are approximations to operator exponentials, such as
	\begin{align}
		\eto{(A+B+\cdots)h} &= \eto{Ah}\,\eto{Bh}\,\cdots\,\eto{\ordnung{h^2}}\,.
	\end{align}
	This simplest splitting of \textit{order} $n=1$ reduces to the \textbf{Lie product formula} for two operators in the limit $h\rightarrow0$. Sometimes ``symplectic integrator'' is used interchangeably with Trotterization, but this can be very misleading. More precisely, symplectic integrators are a subset of Trotter decompositions for the special case of the exponent consisting of exactly two contributions, kinetic and potential energies, and we do not deal with them explicitly in this article.\footnote{Symplectic integrators rely on the property that a large number of high order commutators vanishes~\cite{OMELYAN2003272}. This means that using the coefficients of a symplectic integrator for a general splitting method leads to a reduced efficiency starting from order $n\ge4$ and might not even give the expected order at all for $n\ge6$.}
	
	Deriving precise Trotter decompositions is not only a challenging mathematical problem, their applications are practically ubiquitous. Just to give a few examples from numerical physics, Trotterization is needed from condensed matter~\cite{PhysRevB.103.024203} to high energy physics~\cite{digitising_su2}, allowing real and imaginary time evolution~\cite{PRXQuantum.2.010342} including ground state search~\cite{my_tensor_networks}, it is a corner stone in hybrid Monte Carlo~\cite{Duane:1987de}, tensor networks~\cite{tensor_intro} as well as quantum computing~\cite{qc_with_high_order_st,Endo_2019,Barratt_2021,adaptive_trotter}.
	
	Here we present a variety of established, lesser known and novel Trotterizations, providing a solid overview and a basis to choose the correct method in a given scenario. We also bridge the gap between decompositions for two and arbitrarily many operators.
	
	If you are mainly interested in which decomposition scheme to use, less in the why and especially not in how it works, feel free to jump straight to the flow chart~\ref{fig:flow-chart} answering exactly this question in a simple fashion.
	
	\subsection{Overview}
	
	Throughout this work we are going to use the following naming conventions, motivated by physics. The \textit{time evolution} $\eto{t H}$ over the (imaginary) \textit{time} $t$ defined by the \textit{Hamiltonian} $H$ is split into time \textit{steps} $h$. The Hamiltonian consists of \textit{operators} $H = \sum_{k=1}^{\Lambda}A_k$ and the exponential of an operator over a \textit{sub-step} $a_ih$ (or $b_ih$, $c_ih$, $d_ih$) is called a \textit{stage}\footnote{In literature the word `stage' sometimes refers to the number $q$ or $2q+1$, so some caution is advised. We refer to $q$ as `cycles' because one has to cycle there and back again through all the stages $q$ times for a complete time step in the formulation of eq.~\eqref{eq:any-stage-decomposition}.} $\eto{a_ihA_k}$. A stage can be interpreted as a \textit{gate} if the operator $A_k$ is local. Together all $\Lambda$ stages make a \textit{ramp} in forward $\prod_{k=1}^{\Lambda}\eto{A_kc_ih}$ or backward $\prod_{k=\Lambda}^{1}\eto{A_kd_ih}$ direction. Ramping forwards and backwards once makes a \textit{cycle} and there are $q$ cycles (or $2q+1$ sub-steps for $\Lambda=2$ stages) to a step. \added{Figure~\ref{fig:stages-visualied} provides a visualisation of these concepts.}
	
	A detailed discussion follows in \Cref{sec:2-to-any-stages}, but let it be mentioned here that decompositions for exactly two operators into as many stages are much easier to derive and much better understood than decompositions into a higher number of stages. However, in physical applications the latter case is often necessary and the means we derive to translate from one case to the other is therefore highly valuable. This is particularly true for digital quantum computing where typically gates can only have non-trivial matrix elements for up to two qubits.
	
	So far, splitting methods for two (sometimes three~\cite{Auzinger:2017}) and arbitrarily many operators have been considered separately. Lie~\cite{lie1888theorie} introduced the former, Trotter~\cite{trotter_original} the latter. Suzuki~\cite{Suzuki:1976be} formalised the idea to split exponential operators and later derived a general scheme to construct decomposition formulas of arbitrary order~\cite{Hatano_2005}. These are the \textit{canonical} schemes we use as a reference. They require $q=5^{n/2-1}$ cycles for order $n$ decompositions.
	
	
	
	
	
	Omelyan, Mryglod, and Folk derived some optimised Suzuki-like decompositions in~\cite{Omelyan_2002}, however they assumed too many constraints and did not obtain the optimal decomposition with 5 cycles. In contrast, in Ref.~\cite{OMELYAN2003272} they indeed classify all the optimal algorithms of up to 5 cycles, but only for the specific case of symplectic integrators where some of the commutators can be set to zero.
	
	\added{A method to construct higher order decompositions using fewer cycles than Suzuki's was derived by Yoshida~\cite{YOSHIDA1990262} and highly efficient decomposition up to order $n\le10$ have been derived very recently by Morales et al.~\cite{morales2022greatly} using this method.} Even better Trotterizations for $n\le6$ have been identified in~\cite{BLANES2002313}. Therein Blanes and Moan claim to have identified the single most efficient decompositions of orders $n=4$ and $n=6$ compared over all the different numbers of cycles. To the best of our knowledge this claim has not been challenged for $n=6$. In the case of $n=4$, however, it turns out that their optimisation routine did not succeed in finding the global maxima but only some close to optimal local maxima.
	
	More exotic approaches to improve Trotterization have been proposed, too. In~\cite{trotter_coeff_optimisation} the error compared to the canonical 2nd order decomposition is reduced to ca.\ 60\% by means of evolutionary optimisation.	
	It is found in~\cite{Childs_2019} that some randomisation of the step sizes can significantly reduce the errors made by low order decompositions. This effect, however, is negligible for orders 4 and higher.
	
	Recently, Trotterization with adaptive step sizes has been proposed~\cite{adaptive_trotter}. This development is orthogonal to ours and complements our work. In~\cite{adaptive_trotter} it is derived that the error can be kept independent of the evolution time, as opposed to the typical linear scaling. An alternative approach has been presented in~\cite{Blanes_2019} where embedded error estimation using two Trotter decompositions of different orders is derived.
	
	We only consider operators that are smooth enough here, but there are ways to rigorously expand exponential decompositions to less well-behaved operators~\cite{Wiebe_2010}.
	
	Overall it turns out that Trotter decompositions tend to be more robust in practice than predicted by conservative error bounds. Both references~\cite{PhysRevX.11.011020,Heyl_2019} found that the theoretical error bounds tend to be much larger than the actual errors, especially for local observables. Moreover, in~\cite{PhysRevX.11.011020} tighter error bounds have been derived, but even they readily overestimate the empirical errors by an order of magnitude in some cases. In \Cref{sec:numerical_experiments} we observe that these discrepancies between theory and practice strongly depend on the particular decomposition scheme and this phenomenon has yet to be explained.
	
	At the time of writing Trotterizations for quantum computing are a particularly active area of research.	
	\added{For instance, it has been shown that an arbitrary function can be approximated on a quantum computer using Fourier-based quantum signal processing~\cite{silva2022fourierbased} that relies solely on real-time evolution oracles of the form $\eto{\im t H}$. The methods discussed in this work provide a means to efficiently implement these oracles.}
		
	Quantum devices suffer from physical errors in addition to the algorithmic errors introduced through the decomposition. The physical errors are mostly determined by the number of operations performed, requiring a trade off between physical (as few gates as possible) and algorithmic errors (as many gates as possible)~\cite{Endo_2019,Dborin:2022zdd}. This implies that optimised decompositions necessitating less cycles for a given precision are even more advantageous for simulations on real quantum devices.
	
	A highly optimised time evolution method for quantum computing that technically does not rely on Trotter decompositions at all has been derived in the references~\cite{optimised_circuits} and~\cite{classical_opt_circuits} independently. The authors numerically minimise the Frobenius norm of the difference between the exact operator and the ``brickwall circuit'' formed by quantum gates for a fixed short evolution time. In both cases different numbers of layers are explored and the obtained circuits perform extremely well compared with the canonical Trotter methods. In particular~\cite{classical_opt_circuits} demonstrates that not only the error coefficient but also the order can be improved compared to the canonical Trotterization. However, the optimisation process itself scales badly with the system size (especially in~\cite{optimised_circuits}) and has to be repeated for every given choice of parameters. The optimisation in~\cite{classical_opt_circuits} is sped up significantly by the use of a matrix product operator (MPO) ansatz using only a Taylor expansion up to first order instead of the full exponential operator.
	
	Ref.~\cite{variational_circuits} presents a compromise between optimisation and scalability for systems with translational invariance by tuning quantum circuits on a subsystem and scaling the results up subsequently. An emphasis is put on the restrictions posed by noisy quantum gates that strongly favour a smaller total number of gates.
	
	\subsection{Using complex coefficients}
	
	So far, all the decompositions considered had purely real coefficients $a_i$, $b_i$ and there are good reasons for this. In many types of computations the coefficients have to be real, for instance to avoid the sign problem in Monte Carlo simulations or to guarantee exact unitarity in quantum computing. We speak of unitarity here in a sense motivated by the quantum mechanical real time evolution operator $\eto{\im t H}$ where the Hamiltonian $H$ is hermitian. A Suzuki-Trotter decomposition of such an operator is unitary to all orders if and only if all the coefficients $a_i$, $b_i$ in the decomposition are real. There may also be runtime considerations favouring real coefficients. A multiplication of two complex numbers on a computer requires four multiplications and two additions of real numbers and thus more than four times more runtime. If a computation is compute bound and all the variables can be kept real, then they probably should be.
	
	For many a simulation on a classical computer, however, there is no reason to demand strict unitarity, for instance in finite temperature calculations where time is complex anyway (see \Cref{sec:numerical_experiments} for another example). Instead, it is advantageous to decrease the overall error, real contributions as well as imaginary, as far as possible with a minimal number of parameters. Allowing the coefficients $a_i$, $b_i$ to be complex, can increase the efficiency of a decomposition by an order of magnitude and the opportunity should therefore not be neglected.
	
	The development of symplectic integrators with complex coefficients is an active area of research. To our knowledge the first systematic development of such integrators (up to order $n=6$) has been performed by Chambers~\cite{Chambers_2003}. The original motivation behind using complex coefficients was that all higher order integrators $n\ge4$ with real coefficients have some negative sub-steps whereas complex coefficients allow to avoid negative real parts, yielding a more equally spaced path overall. This smoothness appears to reduce the error. In fact, we find that schemes with similar sub-steps have a more favourable error accumulation over time and fare better with splittings into a higher number of stages.
	
	Just this year Casas, Escorihuela-Tomàs and different collaborators have published several works on higher order symplectic integrators with complex coefficients~\cite{CASAS2022126700,Auzinger:2017,blanes2022symmetric,Blanes:2022}. Again, as Omelyan~\cite{OMELYAN2003272} did with real coefficients, they do not optimise for general Trotter decompositions. Furthermore, similar to Suzuki's approach~\cite{Hatano_2005} they construct higher order schemes starting from the Verlet decomposition~\cite{Verlet:1967} (see eq.\eqref{eq:1-leap-frog}) which is not an atomic building block and comes with a reduced number of tunable parameters. The schemes presented in~\cref{eq:4-non-unitary1,eq:5-non-unitary2} have therefore, to our knowledge, not been derived before and outperform all methods of the same order we found in the literature by at least a factor two, including other non-unitary schemes. 
	
	Especially the results in~\cite{Blanes:2022} are noteworthy nonetheless. There the differences between symmetric (``palindromic'') and so-called ``symmetric-conjugate'' splitting methods are discussed. Symmetric-conjugate means that the reversed scheme is equal to the complex conjugate original. They violate unitarity less than palindromic schemes of the same order and comparable local error. In particular the unitarity violation stays constant over long times for symmetric-conjugate decompositions while it tends to diverge for palindromic ones. There is, however, a simple remedy to this problem with palindromic schemes, namely to complex conjugate all the coefficients in every second time step. Since palindromic decompositions do not have any other disadvantages and we did not find any symmetric-conjugate decomposition with a higher efficiency, we restrain ourselves to truly symmetric schemes throughout the rest of this work.\\
	
	\noindent
	The rest of this paper is structured as follows. In \Cref{sec:2-to-any-stages} we prove a theorem that allows to adapt Trotter decompositions for two operators to an arbitrary number of operators. \Cref{sec:deriving_schemes} contains a list of polynomial Trotterizations with real and complex coefficients including their theoretical efficiency, some practical advice and a useful method to construct higher order schemes. Furthermore, in \Cref{sec:taylor_expansion} we present the non-polynomial Taylor method that has limited applicability but extremely high precision. Finally, we conduct detailed numerical experiments in \Cref{sec:numerical_experiments} comparing decompositions into different numbers of stages and stress-testing theoretical efficiency predictions against de facto performance.

	\section{Adapting 2-stage decompositions to an arbitrary number of stages}\label{sec:2-to-any-stages}
\begin{figure}[bth]
		\centering
		\mbox{\qquad\quad\;\;\,}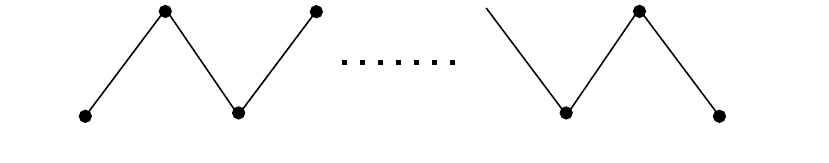\\~\\
		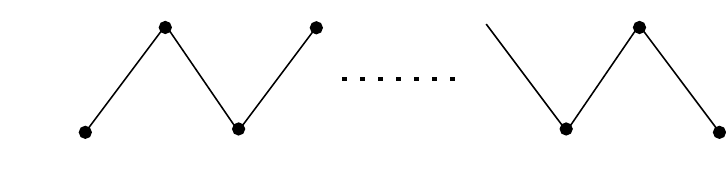\\~\\
		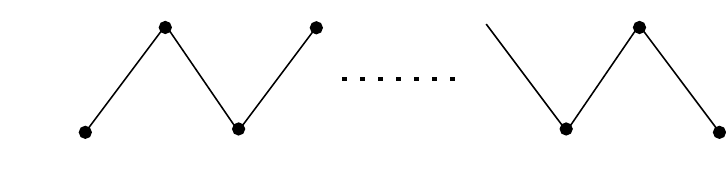
		\caption{Visualisation of the transformation allowing to rewrite a 2-stage decomposition for arbitrarily many stages. In this depiction every dot corresponds to a single operator evaluation or \textit{stage}, a straight line corresponds to a \textit{ramp} and two adjacent ramps form a \textit{cycle}. We start in the top row with a 2-stage decomposition as in eq.~\eqref{eq:2-stage-decomposition} and transform the coefficients $a_i$, $b_i$ into the equivalent $c_i$, $d_i$ as in eq.~\eqref{eq:any-stage-decomposition} using eqs.~\eqref{eq:c1_d1} through~\eqref{eq:cq_dq}. The outcome is shown in the second row. It demonstrates a transition from a stage-based to a ramp-based approach. Every ramp can now be populated with more stages in the last row without changing the validity of the decomposition.}\label{fig:stages-visualied}
	\end{figure}

	Decompositions with two stages, i.e.\ with a Hamiltonian consisting of exactly 2 operators,
	\begin{align}
		\eto{(A+B)h + \ordnung{h^{n+1}}} &= \eto{Aa_1h}\eto{Bb_1h}\eto{Aa_2h}\cdots\eto{Bb_qh}\eto{Aa_{q+1}h}\label{eq:2-stage-decomposition}
	\end{align}
	have been studied and optimised extensively in literature (e.g.~\cite{Omelyan_2002,BLANES2002313,Auzinger:2017} and many more). However, this is only a special case. Often a Hamiltonian consists out of three or more non-commuting operators (an obvious example is the Heisenberg model) and in many cases it is necessary to split it into $\ordnung{L}$ local contributions, e.g.\ to apply gates to tensor network or quantum simulations. In these cases only the schemes devised by Hatano and Suzuki~\cite{Hatano_2005} have been applicable so far. We are now going to show that \textit{every scheme of order $n$ applicable to two stages defines an order $n$ scheme for an arbitrary number of stages}. This allows to adapt the valuable work done on 2-stage decompositions so far and it furthermore provides a simple means to derive efficient new methods in future since it is sufficient to analyse the method's properties in the simple case of two stages.
	\added{The idea behind the transformation required for this adaptation has been visualised in figure~\ref{fig:stages-visualied}.}
	
	\paragraph{Theorem:}
	Let the coefficients $a_i$ and $b_i$ as in equation~\eqref{eq:2-stage-decomposition} define a 2-stage decomposition of order $n$. Then the decomposition
	\begin{align}
		\eto{h\sum\limits_{k=1}^{\Lambda}A_k + \ordnung{h^{n+1}}} &= \left(\prod_{k=1}^{\Lambda}\eto{A_kc_1h}\right) \left(\prod_{k=\Lambda}^{1}\eto{A_kd_1h}\right) \cdots \left(\prod_{k=1}^{\Lambda}\eto{A_kc_qh}\right) \left(\prod_{k=\Lambda}^{1}\eto{A_kd_qh}\right)\label{eq:any-stage-decomposition}
	\end{align}
	of $\Lambda$ non-commuting operators $A_k$ into ordered products $\prod_{k=1}^{\Lambda}X_k\equiv X_1X_2\cdots X_\Lambda$ is exact to order $n$ as well and the coefficients $c_i$, $d_i$ are given by
	\begin{align}
		c_1 &= a_1\,, & d_1 &= b_1 - c_1\,,\label{eq:c1_d1}\\
		c_2 &= a_2 - d_1\,,& d_2 &= b_2 - c_2\,,\\
		&\;\;\vdots & &\;\;\vdots\nonumber\\
		c_i &= a_i - d_{i-1}\,,& d_i &= b_i - c_i\,,\\
		&\;\;\vdots & &\;\;\vdots\nonumber\\
		c_q &= a_q - d_{q-1}\,,& d_q &= b_q - c_q\,.\label{eq:cq_dq}
	\end{align}
	
	\paragraph{Proof:}
	First, we note that for every $i\in\{1,\dots,q\}$ (setting $d_0=0$)
	\begin{align}
		c_i+d_{i-1} &= a_i-d_{i-1}+d_{i-1}\\
		&= a_i\,,\\
		d_i+c_{i} &= b_i-c_{i}+c_{i}\\
		&= b_i\,,
	\end{align}
	i.e.\ we have defined a telescope sum that reduces to the original decomposition~\eqref{eq:2-stage-decomposition} in the case of $\Lambda=2$. This means that the relation~\eqref{eq:any-stage-decomposition} trivially holds for two stages.
	
	We use the Baker–Campbell–Hausdorff formula to rewrite the factors in equation~\eqref{eq:any-stage-decomposition}
	\begin{align}
		\begin{split}
			\left(\prod_{k=1}^{\Lambda}\eto{A_kc_ih}\right) &= \eto{C_1c_ih + C_2\left(c_ih\right)^2 + C_3\left(c_ih\right)^3 + \cdots}\\
			\left(\prod_{k=\Lambda}^{1}\eto{A_kd_ih}\right) &= \eto{D_1d_ih + D_2\left(d_ih\right)^2 + D_3\left(d_ih\right)^3 + \cdots}
		\end{split}\label{eq:bch_of_sum}
	\end{align}
	where $C_1=D_1=\sum\limits_{k=1}^{\Lambda}A_k$ and the remaining operators $C_l$, $D_l$ are some linear combinations of $l$th order commutators $[A_{k_1},[A_{k_2},\cdots[A_{k_{l-1}},A_{k_l}]\cdots]]$. It is crucial that the $C_l$, $D_l$ are independent of $i$ and that for every $\Lambda\ge2$ all the $C_l$ ($D_l$) are mutually non-commuting and linearly independent in general. Thus, even though nothing is known explicitly about the $C_l$, $D_l$ a priori, the right hand side of equation~\eqref{eq:bch_of_sum} is completely independent of $\Lambda$. In particular, none of the terms vanishes exclusively in the $\Lambda=2$ case.
	
	Now we know that
	\begin{align}
		\eto{h\sum\limits_{k=1}^{\Lambda}A_k + \ordnung{h^{n+1}}} &= \eto{C_1c_1h + C_2\left(c_1h\right)^2 + \cdots} \eto{D_1d_1h + D_2\left(d_1h\right)^2 + \cdots} \cdots \eto{C_1c_qh + C_2\left(c_qh\right)^2 + \cdots} \eto{D_1d_qh + D_2\left(d_qh\right)^2 + \cdots}\label{eq:decomposition_forget_stage}
	\end{align}
	holds in the $\Lambda=2$ case because we chose the coefficients $c_i$, $d_i$ accordingly. Since the $c_i$, $d_i$ fulfil this condition for every $C_i$, $D_i$ (they solve the corresponding system of polynomial equations), they automatically fulfil it for every $\Lambda$. This proves the initial claim~\eqref{eq:any-stage-decomposition}.
	
	\qed
	
	\paragraph{Remarks:}
	It might appear counter-intuitive at first that the number of relevant operators $C_l$, $D_l$ does not increase with the order $l$ because the number of conditions on the coefficients $c_i$, $d_i$ certainly does increase with the target order $n$. This, however, is easily resolved when considering that the full product on the right hand side of equation~\eqref{eq:decomposition_forget_stage} accumulates various commutators consisting of $C_{l'}$, $D_{l''}$ terms where $l',l''\le l$.
	
	The operators $C_l$ and $D_l$ are related as $C_l = (-1)^{l+1} D_l$ which is why the even order error terms automatically drop out in symmetric decomposition schemes. This is not relevant for the proof, however. In particular, it is valid for symmetric as well as non-symmetric schemes.
	
	The proof does not offer any information about the higher order terms $\ordnung{h^{n+1}}$. This means that a decomposition optimised for two stages is not guaranteed to be efficient for a larger number of stages as well, it is only guaranteed to have the same order. In practice, usually decompositions perform similarly well for different numbers of stages, but we will encounter exceptions from this rule later in our numerical experiments (see \Cref{sec:numerical_experiments}).
	
	\section{Deriving and optimising decomposition schemes}\label{sec:deriving_schemes}
	For the rest of this work we will consider symmetric decomposition schemes only. Set aside their often useful reversibility, symmetric schemes are simply more efficient for evolutions over long times $t$, i.e.\ when $t/h\gg1$ which is the physically interesting and computationally challenging case. The reason is that any non-symmetric and therefore odd order $n-1$ scheme can be elevated to the even order $n$ without any drawbacks by reverting the sequence of coefficients $a_i$, $b_i$ (or $c_i$, $d_i$) in every second step. The established symmetry guarantees an even order $n$. In case the integer $t/h$ was an even number to start with, the computational effort remains the same. If $t/h$ was odd, it has to be incremented by one at the relative additional computational cost of $h/t\ll1$. In both cases the induced symmetric even order $n$ decomposition is strictly more efficient than the initial non-symmetric one.
	
	\subsection{Higher order schemes}
	\subsubsection{Suzuki's method}
	There are fundamentally different approaches to constructing a higher order ($n\ge4$) decomposition scheme. On the one hand Suzuki~\cite{suzuki_original,Hatano_2005} derived a method that, given a scheme $S_n(h)$ of order $\ordnung{h^n}$, elevates it to order $\ordnung{h^{n+2}}$ by successive application
	\begin{align}
		S_{n+2}(h) &= S_n(s_nh)^p\, S_n\left((1-2ps_n)h\right)\, S_n(s_nh)^p\,,\label{eq:construct_higher_order}\\
		s_n &= \frac{1}{2p-(2p)^\frac{1}{n+1}}\,.
	\end{align}
	Technically $p$ can be any positive integer, but any choice other than $p=2$ leads to very bad performance and should not be used. We will explore this further below. Note that this construction can be applied recursively in order to construct decomposition schemes of any order.
	
	\subsubsection{Yoshida's method}
	A more involved method has been put forward by Yoshida~\cite{YOSHIDA1990262}, advocating to construct higher order schemes with a minimal number of cycles $q=2m+1$ from symmetric building blocks (typically of second order $S_2$) of the form
	\begin{align}
		S_{n}(h) &= S_2(w_mh)\cdots S_2(w_2h)S_2(w_1h)\, S_2\left(w_0h\right)\,S_2(w_1h)S_2(w_2h)\cdots S_2(w_mh)\,,\label{eq:construction_yoshida}\\
		w_0 &= 1-2\sum_{i=1}^m w_m\,.
	\end{align}
	A suitable integer $m$ and tuple $(w_1,\dots,w_m)$ have to be found numerically, so that the composite scheme is of the desired order $n$. Yoshida also provides recursive formulae to construct the system of equations that has to be solved. The system of equations is generally much simpler to construct and to solve than in the general case (without $S_2$ building blocks), allowing Yoshida to find order $n=6$ and $n=8$ decompositions as early as 1990. Significantly more efficient decompositions of orders $n=8$ and $n=10$ have been discovered very recently by Morales et al.~\cite{morales2022greatly}.
	
	When it comes to maximal efficiency, Yoshida's method has two major flaws. The first is that both Yoshida and Morales et al.\ focused their search on schemes with a minimal number of cycles. As is well known (e.g.\ from~\cite{Omelyan_2002,BLANES2002313}) and demonstrated below, the addition of several cycles more than strictly needed pays off because it allows to use the free parameters to minimise the error of the decomposition. This flaw is a mere technical one and Yoshida's method could provide significantly better results than the ones derived so far if more cycles (or equivalently larger $m$) would be taken into account. Morales et al.\ found that they obtained superior results choosing $m$ larger than the minimal value by one for the orders $n=8$ and $n=10$. Extrapolating from the orders $n=4$ and $n=6$, we expect the optimal $m$ somewhere closer to twice the minimal $m$ than to the minimum itself and we therefore see vast potential in further explorations of this direction.
	
	The second flaw of Yoshida's method is inherent and cannot be fixed. By confining the schemes to those built from $S_2$ blocks, the number of potential degrees of freedom is reduced almost by a factor two. The solutions of Yoshida's method are a true subset of the decompositions possible with a given number of cycles, furthermore even numbers of cycles are not possible at all. This problem is exactly quantifiable for $n=4$ where the best possible Yoshida-style decomposition with $q=5$ cycles has been derived by Omelyan et al.~\cite{Omelyan_2002}. It has an efficiency of $\mathrm{Eff}_4=\num{1.4}$ (see definition below), about $\num{7.5}$ times less efficient than the optimal fourth order scheme with $q=5$ cycles~\eqref{eq:5-opt-4th-ord}.
	The problem is also reflected (together with the first flaw) in the extremely poor performance of Yoshida's order $n=6$ schemes. This main flaw of Yoshida's method persists for all orders and numbers of cycles.
	
	\subsubsection{Omelyan's method}
	The alternative approach is to construct a scheme from scratch. This is significantly more complicated and requires some theoretical background, it can however lead to significantly more efficient decompositions. Omelyan, Mryglod and Folk~\cite{Omelyan_2002} as well as Blanes and Moan~\cite{BLANES2002313} derived a number of highly optimised schemes. Closely following the approach by Omelyan et al.\ we will now define what requirements exactly a decomposition scheme has to fulfil to be of order $n$ and what it means for such a scheme to be efficient.
	
	
	Following the derivations in Refs.~\cite{Omelyan_2002,OMELYAN2003272}, we can write down a basis for the leading order errors of a decomposition
	\begin{align}
		\eto{(A+B)h + \ord_1h + \ord_3 h^3 + \ord_5 h^5 + \cdots} &= \eto{Aa_1h}\eto{Bb_1h}\cdots\eto{Bb_qh}\eto{Aa_{q+1}h}
	\end{align}
	in terms of commutators
	\begin{align}
		\ord_1 &= (\nu-1) A + (\sigma-1)B\,,\label{eq:ord_1_errors}\\
		\ord_3 &=  \alpha [A, [A, B]] + \beta [B, [A, B]]\,,\label{eq:ord_3_errors}\\
		\begin{split}
			\ord_5 &= \gamma_1 [A, [A, [A, [A, B]]]] + \gamma_2 [A, [A, [B, [A, B]]]] + \gamma_3 [B, [A, [A, [A, B]]]]\\
			&\quad + \gamma_4 [B, [B, [B, [A, B]]]] + \gamma_5 [B, [B, [A, [A, B]]]] + \gamma_6 [A, [B, [B, [A, B]]]]\,.
		\end{split}\label{eq:ord_5_errors}
	\end{align}

	A decomposition is valid if $\nu=\sigma=1$, that is $\ord_1=0$. This can easily be guaranteed by the intuitive condition $\sum_i a_i=\sum_i b_i=1$ (or equivalently $\sum_i(c_i+d_i)=1$). In order to construct a 4th order decomposition, $\alpha=\beta=0$ has to be satisfied additionally. For a 6th order scheme all the $\gamma_j$ have to vanish, and so on.
	
	\added{We remark that symplectic integrators (or Runge–Kutta–Nyström methods) deal with the special case of $[B, [B, [B, A]]] = 0$ leading to significantly less contributions to all the operators $\ord_k$ with $k\ge 5$ than in the general case discussed here. This allows for more efficient but specialised methods, in particular higher order symplectic decompositions $n\ge 6$ require less cycles $q$ than general Trotterizations. This distinction has been discussed in detail by Blanes and Moan~\cite{BLANES2002313}. A comprehensive list of efficient symplectic integrators up to order $n\le 6$ has been provided by Omelyan~\cite{OMELYAN2003272}, we especially recommend Table~2 therein.}
	
	A decomposition is efficient if its leading order errors are small compared to the number $q$ of cycles\footnote{$q$ is also often called `stages' or `number of force calculations' in the literature.} it requires. We assume all the commutator products in the equations~(\ref{eq:ord_1_errors}-\ref{eq:ord_5_errors}) to be mutually orthogonal so that the euclidean norm over the thus spanned vector space estimates the total errors. Note that orthogonality is not guaranteed, but for high enough dimensional operators the overlap will be negligible in general. In particular we define the efficiencies following Omelyan~\cite{OMELYAN2003272}
	\begin{align}
		\mathrm{Eff}_2 &= \frac{1}{q^2\sqrt{|\alpha|^2+|\beta|^2}}\,,\\
		\mathrm{Eff}_4 &= \frac{1}{q^4\sqrt{\sum_{j=1}^6|\gamma_j|^2}}\,.
	\end{align}
	\added{Higher order efficiencies $\mathrm{Eff}_n$ are defined analogously via the leading order error and the respective number of cycles.}
	\deleted{It does not make sense to directly compare efficiencies of different decompositions across orders. The optimal order to choose depends foremost on the desired accuracy.}
	Within an order the efficiency is directly proportional to the precision obtained at any given computational effort and a higher efficiency is generally better.
	
	\added{A less precise but still useful comparison of efficiencies can be provided across orders as well. In this case, however, the target error $\varepsilon$ and the evolution time $t$ as well as the spectral norm $|H|$ of the evolved operator $H$ have to be taken into account. The cost of an order $n$ decomposition at given precision is proportional to $1/\sqrt[n]{\mathrm{Eff}_n}$. Similarly, since the error scales with the step size as $\varepsilon\sim h^n$, the cost is proportional to $1/\sqrt[n]{\varepsilon}$. Finally, the error accumulates linearly over time $t$ weighted by the operator norm $|H|$, necessitating another adjustment of the step size $h\sim 1/\sqrt[n]{|H|t}$. Thus, the overall total cost of any Trotterization will scale anti-proportionally to the rescaled efficiency
	\begin{align}
		\mathrm{eff}_n(|H|t,\varepsilon) &= \left(\mathrm{Eff}_n\, \frac{\varepsilon}{|H|t}\right)^{\frac 1n}\,.
	\end{align}
	More precisely, the total number of cycles scales as $|H|t/\mathrm{eff}_n(|H|t,\varepsilon)$ (compare Ref.~\cite{morales2022greatly}) where the additional factor $|H|t$ comes simply from the duration of the evolution. As could have been expected, higher order methods become more efficient and therefore preferable with lower target accuracy and longer evolution times $|H|t$.}
	
	There are some deviations from this rule. We have already mentioned in \Cref{sec:2-to-any-stages} that an efficiency derived for two stages $A$, $B$ is not identical to the efficiency of the same decomposition scheme applied to more stages $A_1,A_2,\dots,A_\Lambda$. Furthermore the efficiency only describes a single step in a time evolution usually consisting of many $>\ordnung{100}$ such steps. Even though asymptotically the error always accumulates linearly over time, it might add up more or less favourably. So far, we could not identify a method to quantify these deviations from the predicted efficiency other than empirically, see \Cref{sec:numerical_experiments} for details. Overall, the efficiency is still a good measure for the performance of a decomposition scheme since both deviations are rather small and do not change the order of magnitude of the errors.
	
	\subsection{List of decomposition schemes}\label{sec:list_of_schemes}
	In the following we provide a list of decomposition schemes up to order $n=6$ we find relevant for one or another reason. This does not mean that all of them are efficient, quite the contrary.
	Coming from the 2-stage approach, we provide the coefficients $a_i$, $b_i$ as in equation~\eqref{eq:2-stage-decomposition}, but they can readily be transferred into the $c_i$, $d_i$ formulation of equation~\eqref{eq:any-stage-decomposition} using the relations~(\ref{eq:c1_d1}-\ref{eq:cq_dq}). We also stress that \textit{whenever it is possible to use only two stages, this should be done}. Not only because the theoretical efficiency is applicable in that case and empirically higher than for a different splitting. Using two stages can effectively reduce the computational effort by a factor 2 because every product in the decomposition~\eqref{eq:any-stage-decomposition} ends with the starting term of the next product, so that they can be united, more generally the	computational cost of $s$ stages scales as
	\begin{align}
		\text{cost} &\sim \frac{s-1}{s}\,.\label{eq:cost_with_stages}
	\end{align}
	\added{This result can be obtained by the following considerations. The total number of exponentials that have to be evaluated in a time step is exactly $2q(s-1)+1$ (only $2q(s-1)$ when taking the first-same-as-last property into account), while the cost of any single exponential will likely scale as $1/s$.}
	
	Since we only consider symmetric schemes, there is no need to write down all the coefficients explicitly. The second half is given by
	\begin{align}
		a_{q+2-i} &= a_{i}\,,&b_{q+1-i} &= b_i\,.
	\end{align}
	
	\subsubsection{Verlet or Leapfrog (order $n=2$, $q=1$ cycle)}
	The simplest possible symmetric scheme
	\begin{align}
		\begin{split}
			a_{1} &= \frac 12\,, \\ b_{1} &= 1\,,
		\end{split}\label{eq:1-leap-frog}\\
		\mathrm{Eff}_2 &= \num{10.7}
	\end{align}
	is well known as Verlet decomposition~\cite{Verlet:1967} or Leapfrog algorithm. Taking into account its simplicity, it performs surprisingly well and is a valid choice when programming time is more important than performance or a high precision is explicitly not desired.
	
	\subsubsection{Omelyan (order $n=2$, $q=2$ cycles)}
	Omelyan et al.~\cite{OMELYAN2003272} derived the theoretically most efficient second order decomposition
	\begin{align}
		\begin{split}
			a_{1} &= 0.1931833275037836\,, \\ b_{1} &= \frac 12\,,\\
			a_{2} &= 1 - 2a_1\,,
		\end{split}\label{eq:2-omelyan2}\\
	\mathrm{Eff}_2 &= \num{29.2}\,.
	\end{align}
	It performs extremely well for symplectic integration. For Suzuki-Trotter decompositions we find empirically that this scheme has a rather unfavourable error accumulation over time. Thus, it usually performs similarly well as the simpler Verlet decomposition~\eqref{eq:1-leap-frog}.
	
	\subsubsection{Forest-Ruth (order $n=4$, $q=3$ cycles)}
	Historically, the decomposition
	\begin{align}
		\begin{split}
			a_{1} &= 0.6756035959798288\,, \\ b_{1} &= 1.351207191959658\,,\\
			a_{2} &= \frac 12 - a_1\,, \\ b_{2} &= 1 - 2b_1\,,
		\end{split}\label{eq:3-forest-ruth}\\
	\mathrm{Eff}_4 &= \num{0.315}
	\end{align}
	by Forest and Ruth~\cite{FOREST1990105} was the first scheme of fourth order and it is the only one with the minimal number of $q=3$ cycles. It is still widely used for these reasons even though it is the least efficient $n=4$ scheme available. Note that it is also the simplest example of Suzuki's higher order construction method~\eqref{eq:construct_higher_order} at work using the Verlet scheme~\eqref{eq:1-leap-frog} as a starting point and $p=1$. The scheme's efficiency vividly demonstrates why $p=1$ is a bad choice.
	
	This decomposition should never be used.
	
	\subsubsection{Omelyan's Forest-Ruth-Type (order $n=4$, $q=4$ cycles)}
	In Ref.~\cite{Omelyan_2002} Omelyan et al.\ add a cycle to Forest and Ruth's decomposition~\eqref{eq:3-forest-ruth} and optimise the emergent free parameter. The resulting scheme of Forest-Ruth-type
	\begin{align}
		\begin{split}
			a_{1} &= 0.1720865590295143\,, \\ b_{1} &= 0.5915620307551568\,,\\
			a_{2} &= -0.1616217622107222\,, \\ b_{2} &= \frac 12 - b_1\,,\\
			a_{3} &= 1 - 2\sum_{i=1}^{2}a_i\,,
		\end{split}\label{eq:4-fr-type}\\
	\mathrm{Eff}_4 &= \num{4.24}
	\end{align}
	has the highest possible efficiency for the given choice of order $n=4$ and $q=4$ cycles using real coefficients. When Omelyan et al.\ search for the optimal scheme with $q=5$ cycles, they apply too many constraints and fail to identify the decomposition scheme~\eqref{eq:5-opt-4th-ord} derived here for the first time. They therefore falsely conclude that this FR-type decomposition is the most efficient one for any $q\le5$. We emphasise that it is highly efficient, but it is outshone both by our~\eqref{eq:5-opt-4th-ord} as well as by Blanes and Moan's schemes~\eqref{eq:6-blanes4} and should therefore not be used.
	
	\subsubsection{Omelyan's small $A$ (order $n=4$, $q=4$ cycles)}
	In particular cases of 2-stage decompositions one of the operators, say $A$ w.l.o.g., is small compared to the other one. In this case it is advantageous to eliminate $\gamma_4$ from equation~\eqref{eq:ord_5_errors} completely rather than minimise the sum $\sum_j \gamma_j^2$ because all other contributions are suppressed by an additional factor $A$. The decomposition
	\begin{align}
		\begin{split}
			a_{1} &= 0.5316386245813512\,, \\ b_{1} &= -0.04375142191737413\,,\\
			a_{2} &= -0.3086019704406066\,, \\ b_{2} &= \frac 12 - b_1\,,\\
			a_{3} &= 1 - 2\sum_{i=1}^{2}a_i
		\end{split}\label{eq:4-small-b}
	\end{align}
	does just that. It has been derived together with other similar special cases by Omelyan et al.\ in Appendix A of Ref.~\cite{Omelyan_2002}. We do not quote the efficiency here since it would be misleading.
	
	\subsubsection{Non-Unitary (order $n=4$, $q=4$ cycles)}\label{sec:non-unitary1}
	So far, all the decompositions considered had purely real coefficients $a_i$, $b_i$. Allowing the coefficients $a_i$, $b_i$ to be complex, however, can increase the efficiency of a decomposition by an order of magnitude and the opportunity should therefore not be neglected.
	
	We performed the efficiency optimisation for $q=4$ cycles arriving at
	\begin{align}
		\begin{split}
			a_{1} &= 0.09957801119428374 +0.02359386141367452 \im\,, \\ b_{1} &= 0.2596218597573501 +0.08909472525370253 \im\,,\\
			a_{2} &= 0.2520542187700347 +0.09826170579213035 \im\,, \\ b_{2} &= \frac 12 - b_1\,,\\
			a_{3} &= 1 - 2\sum_{i=1}^{2}a_i\,,
		\end{split}\label{eq:4-non-unitary1}\\
	\mathrm{Eff}_4 &= \num{29.9}
	\end{align}
	and for $q=5$ cycles (see eq.~\eqref{eq:5-non-unitary2}). Both are roughly 7 times more efficient than their respective unitary counterparts. We therefore highly recommend using these novel decompositions when possible. Empirically we find that this scheme~\eqref{eq:4-non-unitary1} is well behaved at long times and therefore more performant in practice than its theoretically more efficient $q=5$ counterpart~\eqref{eq:5-non-unitary2}. Somewhat surprisingly, we find that the gap between our non-unitary schemes to the next best unitary ones is even larger than suggested by the theoretical efficiency in the case of more than two stages.
	
	Note that it is advantageous to complex conjugate all the coefficients in every second time step in order to obtain a ``symmetric-conjugate'' method~\cite{Blanes:2022}. This might be crucial for preserving near-unitarity, even though empirically we find that the improvement in the Frobenius norm induced by this change is negligible ($\ordnung{10\%}$ and not properly resolvable in figure~\ref{fig:err_const_t_ord4}).
	
	\subsubsection{Suzuki (order $n=4$, $q=5$ cycles)}
	Applying Suzuki's~\cite{Hatano_2005} construction~\eqref{eq:construct_higher_order} starting from the simple Verlet scheme~\eqref{eq:1-leap-frog} leads to
	\begin{align}
		\begin{split}
			a_{1} &= 0.2072453858971879\,, \\ b_{1} &= 0.4144907717943757\,,\\
			a_{2} &= 0.4144907717943757\,, \\ b_{2} &= 0.4144907717943757\,,\\
			a_{3} &= \frac 12 - \sum_{i=1}^{2}a_i\,, \\ b_{3} &= 1 - 2\sum_{i=1}^{2}b_i\,,
		\end{split}\label{eq:5-suzuki4}\\
	\mathrm{Eff}_4 &= \num{1.10}\,.
	\end{align}
	This decomposition is highly inefficient for two stages, but it performs very well for many stages. Empirically we find that this property together with its favourable error accumulation over time secures this well established scheme the first place among the unitary order $n=4$ multi-stage decompositions.
	
	\subsubsection{Optimised 4th order (order $n=4$, $q=5$ cycles)}
	The following decomposition scheme is novel to our knowledge
	\begin{align}
		\begin{split}
			a_{1} &= 0.09257547473195787\,, \\ b_{1} &= 0.2540996315529392\,,\\
			a_{2} &= 0.4627160310210738\,, \\ b_{2} &= -0.1676517240119692\,,\\
			a_{3} &= \frac 12 - \sum_{i=1}^{2}a_i\,, \\ b_{3} &= 1 - 2\sum_{i=1}^{2}b_i\,,
		\end{split}\label{eq:5-opt-4th-ord}\\
	\mathrm{Eff}_4 &= \num{10.5}\,.
	\end{align}
	We derived it in the usual way by optimising the efficiency for given order and number of cycles. We are confident that no $q=5$ scheme has a higher efficiency and we are not aware of any order $n=4$ scheme with a higher efficiency in the literature.
	
	That said, we find empirically that this scheme has a rather unfavourable error accumulation over time. Thus, it is usually surpassed at long times by Blanes and Moan's scheme~\eqref{eq:6-blanes4} with similar theoretical efficiency.
	
%
	
	\subsubsection{Non-Unitary (order $n=4$, $q=5$ cycles)}
	Similarly to the decomposition scheme~\eqref{eq:4-non-unitary1} this choice of complex coefficients
	\begin{align}
		\begin{split}
			a_{1} &= 0.07613272445178274 -0.03518797331257356 \im\,, \\ b_{1} &= 0.1658339349217486 -0.07090293766092534 \im\,,\\
			a_{2} &= 0.2017183745725757 +0.02597491015915232 \im\,, \\ b_{2} &= 0.2137425142256234 +0.1386193640914034 \im\,,\\
			a_{3} &= \frac 12 - \sum_{i=1}^{2}a_i\,, \\ b_{3} &= 1 - 2\sum_{i=1}^{2}b_i\,,
		\end{split}\label{eq:5-non-unitary2}\\
	\mathrm{Eff}_4 &= \num{67.4}
	\end{align}
	is not exactly unitary at all orders. There are cases in which it cannot be applied for this reason. Whenever it is applicable, however, it is the most efficient fourth order decomposition known to us. In some cases, due to its rather unfavourable error accumulation over time, it is outperformed by our non-unitary schemes~\eqref{eq:4-non-unitary1} and~\eqref{eq:5-non-unitary-const}, but no unitary decomposition comes even close. Similarly to scheme~\eqref{eq:4-non-unitary1}, the performance gap between this decomposition and the next best unitary one increases with the number of stages.
	
	Same as for scheme~\eqref{eq:4-non-unitary1} it is advantageous to complex conjugate all the coefficients in every second time step in order to obtain a ``symmetric-conjugate'' method~\cite{Blanes:2022}. This might be crucial for preserving near-unitarity, even though empirically we find that the improvement in the Frobenius norm induced by this change is negligible ($\ordnung{1\%}$ and not properly resolvable in figure~\ref{fig:err_const_t_ord4}).
	
	\subsubsection{Uniform non-unitary (order $n=4$, $q=5$ cycles)}
	With $q=5$ cycles it is just possible to construct a non-unitary decomposition scheme with constant real parts of the sub-steps. Among these schemes the coefficients
	\begin{align}
		\begin{split}
			a_{1} &= 0.1 +0.02523113193557069 \im\,, \\ b_{1} &= 0.2 +0.05046226387114138 \im\,,\\
			a_{2} &= 0.2 -0.04082482904638631 \im\,, \\ b_{2} &= 0.2 -0.132111921963914 \im\,,\\
			a_{3} &= \frac 12 - \sum_{i=1}^{2}a_i\,, \\ b_{3} &= 1 - 2\sum_{i=1}^{2}b_i\,,
		\end{split}\label{eq:5-non-unitary-const}\\
	\mathrm{Eff}_4 &= \num{6.38}
	\end{align}
	define the one with the highest efficiency. It has been noted that large fluctuations in the sub-step sizes and especially negative (real parts of the) sub-steps can affect the error accumulation rather unfavourably. For this reason decomposition~\eqref{eq:5-non-unitary-const} promises to perform well compared to its theoretical efficiency. Empirically we find that this indeed is the case, especially for a large number of stages $\Lambda\gg2$.

	Same as for scheme~\eqref{eq:4-non-unitary1} it is advantageous to complex conjugate all the coefficients in every second time step in order to obtain a ``symmetric-conjugate'' method~\cite{Blanes:2022}. This might be crucial for preserving near-unitarity, even though empirically we find that the improvement in the Frobenius norm induced by this change is negligible ($\ordnung{1\%}$ and not properly resolvable in figure~\ref{fig:err_const_t_ord4}).
	
	\subsubsection{Blanes \& Moan (order $n=4$, $q=6$ cycles)}
	In Ref.~\cite{BLANES2002313} Blanes and Moan perform an extensive numerical search among different numbers of cycles for the most efficient orders $n=4$ and $n=6$ decomposition schemes. Apparently their optimisation algorithm got stuck in some local maxima, as they failed to notice the theoretically more efficient scheme~\eqref{eq:5-opt-4th-ord}. Even so, the decomposition they propose
	\begin{align}
		\begin{split}
			a_{1} &= 0.07920369643119569\,, \\ b_{1} &= 0.209515106613362\,,\\
			a_{2} &= 0.353172906049774\,, \\ b_{2} &= -0.143851773179818\,,\\
			a_{3} &= -0.0420650803577195\,, \\ b_{3} &= \frac 12 - \sum_{i=1}^{2}b_i\,,\\
			a_{4} &= 1 - 2\sum_{i=1}^{3}a_i\,,
		\end{split}\label{eq:6-blanes4}\\
	\mathrm{Eff}_4 &= \num{10.2}\,,
	\end{align}
	is highly efficient and it has a quite favourable error accumulation over time. It is therefore the best choice among the unitary order $n=4$ schemes for a 2-stage decomposition and still a good choice in the case of more stages.
	
	\subsubsection{Yoshida (order $n=6$, $q=7$ cycles)}
	\added{In~\cite{YOSHIDA1990262}, besides presenting the method~\eqref{eq:construction_yoshida}, Yoshida derived three different 6th order decompositions of which `Solution A' is the most efficient one and presented here:}
	\begin{align}
		a_{1} &= 0.39225680523878\,, & b_{1} &= 0.78451361047756\,,\nonumber\\
		a_{2} &= 0.5100434119184585\,, & b_{2} &= 0.235573213359357\,,\nonumber\\
		a_{3} &= -0.4710533854097566\,, & b_{3} &= -1.17767998417887\,,\nonumber\\
		a_{4} &= \frac 12 - \sum_{i=1}^{3}a_i\,, & b_{4} &= 1 - 2\sum_{i=1}^{3}b_i\,,\label{eq:7-yoshida}\\
		\mathrm{Eff}_6 &= \num{0.0036}\,.
	\end{align}
	\added{These coefficients are related to the original tuple $w_{1,2,3}$ via}
	\begin{align}
		a_{1} &= \frac 12 w_3\,, & b_{1} &= w_3\,,\label{eq:w_to_a_1}\\
		a_{2} &= \frac 12 \left(w_2+w_3\right)\,, & b_{2} &= w_2\,,\\
		a_{3} &= \frac 12 \left(w_1+w_2\right)\,, & b_{3} &= w_1\,.\label{eq:w_to_a_3}
	\end{align}
	\added{Note the reverted order.}

	\added{This decomposition is mainly interesting for historical reasons and its minimal number of cycles (very similar to Forest-Ruth~\eqref{eq:3-forest-ruth}). It should never be used because of its extremely poor efficiency.}
	
	\subsubsection{Blanes \& Moan (order $n=6$, $q=10$ cycles)}
	Similarly to its fourth order counterpart this decomposition
	\begin{align}
			a_{1} &= 0.0502627644003922\,, & b_{1} &= 0.148816447901042\,,\nonumber\\
			a_{2} &= 0.413514300428344\,, & b_{2} &= -0.132385865767784\,,\nonumber\\
			a_{3} &= 0.0450798897943977\,, & b_{3} &= 0.067307604692185\,,\nonumber\\
			a_{4} &= -0.188054853819569\,, & b_{4} &= 0.432666402578175\,,\nonumber\\
			a_{5} &= 0.54196067845078\,, & &\nonumber\\
			a_{6} &= 1 - 2\sum_{i=1}^{5}a_i\,, & b_{5} &= \frac 12 - \sum_{i=1}^{4}b_i\,,\label{eq:10-blanes6}\\
	\mathrm{Eff}_6 &= \num{0.51}\,,
	\end{align}
	has been derived by Blanes and Moan~\cite{BLANES2002313} in an extensive search over a large number of cycles. We are not aware of any better order $n=6$ scheme.

	\subsubsection{Suzuki (order $n=6$, $q=25$ cycles)}\label{sec:suzuki6}
	Apply the construction~\eqref{eq:construct_higher_order} starting from Suzuki's order $n=4$ scheme~\cite{Hatano_2005}. We quote this decomposition as a reference, but it is easily outperformed by Blanes and Moan's scheme of the same order~\eqref{eq:10-blanes6}.
	
	\subsubsection{Morales et al.\ (order $n=8$, $q=17$ cycles)}
	\added{Following Yoshida's method~\eqref{eq:construction_yoshida}, Morales et al.~\cite{morales2022greatly} derived the highly optimised order scheme}
	\begin{align}
			a_{1} &= 0.06391680493142055\,, & b_{1} &= 0.1278336098628411\,,\nonumber\\
			a_{2} &= 0.3446610312632028\,, & b_{2} &= 0.5614884526635645\,,\nonumber\\
			a_{3} &= 0.08874135982432522\,, & b_{3} &= -0.384005733014914\,,\nonumber\\
			a_{4} &= -0.1120890554644074\,, & b_{4} &= 0.1598276220860992\,,\nonumber\\
			a_{5} &= -0.1203317410978509\,, & b_{5} &= -0.4004911042818011\,,\nonumber\\
			a_{6} &= -0.1068973113931971\,, & b_{6} &= 0.1866964814954069\,,\nonumber\\
			a_{7} &= 0.2234502119222242\,, & b_{7} &= 0.2602039423490415\,,\nonumber\\
			a_{8} &= 0.2757888950144541\,, & b_{8} &= 0.2913738476798666\,,\nonumber\\
			a_{9} &= \frac 12 - \sum_{i=1}^{8}a_i\,, & b_{9} &= 1 - 2\sum_{i=1}^{8}b_i\,,\label{eq:17-morales}\\
	\mathrm{Eff}_8 &= \num{0.00018}\,, &&
	\end{align}
	where the coefficients have been converted analogously to equations~\eqref{eq:w_to_a_1} through~\eqref{eq:w_to_a_3}. We have translated its efficiency into our system using
	\begin{align}
		\mathrm{Eff}_n &= \frac{1}{\sqrt{2}\, q^n\, \chi}\,,\label{eq:eff_conversion}
	\end{align}
	with the error $\chi$ derived in~\cite{morales2022greatly} by means of numerical experiments.
	
	\added{This decomposition outclasses Yoshida's original 8th order ones by orders of magnitude. It is, in fact, so performant that Morales et al.\ conclude that 6th order schemes are never to be used, instead advocating to jump from 4th order straight to 8th. We find that in practice this decomposition has a very favourable error accumulation for high numbers of stages making 6th order methods obsolete in this case. However, in the case of fewer stages there is a significant region in which Blanes and Moan's order $n=6$ scheme~\eqref{eq:10-blanes6} is preferable.}

	\subsubsection{Order $n\ge8$, $q\ge 27$ cycles}\label{sec:very_high_orders}
	\added{We did not find any other algorithms of higher order $n\ge8$ worth mentioning explicitly (the schemes Yoshida~\cite{YOSHIDA1990262} derived originally are highly inefficient)}. However, the construction~\eqref{eq:construct_higher_order} allows to recursively elevate any of the algorithms above to the desired order. The smallest number of cycles achievable with this method is $q=3^{n/2-1}$ using $p=1$ in every iteration of equation~\eqref{eq:construct_higher_order}. We strongly discourage this construction due to its extremely poor performance. Instead we recommend taking an efficient high order scheme as a starting point and $p=2$ at every iteration (e.g.\ Blanes and Moan's order $n=6$ decomposition leads to $q=50$ for $n=8$). To our knowledge this particular combination has not been suggested before and it is significantly more efficient than Suzuki's recursive approach with $p=2$ requiring $q=125$ cycles for order $n=8$. We demonstrate below that for few stages it is even competitive with Morales et al.'s highly optimised formula~\eqref{eq:17-morales}.
	
	\added{Morales et al.~\cite{morales2022greatly} have derived the potentially most efficient scheme of order $n=10$ as yet. Again, they followed Yoshida's method and obtained a scheme with $q=33$ cycles ($m=16$ in their notation) that has an efficiency of $\mathrm{Eff}_{10}=\num{2.4e-8}$ (calculated via eq.~\eqref{eq:eff_conversion}). It stands to reason that elevating this scheme in a similar manner as we do with Blanes and Moan's~\eqref{eq:10-blanes6} will lead to order $n\ge12$ decompositions of unrivalled precision.}
	
	\subsection{Taylor expansion}\label{sec:taylor_expansion}
	Let us present a totally different, not strictly unitary, approach based on the truncated Taylor expansion
	\begin{align}
		\eto{H h} &= \lim_{k\rightarrow\infty}\sum_{i=0}^k\frac{\left(H h\right)^i}{i!}\,.
	\end{align}
	\added{This is a well-known technique (see e.g.\ Ref.~\cite{PhysRevLett.114.090502}) and in many cases it can be further enhanced, for instance through the use of a Chebyshev expansion~\cite{RevModPhys.78.275,PhysRevLett.118.196801}
	\begin{align}
		\eto{H h} &= \lim_{k\rightarrow\infty}\sum_{i=0}^k \mu_i T_i\left(\frac{H}{\Gamma}\right)\,,\quad \mu_0 = I_0(\Gamma h)\,,\; \mu_i = 2I_i(\Gamma h)
	\end{align}
	replacing the simple Taylor truncation, where the $T_i$ denote the Chebyshev polynomials, $I_i$ the modified Bessel functions of first kind and $\Gamma>0$ large enough (as defined below). As a rule, Chebyshev expansions are numerically more stable than their Taylor counterparts and need a smaller cutoff $k$ at a given step size $h$. On the other hand, Taylor expansions are easier to implement and generally more versatile. In particular, the Taylor expansion provides a good approximation for general operators $H$, whereas all eigenvalues of $H$ have to be real for a good convergence with Chebyshev. This also implies that the coefficients $\mu_i$ are different for real- and imaginary time evolution (use $I_i(\im \Gamma h) = \im^{-i} J_i(\Gamma h)$ with the (unmodified) Bessel functions $J_i$ in the former case).}
	
	The cutoff $k$ has to be chosen so that all eigenvalues $\lambda$ of $H$ are sufficiently suppressed
	\begin{align}
		\left|\frac{\left(\lambda_\text{max}(H) h\right)^{k}}{(k+1)!}\right| < \varepsilon\,,\label{eq:prec_req}
	\end{align}
	where $\varepsilon$ is the desired relative precision and the exponent is $k$ rather than $k+1$ because the error accumulates with every step and $\ordnung{h^{-1}}$ steps are needed. Technically a cutoff $k$ large enough guarantees super-exponential convergence independently of the step size $h$. In practice, however, the addends can span many orders of magnitude so that finite precision arithmetics leads to significant errors. It is therefore crucial to have at least some minimal information about the spectrum of $H$ before using this method, so that $h$ can be chosen such that all eigenvalues of $Hh$ have modulus not much larger than 1.
	
	In practice we recommend finding an upper bound $\Gamma\ge |\lambda_\text{max}(H)|$ for the absolute value of the largest eigenvalue of $H$. Then choose
	\begin{align}
		h &= \frac{1}{\Gamma}\,.\label{eq:choose_h_taylor}
	\end{align}
	Of course, a tighter upper bound increases the algorithm's efficiency.
	
	Once $h$ is chosen, $k$ has to be found accordingly. In the above case of the spectrum dominated by 1, double machine precision is reached for
	\begin{align}
		k &= 17\,.
	\end{align}

	More details on the derivation of the optimal step size $h$ and cutoff $k$ are provided in \Cref{sec:optimal_taylor_cutoff}.

	Note that the number $k$ is not equivalent with the number of cycles. This relation strongly depends on the implementation. For instance in the case we are discussing below (see \Cref{sec:numerical_experiments}) the implementation with $k=17$ has approximately the runtime of $q=3$ cycles.
	
	In this work we restrict ourselves to the direct classical application of the Taylor decomposition. However, it can also be used for quantum circuit optimisation, as was done in~\cite{classical_opt_circuits} up to first order $k=1$. It was also shown in~\cite{classical_opt_circuits} that the thus obtained circuits are close to optimal and significantly outperform the usual Trotter decompositions. We expect that a larger cutoff $k$ (depending on the target precision) can further improve the quantum circuit optimisation.
	
	\section{Numerical experiments}\label{sec:numerical_experiments}
	We use the Heisenberg model~\cite{PhysRevX.5.041047,10.21468/SciPostPhys.5.5.045,PhysRevLett.125.156601,ABANIN2021168415,PhysRevE.104.054105,KIEFEREMMANOUILIDIS2021168481} as a test for the decomposition schemes. In physics, it is an important and challenging task to compute the real time evolution of a quantum mechanical system as efficiently as possible.
	
	The Heisenberg model in particular has the additional advantage that it allows to directly estimate the efficiency a given decomposition would have on a quantum computer. Essentially our model consists of a spin-$\frac12$ chain of length $L$ with nearest neighbour coupling and small local perturbations. The spins are identical to qubits on a quantum device, so that the time evolution of the Heisenberg model can be simulated directly using a quantum circuit with $L$ qubits. The nearest neighbour interactions are then realised by two-qubit gates. As mentioned before, this also implies that a single time step requires $\ordnung{L}$ gates and therefore a Trotter decomposition into as many stages.
	
	In practice the Trotterizations introduced in \Cref{sec:deriving_schemes} have to provide a good approximation to the time evolution operator
	\begin{align}
		U(t) &= \eto{\im H t}\,,
	\end{align}
	where the Heisenberg Hamiltonian over a chain of length $L$ is defined as
	\begin{align}
		H &= \sum_{i=1}^L\left(J^x\sigma^x_i\sigma^x_{i+1} + J^y\sigma^y_i\sigma^y_{i+1} + J^z\sigma^z_i\sigma^z_{i+1} + h_i\sigma^z_{i}\right)\,.
	\end{align}
	The $\sigma^{x,y,z}$ denote the Pauli spin matrices, $J$ can be interpreted as a coupling vector and $h$ as a local external magnetic field in $z$-direction. Periodic boundary conditions are applied. For later convenience we define
	\begin{align}
		H^a_i &= J^a\sigma^a_i\sigma^a_{i+1} + \delta_{az}h_i\sigma^a_i\,.
	\end{align}
	
	In the following we distinguish two cases. One is the XZ-model where we set $J^x=J^z=1$ and $J^y=0$, the other the XXZ-model with $J^x=J^y=J^z=1$. The crucial difference from our algorithmic point of view is that the XZ-model can be split into just two stages, whereas the XXZ-model cannot. In both cases we randomly choose the $h_i\in[-0.1,0.1]$ from a uniform distribution. The non-zero values of $h$ guarantee that we do not measure any properties arising just in the case of high symmetry, it is kept small though in order to minimize the effect of a particular realisation.
	
	Given a decomposition scheme $S$, the time evolution operator can be approximated
	\begin{align}
		U(t) &= U(h)^{t/h}\\
		&= S(h)^{t/h} + \ordnung{h^n}\,,
	\end{align}
	where a small step size $h$ has to be chosen appropriately so that the error of order $n$ is small enough. The error is not uniquely defined since we are approximating a matrix and can in principle choose any matrix norm. Here we settled for the Frobenius norm
	\begin{align}
		\mathrm{error} &= \frac 1{\sqrt{N}} \left|\left|U(t)-S(h)^{t/h}\right|\right|_\text{F}\label{eq:err_definition}\\
		&\equiv \frac 1{\sqrt{N}} \sqrt{\tr\left[\left(U(t)-S(h)^{t/h}\right)^\dagger\left(U(t)-S(h)^{t/h}\right)\right]}\\
		&= \frac 1{\sqrt{N}} \sqrt{\sum_v \left|U(t)\cdot v - S(h)^{t/h}\cdot v\right|^2}\,,
	\end{align}
	where the sum runs over a basis $\{v\}$ of the corresponding $N$-dimensional vector space and $|\cdot|$ denotes the Euclidean norm. This allows us to calculate the error by means of successive application of $S(h)$ to a vector which is computationally much more efficient than calculating the explicit matrix form of $S(h)^{t/h}$. All the results presented in this section have been obtained for $L=6$ and $U(t)$ has been calculated using exact diagonalisation.
	
	It is not clear a priori which choice of stages provides the best approximation and depending on the type of simulation one might or might not be in the position to choose the stages. So we investigate the different approaches to split $H$ in addition to the influence of the different decompositions. The fundamental decision one has to make is whether to apply local gates (necessary e.g.\ for quantum computing) and to use the order
	\begin{align}
		S_{3L}(h) &= \eto{\im H^x_1c_1h}\eto{\im H^y_1c_1h}\eto{\im H^z_1c_1h}\eto{\im H^x_2c_1h}\eto{\im H^y_2c_1h}\eto{\im H^z_2c_1h}\cdots\eto{\im H^z_1d_q1h}\eto{\im H^y_1d_qh}\eto{\im H^x_1d_qh}\,,
	\end{align}
	or instead to group all the operators of similar type together (typically more efficient for classical computing, see eq.~\eqref{eq:cost_with_stages})
	\begin{align}
		\begin{split}
			S_3(h) &= \left(\prod_{i=1}^L\eto{\im H^x_ic_1h}\right)\left(\prod_{i=1}^L\eto{\im H^y_ic_1h}\right)\left(\prod_{i=1}^L\eto{\im H^z_ic_1h}\right)\dots\\
			&\quad\times\left(\prod_{i=1}^L\eto{\im H^z_id_qh}\right)\left(\prod_{i=1}^L\eto{\im H^y_id_qh}\right)\left(\prod_{i=1}^L\eto{\im H^x_id_qh}\right)\,.
		\end{split}
	\end{align}
	Note that all the exponentials above can be calculated exactly analytically and that operators of similar type commute
	\begin{align}
		\left[\eto{\im H^a_ih},\,\eto{\im H^a_jh}\right] &= 0\,.
	\end{align}
	Thus, $S_{3L}$ is a $3L$-stage decomposition and $S_3$ is a 3-stage decomposition. In the case of the XZ-model all the $y$-contributions drop out so that we obtain $2L$-stage and $2$-stage decompositions respectively.
	
	\added{Let us remark that we focus on real time evolution in this work for a number of reasons, including its relevance in quantum computing and its usefulness in providing a direct error estimate for a method at hand. However, all the presented time evolution methods can also be used for ground state search employing imaginary time evolution, relevant in classical computations such as tensor network based methods. Assuming an initial (random) state $v$ had non-zero overlap with the ground state, imaginary time evolution $\eto{-H t}v$ is guaranteed to converge to the correct ground state $v_0=\lim_{t\rightarrow\infty}\eto{-H t}v$ of $H$ with minimal eigenvalue (energy) $E_0$, i.e.\ $Hv_0=E_0v_0$. The rate of this convergence will be dominated by the gap to the first excited state $v_1$ with eigenvalue $E_1$ and scale as $\varepsilon_E(t) = \eto{-(E_1-E_0) t}$. For high accuracy, the error $\varepsilon_T(h)$ introduced by the (Trotter) decomposition scheme at chosen step size $h$ has to be small $\varepsilon_T(h) \ll \varepsilon_E(t)$ at any given time $t$, until both errors decrease below a given target error. Thus, again higher performance can be achieved with schemes featuring smaller discretisation error $\varepsilon_T(h)$. An adaptive method for the best choice of the step size $h$ is presented e.g.\ in Ref.~\cite{my_tensor_networks} (Sec.\ III.B.2).}
	
	\subsection{Real time evolution of the Heisenberg model}
	
	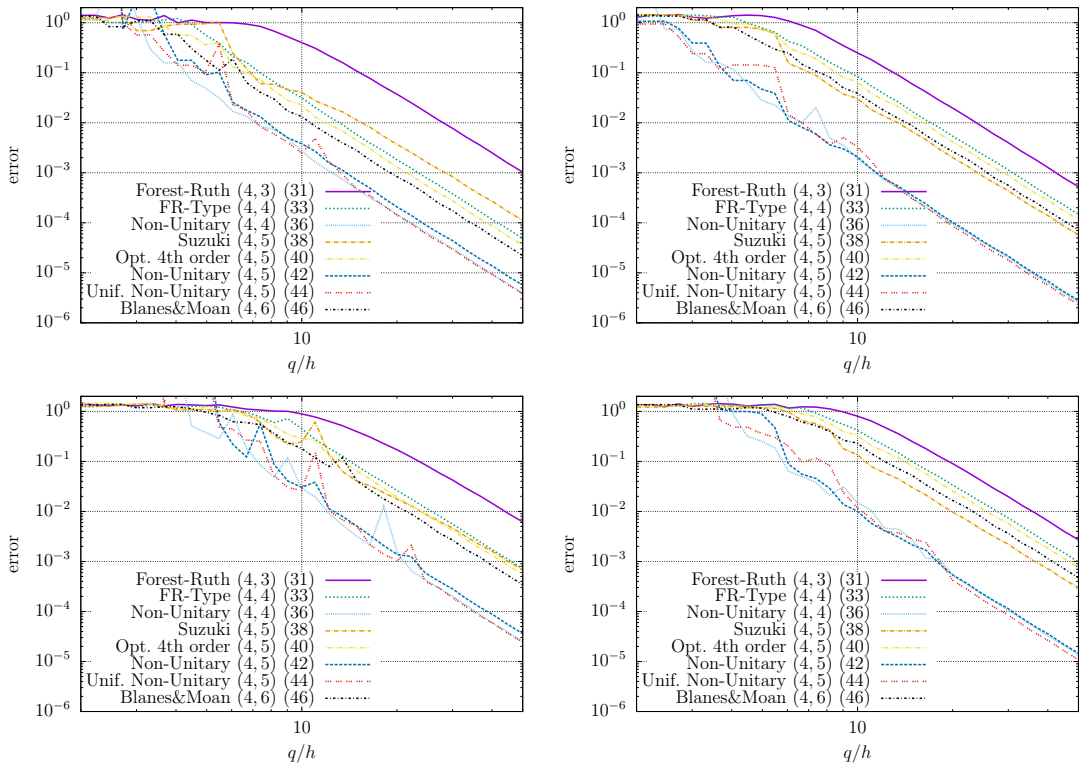
\begin{figure}[ht]
		\centering
		\resizebox{0.98\textwidth}{!}{{\large\input{simulation/benchmark/2-stage_fix-t_ord4}\input{simulation/benchmark/2L-stage_fix-t_ord4}}}\\
		\resizebox{0.98\textwidth}{!}{{\large\input{simulation/benchmark/3-stage_fix-t_ord4}\input{simulation/benchmark/3L-stage_fix-t_ord4}}}
		\caption{Collection of order $n=4$ schemes. Error~\eqref{eq:err_definition} of the difference between the exact time evolution operator and the respective decomposition as a function of computational cost, i.e.\ number of cycles $q$ divided by the time step $h$. Behind the name of the scheme we write the order / cycles tuple $(n,q)$. Top: XZ-model split into 2 stages (left) and $2L$ stages (right). Bottom: XXZ-model split into 3 stages (left) and $3L$ stages (right).}\label{fig:err_const_t_ord4}
	\end{figure}

	\begin{figure}[ht]
		\centering
		\resizebox{0.98\textwidth}{!}{{\large\input{simulation/benchmark/2-stage_fix-t_ordRest}\input{simulation/benchmark/2L-stage_fix-t_ordRest}}}\\
		\resizebox{0.98\textwidth}{!}{{\large\input{simulation/benchmark/3-stage_fix-t_ordRest}\input{simulation/benchmark/3L-stage_fix-t_ordRest}}}
		\caption{Collection of order $n\neq 4$ schemes. Error~\eqref{eq:err_definition} of the difference between the exact time evolution operator and the respective decomposition as a function of computational cost, i.e.\ number of cycles $q$ divided by the time step $h$. Behind the name of the scheme we write the order / cycles tuple $(n,q)$. The abbreviation `BM6+S' stands for the Blanes and Moan scheme~\eqref{eq:10-blanes6} of order $n=6$ elevated to order $n=8$ using Suzuki's method~\eqref{eq:construct_higher_order}. To guarantee a fair runtime comparison, we set $q=3$ for the Taylor decomposition. Top: XZ-model split into 2 stages (left) and $2L$ stages (right). Bottom: XXZ-model split into 3 stages (left) and $3L$ stages (right).}\label{fig:err_const_t_ord_rest}
	\end{figure}
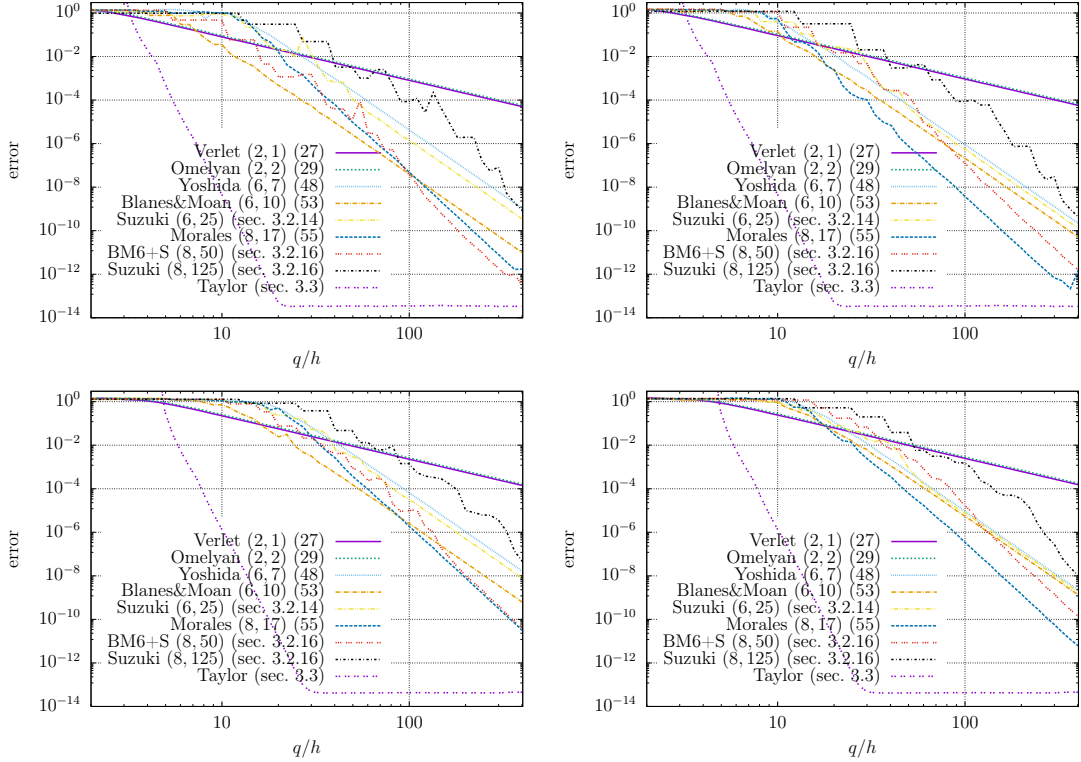

\begin{figure}[ht]
		\centering
		\resizebox{0.98\textwidth}{!}{{\large\input{simulation/benchmark/2-stage_fix-t_opt}\input{simulation/benchmark/3L-stage_fix-t_opt}}}
		\caption{Collection of the best schemes of each order. Error~\eqref{eq:err_definition} of the difference between the exact time evolution operator and the respective decomposition as a function of computational cost, i.e.\ number of cycles $q$ divided by the time step $h$. Behind the name of the scheme we write the order / cycles tuple $(n,q)$. The abbreviation `BM6+S' stands for the Blanes and Moan scheme~\eqref{eq:10-blanes6} of order $n=6$ elevated to order $n=8$ using Suzuki's method~\eqref{eq:construct_higher_order}. To guarantee a fair runtime comparison, we set $q=3$ for the Taylor decomposition. XZ-model split into 2 stages (left) and XXZ-model split into $3L$ stages (right).}\label{fig:err_const_t_opt}
	\end{figure}
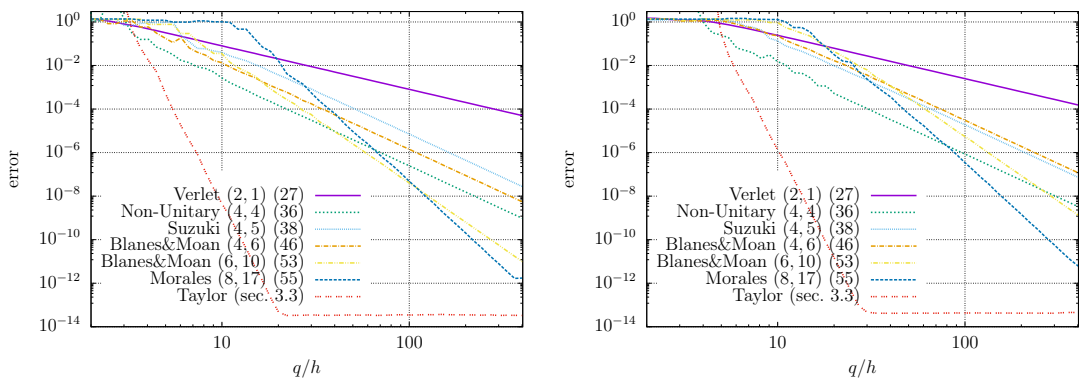

	For now we set $t=10$ constant and analyse the error of the different decomposition schemes as a function of computational effort $\frac qh$ (it grows linearly with the number of cycles $q$ per step and is inversely proportional to the step size $h$). The results have been summarised in figure~\ref{fig:err_const_t_ord4} for the order $n=4$ schemes and in figure~\ref{fig:err_const_t_ord_rest} for all the other schemes. In addition, figure~\ref{fig:err_const_t_opt} shows a comparison of the best respective schemes of each type.
	
	Asymptotically with small time steps, i.e.\ large computational cost, all the schemes display the theoretically predicted decay governed by their respective order $n$.
	The only exception is the Taylor decomposition that is not polynomial (see sec.~\ref{sec:taylor_expansion}). After a very steep drop it reaches a constant error dominated by machine precision. In fact, exact diagonalisation and Taylor expansion have errors in the same order of magnitude, both dictated purely by rounding errors due to the double precision arithmetics used here.
	
	Towards large time steps the errors of all the unitary decomposition schemes plateau out close to 1. This is due to all eigenvalues of a unitary matrix having modulus 1 and thus the matrix having Frobenius norm $\sqrt N$ (following the central limit theorem). A decomposition using very large time steps will be completely wrong and can be treated like a pseudorandom matrix. But as long as it remains unitary, the Frobenius norm of its difference to the correct matrix will still be of order $\sqrt N$ and the normalised error therefore close to 1. The non-unitary schemes in contrast are not bounded in such a way, so that their error diverges as $q/h\rightarrow0$.
	
	Let us now compare the empirical evidence to the respective efficiencies predicted in \Cref{sec:list_of_schemes}. It makes sense to start with the top left panels of figures~\ref{fig:err_const_t_ord4} and~\ref{fig:err_const_t_ord_rest} because the theoretical efficiency is only directly applicable to decompositions with two stages.
	
\begin{figure}[ht]
		\centering
		\resizebox{0.98\textwidth}{!}{{\large\input{simulation/benchmark/2-stage_fix-dt_ord4}\input{simulation/benchmark/3L-stage_fix-dt_ord4}}}\\
		\resizebox{0.98\textwidth}{!}{{\large\input{simulation/benchmark/2-stage_fix-dt_ordRest}\input{simulation/benchmark/3L-stage_fix-dt_ordRest}}}
		\caption{Error~\eqref{eq:err_definition} per time of the difference between the exact time evolution operator and the respective decomposition as a function of time. The respective step size has been chosen such that all the schemes have the same runtime per evolution time. Behind the name of the scheme we write the order / cycles tuple $(n,q)$. The abbreviation `BM6+S' stands for the Blanes and Moan scheme~\eqref{eq:10-blanes6} of order $n=6$ elevated to order $n=8$ using Suzuki's method~\eqref{eq:construct_higher_order}. To guarantee a fair runtime comparison, we set $q=3$ for the Taylor decomposition. Top: order $n=4$ schemes, bottom: $n\neq 4$ schemes. XZ-model split into 2 stages (left) and XXZ-model split into $3L$ stages (right).}\label{fig:err_const_dt}
	\end{figure}
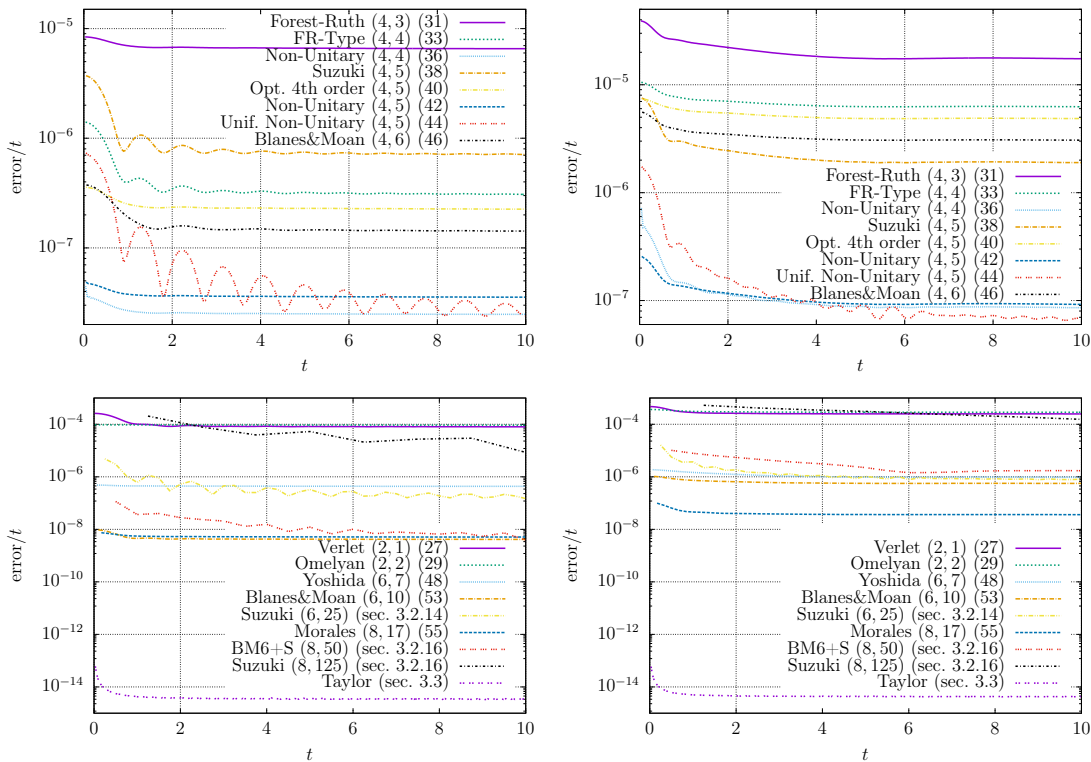

	From \Cref{sec:list_of_schemes} we expect the following order for the order $n=4$ schemes from least to most efficient, that is from largest to smallest asymptotic error: Forest-Ruth~\eqref{eq:3-forest-ruth}, Suzuki~\eqref{eq:5-suzuki4}, Omelyan's Forest-Ruth-type~\eqref{eq:4-fr-type}, uniform non-unitary~\eqref{eq:5-non-unitary-const}, Blanes and Moan~\eqref{eq:6-blanes4}, our optimised 4th order~\eqref{eq:5-opt-4th-ord}, non-unitary ($q=4$)~\eqref{eq:4-non-unitary1}, and non-unitary ($q=5$)~\eqref{eq:5-non-unitary2}. Furthermore we expect large gaps between Forest-Ruth, the other unitary schemes, and the non-unitary schemes. Most of these predictions are confirmed by the data, in particular the gaps are very pronounced, emphasising once more that Forest-Ruth should never be used whereas the non-unitary schemes should be used whenever possible. There are, however, two notable deviations from the predictions. For one, in practice Blanes and Moan's decomposition~\eqref{eq:6-blanes4} outperforms the theoretically most efficient scheme~\eqref{eq:5-opt-4th-ord}. Similarly, the $q=4$ cycles non-unitary~\eqref{eq:4-non-unitary1} and even the uniform non-unitary~\eqref{eq:5-non-unitary-const} decompositions outperforms the one with $q=5$ cycles~\eqref{eq:5-non-unitary2}. Both deviations from the expected order become clear when looking at the early time $t<2$ evolution of the error in the top left panel of figure~\ref{fig:err_const_dt}. The errors start out as expected from the theoretical efficiencies, but some decompositions accumulate their errors almost linearly immediately while others have a first sub-linear region before reaching the universal linear asymptotic behaviour. Blanes and Moan as well as the two former non-unitary schemes accumulate their error particularly favourably and therefore outperform their respective counterparts.
	
	The order of the remaining $n\neq 4$ schemes offers fewer surprises. Quite expectedly, the 2nd order decompositions are good at minimal computational effort, they are quickly outperformed by the 6th order schemes (Blanes and Moan~\eqref{eq:10-blanes6} being better than Suzuki, sec.~\ref{sec:suzuki6}, better than Yoshida~\eqref{eq:7-yoshida}), which again yield to the 8th order schemes at very small time steps. As explained in \Cref{sec:very_high_orders}, the order $n=8$ decompositions apart from Morales et al.~\eqref{eq:17-morales} are constructed from order $n=6$ decompositions using Suzuki's method~\eqref{eq:construct_higher_order} and the version starting out with Blanes and Moan's 6th order scheme is clearly superior to the canonical one starting out with Suzuki's scheme. The Taylor expansion outperforms all the polynomial decompositions by many orders of magnitude, practically over the complete range of time steps that yield any kind of useful results. The only exception from the behaviour predicted by the theoretical efficiency is that Verlet~\eqref{eq:1-leap-frog} and Omelyan~\eqref{eq:2-omelyan2} have practically identical errors. Again, this is explained by the more favourable error accumulation of Verlet's decomposition as shown in the bottom left panel of figure~\ref{fig:err_const_dt}.
	
	For more than two stages (all but the top left panels of figs.~\ref{fig:err_const_t_ord4} and~\ref{fig:err_const_t_ord_rest}) the picture does not change dramatically. As a general rule, the error of the non-unitary schemes decreases while the error of the unitary schemes increases, but for some more than for others. This broadens the gap between non-unitary and unitary decompositions and it makes the unitary schemes move closer together. There are only two changes worth mentioning explicitly. For many stages $\Lambda\sim L$ Suzuki's 4th order decomposition features such a favourable error accumulation (top right panel of fig.~\ref{fig:err_const_dt}) that it outperforms all the other unitary schemes. \added{Similarly, Morales et al.'s 8th order has very favourable behaviour in the case of many stages, significantly outperforming the elevated 8th order `BM6+S' scheme and even all the 6th order decompositions.}
	
	\section{Conclusion}
	In this work, we addressed three important questions concerning the Suzuki-Trotter decomposition (or splitting methods) of exponential operators. First, we proved in \Cref{sec:2-to-any-stages} that a decomposition scheme for two stages automatically defines a decomposition of the same order for any number of stages and we provided a recipe for the corresponding transformation in the equations~\eqref{eq:any-stage-decomposition} to~\eqref{eq:cq_dq}. Next, in \Cref{sec:deriving_schemes} we presented a comprehensive overview of different decomposition schemes (both well-known and efficient ones) with real and complex coefficients up to order $n\le4$ as well as some higher order schemes and the highly efficient Taylor expansion method. Last but not least, in \Cref{sec:numerical_experiments} we demonstrated on the example of the Heisenberg model real time evolution how the different Trotterizations perform in practice noting few but significant deviations from the theoretically predicted efficiency.
	
	The theorem from \Cref{sec:2-to-any-stages} is essential for the construction of efficient higher order Trotter schemes applicable to problems that require gates including tensor networks and quantum computing, or more generally splitting of the exponent in $\eto{tH}$ into more than two operators $H=\sum_kA_k$. The outstanding performance of decompositions developed for two stages like our non-unitary \cref{eq:4-non-unitary1,eq:5-non-unitary2,eq:5-non-unitary-const} and Blanes \& Moan's schemes \cref{eq:6-blanes4,eq:10-blanes6} applied to $2L$- and $3L$-stage splittings demonstrate its usefulness.
	
	We recall that for instance the quantum spin chain simulated in \Cref{sec:numerical_experiments} can be identified with qubits on a quantum device where the splitting into $\ordnung{L}$ stages canonically defines the corresponding quantum gates. Thus any improvement in multi-stage simulations directly translates to improved quantum computations.
		
	From the decompositions presented in \Cref{sec:list_of_schemes,sec:taylor_expansion}, we summarise the best ones of every order in figure~\ref{fig:err_const_t_opt} for the two extreme cases of two stages (left) and $3L$ stages (right). For the optimal algorithm choice one should always decide on a desired target precision first, and then choose the leftmost decomposition at the given error level, that is the one reaching this precision with the least computational effort. Of course, constraints like unitarity have to be considered as well.
	
	On a general note, it makes sense to opt for a scheme with less cycles if two schemes perform similarly well otherwise, because fewer cycles allow for a finer sampling of time and a more precise tuning of the time step size. It is also always advised to use the minimal possible number of stages, simply to avoid redundant computations (see eq.~\eqref{eq:cost_with_stages}).
	
	Once again we emphasise the super-exponential superiority of the Taylor decomposition method proposed in \Cref{sec:taylor_expansion} that should be used whenever possible.
	
\begin{figure}
		\centering
		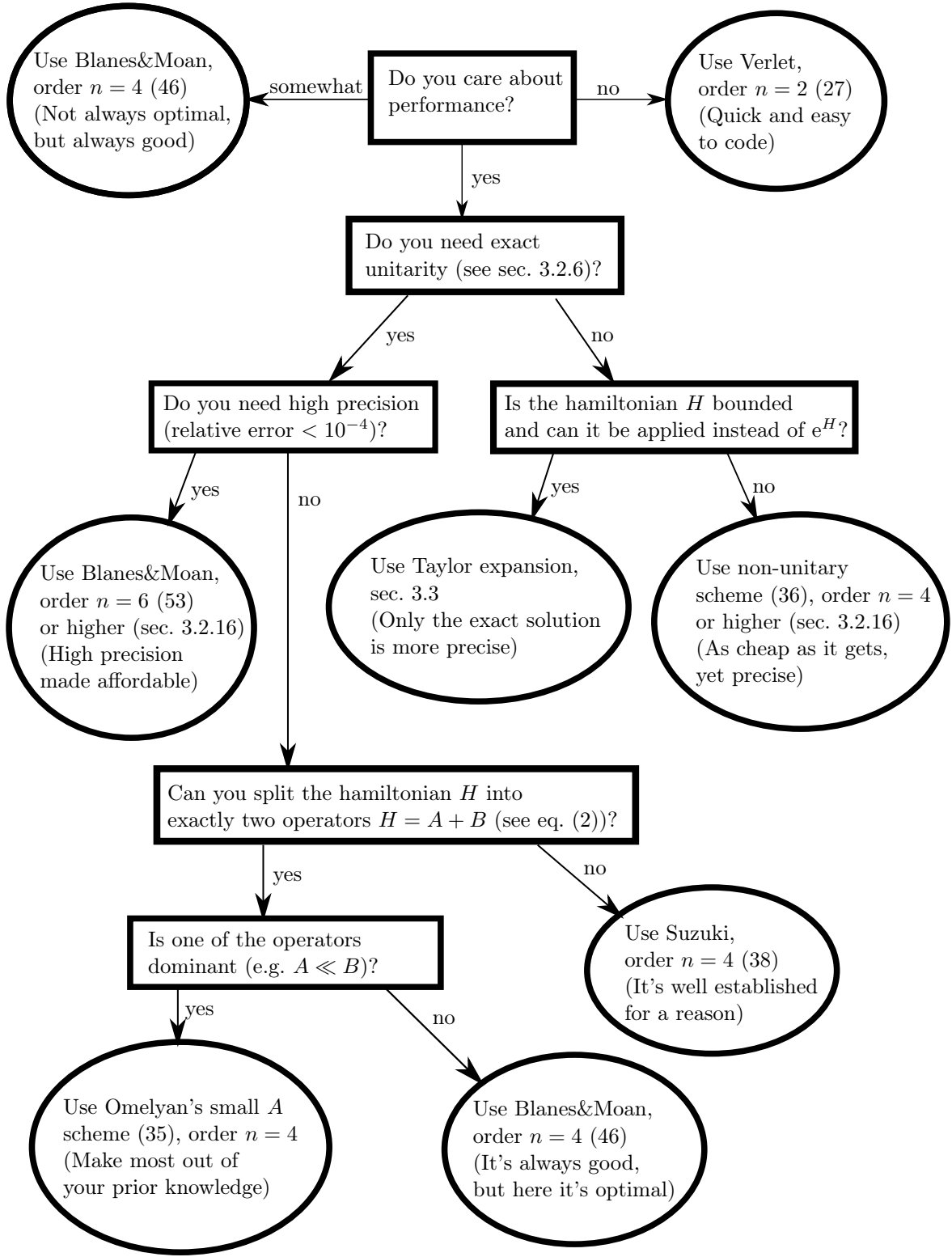
		\caption{This flow-chart provides a simple way to find the algorithm (close to) optimal for the particular requirements of your Trotterization problem at hand. We strongly advise to read the complete section describing the algorithm you arrived at, as it might contain hints allowing further optimisation. As presented, the decompositions are designed for exactly 2 operators/stages, but you can use eqs.~(\ref{eq:any-stage-decomposition}-\ref{eq:cq_dq}) to adapt them for any number of stages.}\label{fig:flow-chart}
	\end{figure}

	We have summarised the decision tree leading to the best decomposition scheme in a given scenario in the flow-chart~\ref{fig:flow-chart}. Have a look at it. It might be useful for you, even (or especially) if you are not interested in reading this paper completely.
	
	The numerical experiments on the Heisenberg model we conducted in \Cref{sec:numerical_experiments} are qualitatively representative, but the results might look quantitatively different for other models. More crucially, the Frobenius norm we analysed in this work is the least forgiving quality measure capturing every kind of deviations from the exact result. It was chosen for exactly this reason, but in many cases it significantly overestimates the error that is physically relevant, for instance of observables like the energy~\cite{PhysRevX.11.011020,Blanes:2022}. This means that our results do not replace case by case studies of a particular problem and its observables if one needs to find the single best decomposition scheme and the optimal step size.
	
	While in \Cref{sec:list_of_schemes} we focused mostly on the theoretical efficiency of the decomposition schemes, in \Cref{sec:numerical_experiments} we observed that it is not always sufficient to predict the performance of a Trotterization scheme in practice. The theoretical efficiency relies purely on the norm of the leading order error in a vector space spanned by the different commutators of exactly two operators (2-stage splitting). It does not take into account how the error accumulates over time, nor how it changes for a different number of stages. Therefore the theoretical efficiency predicts the error of a 2-stage splitting after a single time step very accurately, but it only correlates with the de facto performance over longer times and with more stages. We find that both, longer times and more stages, strongly favour decomposition schemes with similar (real parts of the) coefficients in every sub-step. This is particularly vividly demonstrated by the uniform non-unitary scheme~\eqref{eq:5-non-unitary-const} with constant real parts of the coefficients $c_i$, $d_i$. Said decomposition can outperform the theoretically more than ten times superior non-unitary scheme~\eqref{eq:5-non-unitary2} due to its peculiar error accumulation.
	
	We expect there to be a sweet-spot combining high efficiency with favourable error accumulation that outperforms both the theoretically most efficient and the uniform coefficient schemes. A possible approach for finding a decomposition of this kind would be to assign a penalty to the non-uniformity of the sub-steps and maximise the efficiency including that penalty. However, more research on this topic is needed from the theoretical as well as the numerical directions. We intend to pursue this line of investigations.
	
	\section*{Code and Data}
	The entire code and data required to reproduce the results and plots of this paper have been published under open access and can be found in~\cite{johann_ostmeyer_2022_7268893}. We used \texttt{Mathematica}~\cite{Mathematica} for the efficiency calculations of the different Suzuki-Trotter decomposition schemes and for their optimisation. The numerical experiments have been implemented in \texttt{C} with a front-end in \texttt{R}~\cite{r_language}.
	
	\section*{Acknowledgements}
	This work was funded in part by the STFC Consolidated Grant ST/T000988/1.
	The author thanks Evan Berkowitz, Pavel Buividovich, Tom Luu, and Carsten Urbach for their helpful comments.
	Special thank goes to David Luitz without whom the author would not have found out that significant results of this work had been published before.
	No thanks whatsoever to Wikipedia for not having an article on Trotterization. Anyone reading this is strongly encouraged to write it.
	
	\printbibliography
	
	\appendix
	
	\section{Choosing correct step size and cutoff in the Taylor expansion}\label{sec:optimal_taylor_cutoff}
	
	In the following we will understand the super-exponential convergence of the Taylor expansion in more detail. Explicitly we will show that \textit{the optimal Taylor decomposition at a given level of precision maximises the step size $h$ and the cutoff $k$}, provided additional constraints like finite precision arithmetics are taken into account. For this we will derive the computational cost and show that it decreases monotonously with growing $h$ at fixed precision. Then we will show how the situation changes with finite precision arithmetics and derive the most efficient tuple ($h$, $k$) in that case. Finally we will consider the alternative problem when powers of the Hamiltonian have to be calculated explicitly, i.e.\ $H^k$ is required and not just the $k$-fold application to a vector $v$ of the form $H\cdot(H\cdots(H\cdot v)\cdots)$.
	A summary provides a recipe to obtain the optimal tuple ($h$, $k$).
	
	\subsection{Cost function}
	The cost of using a $k$-th order Taylor expansion is readily given by the number of iterations per step divided by the step size $h$, that is
	\begin{align}
		\text{cost}_1 &\coloneqq \frac kh\label{eq:cost_1_taylor}
	\end{align}
	for equally expensive iterations. More generally, the sum of all the iterations' computational costs divided by $h$ defines the total cost. In particular, if $H^k$ has to be calculated directly and $H$ is sparse, i.e.\ some $M$ entries per row are non-zero where usually $M$ is of order $L$, then the $i$-th iteration's cost scales as $\min(M^i,N)$. We neglect all but the last iteration and obtain
	\begin{align}
		\text{cost}_M & \coloneqq \frac{M^k}{h}\;,\quad \text{for } 1\ll M\ll N\,.\label{eq:cost_m_taylor}
	\end{align}

	In order to satisfy the precision requirements from equation~\eqref{eq:prec_req}, we find that for any given $k$ the step size has to be chosen at most
	\begin{align}
		\tilde h\coloneqq \left|\lambda_\text{max}(H)h\right| &= \sqrt[k]{(k+1)!\,\varepsilon}\label{eq:h_exact}\\
		&\sim \frac {k+1}{e} \left(\frac{\sqrt{2\pi}}{e}(k+1)^{\frac{3}{2}}\varepsilon\right)^{\frac 1k}\\
		&\sim \frac {k+1}{e} \sqrt[k]{\varepsilon}\,,\label{eq:h_monotonous}
	\end{align}
	where we applied Stirling's formula in the last step and used the Bachmann-Landau notation with `$\sim$' as `in the order of' implying a ratio converging to 1 with large $k$.
	
	\subsection{Repeated matrix-vector multiplications}
	Plugging this result back into the first cost function~\eqref{eq:cost_1_taylor} yields
	\begin{align}
		\text{cost}_1 &\propto \frac{k}{k+1}\,\varepsilon^{-\frac{1}{k}}\,,
	\end{align}
	where irrelevant factors of eigenvalues and $e$ have been absorbed in the proportionality.
	The logarithm of this result is easier to understand
	\begin{align}
		\ln \text{cost}_1 &= -\frac 1k \left(1+\ln\varepsilon\right) + \ordnung{k^{-2}} + \text{const.}
	\end{align}
	For large enough $k$ the cost will decrease monotonously with $k$ as long as $\varepsilon<1/e$ since then the $k$-th root dominates the expression. Therefore larger $k$ and thus larger $h$ always reduce the cost.\footnote{This is not true any more when corrections of order $k^{\frac{3}{2k}}$ to eq.~\eqref{eq:h_monotonous} are taken into account, but these corrections only change the picture for $k\gtrsim\varepsilon^{-2/3}$ much larger than required in any realistic scenario.} Note that the dominance of the $k$-th root term is a direct consequence of the super-exponential behaviour of the factorial.
	
	Now, as mentioned in \Cref{sec:taylor_expansion}, machine precision $\varepsilon_\text{MP}$ becomes crucially important at some point because the intermediate iterations might feature very large terms compared to the zeroth addend which is always identity. To quantify this, we first have to identify the iteration $i_\text{max}$ with largest absolute contribution. We consider the logarithm of the $i$-th iteration corresponding to the largest eigenvalue
	\begin{align}
		I(i) &\coloneqq\ln\left(\frac{\tilde{h}^i}{i!}\right)\\
		&= i\left(1+\ln \tilde h - \ln i\right) + \ordnung{\ln i}\\
		\Rightarrow I'(i) &\approx \ln\tilde{h} - \ln i\\
		\Rightarrow i_\text{max} &\approx \tilde h\,.
	\end{align}
	Once more we used Stirling's formula and we obtained the maximum from the zero of the first derivative. In the first approximation we dropped the logarithmic corrections and the last approximation also includes the constraint that $i_\text{max}$ has to be an integer.
	
	Overall the relative error due to rounding effects is then
	\begin{align}
		\text{err}_\text{MP} &\equiv \varepsilon_\text{MP}\, \frac{\tilde{h}^{i_\text{max}}}{i_\text{max}!}\\
		&\approx \epsilon_\text{MP}\, \eto{\tilde{h}}
	\end{align}
	and we require it to be smaller than the target precision as well. Thus we demand
	\begin{align}
		\tilde h &= \max\left(1,\ln\frac{\varepsilon}{\varepsilon_\text{MP}}\right)\,,\label{eq:chose_h_tilde}
	\end{align}
	where the maximum is inserted because Stirling's formula has large deviations from the factorial for small arguments and the zeroth addend of the Taylor expansion is always largest if $\tilde h\leq 1$. We chose $\tilde h = 1$ in equation~\eqref{eq:choose_h_taylor} for exactly the reason that it is the largest step size (i.e.\ minimising cost) that allows machine precise calculations.
	
	After $\tilde{h}$ has been chosen according to equation~\eqref{eq:chose_h_tilde}, the optimal cutoff $k$ is provided by the (numerical) solution of the equation
	\begin{align}
		e\max\left(1,\ln\frac{\varepsilon}{\varepsilon_\text{MP}}\right) &= (k+1)\sqrt[k]{\varepsilon}\label{eq:get_opt_k}
	\end{align}
	obtained by setting equal the two relations~\eqref{eq:h_monotonous} and~\eqref{eq:chose_h_tilde} for $\tilde h$.

	\Cref{sec:taylor_expansion} can be considered as an example for this procedure where we set $\varepsilon=\varepsilon_\text{MP}\equiv2^{-52}$ (in double precision), therefore according to equation~\eqref{eq:chose_h_tilde} we obtained $\tilde h = 1$ or $h=1/\Gamma\le|1/\lambda_\text{max}(H)|$. Finally, the numerical solution of equation~\eqref{eq:get_opt_k} yields $k\approx\num{18.4}$ (which of course has to be rounded to an integer) not too far from the exact solution of equation~\eqref{eq:prec_req} that allows to pinpoint $k=17$.
	
	\subsection{Explicit matrix powers}
	The situation is crucially different when powers of $H$ are needed explicitly and thus $\text{cost}_M$ has to be minimised. Machine precision hardly matters in this case, instead the relation~\eqref{eq:h_monotonous} for $\tilde h$ has to be plugged into $\text{cost}_M$. Again, we consider the cost's logarithm
	\begin{align}
		\ln \text{cost}_M &= k\ln M - \ln\left(k+1\right) - \frac{1}{k}\ln\varepsilon + \ordnung{\frac 1k\ln k} + \text{const.}\label{eq:cost_m_plugged_in}
	\end{align}
	with a unique positive minimum at
	\begin{align}
		k_\text{min} &= \sqrt{\log_M\left(\varepsilon^{-1}\right)} + \ordnung{1/\ln M}\,.\label{eq:min_k_power_H}
	\end{align}
	In practice it is advisable to evaluate the cost~\eqref{eq:cost_m_plugged_in} explicitly at some integer values because rounding might not provide the best result.
	
	It turns out that the optimal cutoff is very close to one when powers of $H$ are required. For instance even with a small value of $M=10$ and high desired precision $\varepsilon=\varepsilon_\text{MP}\equiv2^{-52}$ the optimal cutoff is $k=4$. A low target precision of say $1\%$ and some intermediate $M=100$ clearly results in $k=1$.
	
	\subsection{Summary}
	Two cases are to be distinguished:
	
	\begin{enumerate}[a)]
		\item Repeated matrix-vector multiplications, then proceed as follows:
		\begin{enumerate}[1.]
			\item Calculate $\tilde h$ from eq.~\eqref{eq:chose_h_tilde} for given target precision $\varepsilon$ and machine precision $\varepsilon_\text{MP}$.
			\item Set the step size $h=\tilde h/\Gamma$, where $\Gamma$ is a (close) upper bound for the modulus of the eigenvalues of the matrix $H$.
			\item Solve eq.~\eqref{eq:get_opt_k} numerically to obtain the cutoff $k$.
		\end{enumerate}
		\item Explicit matrix power calculations, then proceed as follows:
		\begin{enumerate}[1.]
			\item Calculate the cutoff $k$ from eq.~\eqref{eq:min_k_power_H} for given target precision $\varepsilon$ and $M$ non-zero entries per row of the matrix $H$.
			\item Obtain $\tilde h$ from eq.~\eqref{eq:h_exact}.
			\item Set the step size $h=\tilde h/\Gamma$, where $\Gamma$ is a (close) upper bound for the modulus of the eigenvalues of the matrix $H$.
		\end{enumerate}
	\end{enumerate}
\end{document}

%% file: inkscape/2_stage_a-b.pdf_tex
\begingroup%
  \makeatletter%
  \providecommand\color[2][]{%
    \errmessage{(Inkscape) Color is used for the text in Inkscape, but the package 'color.sty' is not loaded}%
    \renewcommand\color[2][]{}%
  }%
  \providecommand\transparent[1]{%
    \errmessage{(Inkscape) Transparency is used (non-zero) for the text in Inkscape, but the package 'transparent.sty' is not loaded}%
    \renewcommand\transparent[1]{}%
  }%
  \providecommand\rotatebox[2]{#2}%
  \newcommand*\fsize{\dimexpr\f@size pt\relax}%
  \newcommand*\lineheight[1]{\fontsize{\fsize}{#1\fsize}\selectfont}%
  \ifx\svgwidth\undefined%
    \setlength{\unitlength}{397.60184867bp}%
    \ifx\svgscale\undefined%
      \relax%
    \else%
      \setlength{\unitlength}{\unitlength * \real{\svgscale}}%
    \fi%
  \else%
    \setlength{\unitlength}{\svgwidth}%
  \fi%
  \global\let\svgwidth\undefined%
  \global\let\svgscale\undefined%
  \makeatother%
  \begin{picture}(1,0.19642491)%
    \lineheight{1}%
    \setlength\tabcolsep{0pt}%
    \put(-0.00208771,0.17731621){\color[rgb]{0,0,0}\makebox(0,0)[lt]{\lineheight{1.25}\smash{\begin{tabular}[t]{l}$B$\end{tabular}}}}%
    \put(-0.00208771,0.05281971){\color[rgb]{0,0,0}\makebox(0,0)[lt]{\lineheight{1.25}\smash{\begin{tabular}[t]{l}$A$\end{tabular}}}}%
    \put(0,0){\includegraphics[width=\unitlength,page=1]{2_stage_a-b.pdf}}%
    \put(0.08737721,0.00593156){\color[rgb]{0,0,0}\makebox(0,0)[lt]{\lineheight{1.25}\smash{\begin{tabular}[t]{l}$a_1$\end{tabular}}}}%
    \put(0.2760083,0.00593156){\color[rgb]{0,0,0}\makebox(0,0)[lt]{\lineheight{1.25}\smash{\begin{tabular}[t]{l}$a_2$\end{tabular}}}}%
    \put(0.18169276,0.00593156){\color[rgb]{0,0,0}\makebox(0,0)[lt]{\lineheight{1.25}\smash{\begin{tabular}[t]{l}$b_1$\end{tabular}}}}%
    \put(0.37032385,0.00593156){\color[rgb]{0,0,0}\makebox(0,0)[lt]{\lineheight{1.25}\smash{\begin{tabular}[t]{l}$b_2$\end{tabular}}}}%
    \put(0.76267629,0.00593156){\color[rgb]{0,0,0}\makebox(0,0)[lt]{\lineheight{1.25}\smash{\begin{tabular}[t]{l}$b_q$\end{tabular}}}}%
    \put(0.66836069,0.00593156){\color[rgb]{0,0,0}\makebox(0,0)[lt]{\lineheight{1.25}\smash{\begin{tabular}[t]{l}$a_q$\end{tabular}}}}%
    \put(0.85699173,0.00593156){\color[rgb]{0,0,0}\makebox(0,0)[lt]{\lineheight{1.25}\smash{\begin{tabular}[t]{l}$a_{q+1}$\end{tabular}}}}%
  \end{picture}%
\endgroup%

%% file: inkscape/2_stage_c-d.pdf_tex
\begingroup%
  \makeatletter%
  \providecommand\color[2][]{%
    \errmessage{(Inkscape) Color is used for the text in Inkscape, but the package 'color.sty' is not loaded}%
    \renewcommand\color[2][]{}%
  }%
  \providecommand\transparent[1]{%
    \errmessage{(Inkscape) Transparency is used (non-zero) for the text in Inkscape, but the package 'transparent.sty' is not loaded}%
    \renewcommand\transparent[1]{}%
  }%
  \providecommand\rotatebox[2]{#2}%
  \newcommand*\fsize{\dimexpr\f@size pt\relax}%
  \newcommand*\lineheight[1]{\fontsize{\fsize}{#1\fsize}\selectfont}%
  \ifx\svgwidth\undefined%
    \setlength{\unitlength}{348.4532983bp}%
    \ifx\svgscale\undefined%
      \relax%
    \else%
      \setlength{\unitlength}{\unitlength * \real{\svgscale}}%
    \fi%
  \else%
    \setlength{\unitlength}{\svgwidth}%
  \fi%
  \global\let\svgwidth\undefined%
  \global\let\svgscale\undefined%
  \makeatother%
  \begin{picture}(1,0.26559953)%
    \lineheight{1}%
    \setlength\tabcolsep{0pt}%
    \put(-0.00238218,0.22167706){\color[rgb]{0,0,0}\makebox(0,0)[lt]{\lineheight{1.25}\smash{\begin{tabular}[t]{l}$B$\end{tabular}}}}%
    \put(-0.00238218,0.07962061){\color[rgb]{0,0,0}\makebox(0,0)[lt]{\lineheight{1.25}\smash{\begin{tabular}[t]{l}$A$\end{tabular}}}}%
    \put(0,0){\includegraphics[width=\unitlength,page=1]{2_stage_c-d.pdf}}%
    \put(0.15135842,0.02611899){\color[rgb]{0,0,0}\makebox(0,0)[lt]{\lineheight{1.25}\smash{\begin{tabular}[t]{l}$c_1$\end{tabular}}}}%
    \put(0.3665955,0.02611899){\color[rgb]{0,0,0}\makebox(0,0)[lt]{\lineheight{1.25}\smash{\begin{tabular}[t]{l}$c_2$\end{tabular}}}}%
    \put(0.25897696,0.02611899){\color[rgb]{0,0,0}\makebox(0,0)[lt]{\lineheight{1.25}\smash{\begin{tabular}[t]{l}$d_1$\end{tabular}}}}%
    \put(0.80137426,0.02611899){\color[rgb]{0,0,0}\makebox(0,0)[lt]{\lineheight{1.25}\smash{\begin{tabular}[t]{l}$c_q$\end{tabular}}}}%
    \put(0.69375565,0.02611899){\color[rgb]{0,0,0}\makebox(0,0)[lt]{\lineheight{1.25}\smash{\begin{tabular}[t]{l}$d_{q-1}$\end{tabular}}}}%
    \put(0.90899261,0.02611899){\color[rgb]{0,0,0}\makebox(0,0)[lt]{\lineheight{1.25}\smash{\begin{tabular}[t]{l}$d_q$\end{tabular}}}}%
    \put(0,0){\includegraphics[width=\unitlength,page=2]{2_stage_c-d.pdf}}%
  \end{picture}%
\endgroup%

%% file: inkscape/n_stage_c-d.pdf_tex
\begingroup%
  \makeatletter%
  \providecommand\color[2][]{%
    \errmessage{(Inkscape) Color is used for the text in Inkscape, but the package 'color.sty' is not loaded}%
    \renewcommand\color[2][]{}%
  }%
  \providecommand\transparent[1]{%
    \errmessage{(Inkscape) Transparency is used (non-zero) for the text in Inkscape, but the package 'transparent.sty' is not loaded}%
    \renewcommand\transparent[1]{}%
  }%
  \providecommand\rotatebox[2]{#2}%
  \newcommand*\fsize{\dimexpr\f@size pt\relax}%
  \newcommand*\lineheight[1]{\fontsize{\fsize}{#1\fsize}\selectfont}%
  \ifx\svgwidth\undefined%
    \setlength{\unitlength}{348.4532983bp}%
    \ifx\svgscale\undefined%
      \relax%
    \else%
      \setlength{\unitlength}{\unitlength * \real{\svgscale}}%
    \fi%
  \else%
    \setlength{\unitlength}{\svgwidth}%
  \fi%
  \global\let\svgwidth\undefined%
  \global\let\svgscale\undefined%
  \makeatother%
  \begin{picture}(1,0.26559953)%
    \lineheight{1}%
    \setlength\tabcolsep{0pt}%
    \put(-0.00238218,0.22167706){\color[rgb]{0,0,0}\makebox(0,0)[lt]{\lineheight{1.25}\smash{\begin{tabular}[t]{l}$A_\Lambda$\end{tabular}}}}%
    \put(-0.00238218,0.07962061){\color[rgb]{0,0,0}\makebox(0,0)[lt]{\lineheight{1.25}\smash{\begin{tabular}[t]{l}$A_1$\end{tabular}}}}%
    \put(0,0){\includegraphics[width=\unitlength,page=1]{n_stage_c-d.pdf}}%
    \put(0.15135842,0.02611899){\color[rgb]{0,0,0}\makebox(0,0)[lt]{\lineheight{1.25}\smash{\begin{tabular}[t]{l}$c_1$\end{tabular}}}}%
    \put(0.3665955,0.02611899){\color[rgb]{0,0,0}\makebox(0,0)[lt]{\lineheight{1.25}\smash{\begin{tabular}[t]{l}$c_2$\end{tabular}}}}%
    \put(0.25897696,0.02611899){\color[rgb]{0,0,0}\makebox(0,0)[lt]{\lineheight{1.25}\smash{\begin{tabular}[t]{l}$d_1$\end{tabular}}}}%
    \put(0.80137426,0.02611899){\color[rgb]{0,0,0}\makebox(0,0)[lt]{\lineheight{1.25}\smash{\begin{tabular}[t]{l}$c_q$\end{tabular}}}}%
    \put(0.69375565,0.02611899){\color[rgb]{0,0,0}\makebox(0,0)[lt]{\lineheight{1.25}\smash{\begin{tabular}[t]{l}$d_{q-1}$\end{tabular}}}}%
    \put(0.90899261,0.02611899){\color[rgb]{0,0,0}\makebox(0,0)[lt]{\lineheight{1.25}\smash{\begin{tabular}[t]{l}$d_q$\end{tabular}}}}%
    \put(0,0){\includegraphics[width=\unitlength,page=2]{n_stage_c-d.pdf}}%
    \put(-0.00238218,0.12697277){\color[rgb]{0,0,0}\makebox(0,0)[lt]{\lineheight{1.25}\smash{\begin{tabular}[t]{l}$A_2$\end{tabular}}}}%
    \put(0.01053202,0.17432493){\color[rgb]{0,0,0}\makebox(0,0)[lt]{\lineheight{1.25}\smash{\begin{tabular}[t]{l}$\vdots$\end{tabular}}}}%
    \put(0,0){\includegraphics[width=\unitlength,page=3]{n_stage_c-d.pdf}}%
  \end{picture}%
\endgroup%

%% file: simulation/benchmark/2-stage_fix-t_ord4.tex
\begingroup
  \inputencoding{latin1}%
  \makeatletter
  \providecommand\color[2][]{%
    \GenericError{(gnuplot) \space\space\space\@spaces}{%
      Package color not loaded in conjunction with
      terminal option `colourtext'%
    }{See the gnuplot documentation for explanation.%
    }{Either use 'blacktext' in gnuplot or load the package
      color.sty in LaTeX.}%
    \renewcommand\color[2][]{}%
  }%
  \providecommand\includegraphics[2][]{%
    \GenericError{(gnuplot) \space\space\space\@spaces}{%
      Package graphicx or graphics not loaded%
    }{See the gnuplot documentation for explanation.%
    }{The gnuplot epslatex terminal needs graphicx.sty or graphics.sty.}%
    \renewcommand\includegraphics[2][]{}%
  }%
  \providecommand\rotatebox[2]{#2}%
  \@ifundefined{ifGPcolor}{%
    \newif\ifGPcolor
    \GPcolortrue
  }{}%
  \@ifundefined{ifGPblacktext}{%
    \newif\ifGPblacktext
    \GPblacktexttrue
  }{}%
  \let\gplgaddtomacro\g@addto@macro
  \gdef\gplbacktext{}%
  \gdef\gplfronttext{}%
  \makeatother
  \ifGPblacktext
    \def\colorrgb#1{}%
    \def\colorgray#1{}%
  \else
    \ifGPcolor
      \def\colorrgb#1{\color[rgb]{#1}}%
      \def\colorgray#1{\color[gray]{#1}}%
      \expandafter\def\csname LTw\endcsname{\color{white}}%
      \expandafter\def\csname LTb\endcsname{\color{black}}%
      \expandafter\def\csname LTa\endcsname{\color{black}}%
      \expandafter\def\csname LT0\endcsname{\color[rgb]{1,0,0}}%
      \expandafter\def\csname LT1\endcsname{\color[rgb]{0,1,0}}%
      \expandafter\def\csname LT2\endcsname{\color[rgb]{0,0,1}}%
      \expandafter\def\csname LT3\endcsname{\color[rgb]{1,0,1}}%
      \expandafter\def\csname LT4\endcsname{\color[rgb]{0,1,1}}%
      \expandafter\def\csname LT5\endcsname{\color[rgb]{1,1,0}}%
      \expandafter\def\csname LT6\endcsname{\color[rgb]{0,0,0}}%
      \expandafter\def\csname LT7\endcsname{\color[rgb]{1,0.3,0}}%
      \expandafter\def\csname LT8\endcsname{\color[rgb]{0.5,0.5,0.5}}%
    \else
      \def\colorrgb#1{\color{black}}%
      \def\colorgray#1{\color[gray]{#1}}%
      \expandafter\def\csname LTw\endcsname{\color{white}}%
      \expandafter\def\csname LTb\endcsname{\color{black}}%
      \expandafter\def\csname LTa\endcsname{\color{black}}%
      \expandafter\def\csname LT0\endcsname{\color{black}}%
      \expandafter\def\csname LT1\endcsname{\color{black}}%
      \expandafter\def\csname LT2\endcsname{\color{black}}%
      \expandafter\def\csname LT3\endcsname{\color{black}}%
      \expandafter\def\csname LT4\endcsname{\color{black}}%
      \expandafter\def\csname LT5\endcsname{\color{black}}%
      \expandafter\def\csname LT6\endcsname{\color{black}}%
      \expandafter\def\csname LT7\endcsname{\color{black}}%
      \expandafter\def\csname LT8\endcsname{\color{black}}%
    \fi
  \fi
    \setlength{\unitlength}{0.0500bp}%
    \ifx\gptboxheight\undefined%
      \newlength{\gptboxheight}%
      \newlength{\gptboxwidth}%
      \newsavebox{\gptboxtext}%
    \fi%
    \setlength{\fboxrule}{0.5pt}%
    \setlength{\fboxsep}{1pt}%
\begin{picture}(7200.00,5040.00)%
    \gplgaddtomacro\gplbacktext{%
      \csname LTb\endcsname
      \put(946,704){\makebox(0,0)[r]{\strut{}$10^{-6}$}}%
      \csname LTb\endcsname
      \put(946,1357){\makebox(0,0)[r]{\strut{}$10^{-5}$}}%
      \csname LTb\endcsname
      \put(946,2010){\makebox(0,0)[r]{\strut{}$10^{-4}$}}%
      \csname LTb\endcsname
      \put(946,2663){\makebox(0,0)[r]{\strut{}$10^{-3}$}}%
      \csname LTb\endcsname
      \put(946,3316){\makebox(0,0)[r]{\strut{}$10^{-2}$}}%
      \csname LTb\endcsname
      \put(946,3969){\makebox(0,0)[r]{\strut{}$10^{-1}$}}%
      \csname LTb\endcsname
      \put(946,4622){\makebox(0,0)[r]{\strut{}$10^{0}$}}%
      \csname LTb\endcsname
      \put(3941,484){\makebox(0,0){\strut{}$10$}}%
    }%
    \gplgaddtomacro\gplfronttext{%
      \csname LTb\endcsname
      \put(209,2761){\rotatebox{-270}{\makebox(0,0){\strut{}error}}}%
      \put(3940,154){\makebox(0,0){\strut{}$q/h$}}%
      \csname LTb\endcsname
      \put(4114,2417){\makebox(0,0)[r]{\strut{}Forest-Ruth~$(4,3)$~\eqref{eq:3-forest-ruth}}}%
      \csname LTb\endcsname
      \put(4114,2197){\makebox(0,0)[r]{\strut{}FR-Type~$(4,4)$~\eqref{eq:4-fr-type}}}%
      \csname LTb\endcsname
      \put(4114,1977){\makebox(0,0)[r]{\strut{}Non-Unitary~$(4,4)$~\eqref{eq:4-non-unitary1}}}%
      \csname LTb\endcsname
      \put(4114,1757){\makebox(0,0)[r]{\strut{}Suzuki~$(4,5)$~\eqref{eq:5-suzuki4}}}%
      \csname LTb\endcsname
      \put(4114,1537){\makebox(0,0)[r]{\strut{}Opt.~4th~order~$(4,5)$~\eqref{eq:5-opt-4th-ord}}}%
      \csname LTb\endcsname
      \put(4114,1317){\makebox(0,0)[r]{\strut{}Non-Unitary~$(4,5)$~\eqref{eq:5-non-unitary2}}}%
      \csname LTb\endcsname
      \put(4114,1097){\makebox(0,0)[r]{\strut{}Unif.~Non-Unitary~$(4,5)$~\eqref{eq:5-non-unitary-const}}}%
      \csname LTb\endcsname
      \put(4114,877){\makebox(0,0)[r]{\strut{}Blanes\&Moan~$(4,6)$~\eqref{eq:6-blanes4}}}%
    }%
    \gplbacktext
    \put(0,0){\includegraphics{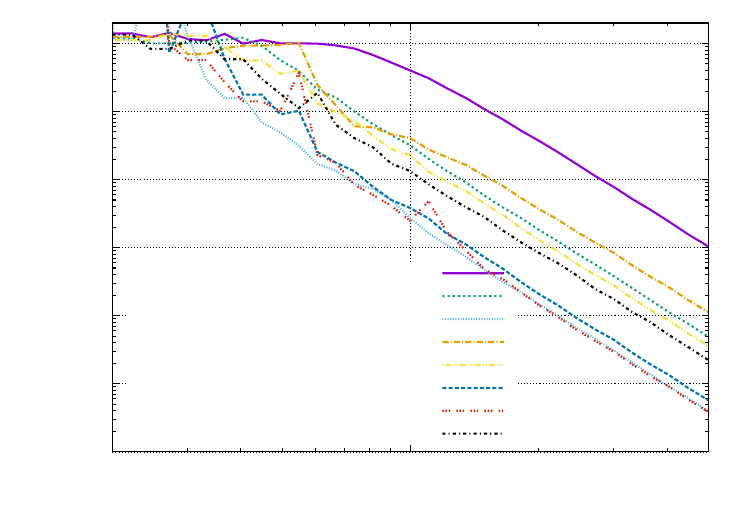}}%
    \gplfronttext
  \end{picture}%
\endgroup

%% file: simulation/benchmark/2L-stage_fix-t_ord4.tex
\begingroup
  \inputencoding{latin1}%
  \makeatletter
  \providecommand\color[2][]{%
    \GenericError{(gnuplot) \space\space\space\@spaces}{%
      Package color not loaded in conjunction with
      terminal option `colourtext'%
    }{See the gnuplot documentation for explanation.%
    }{Either use 'blacktext' in gnuplot or load the package
      color.sty in LaTeX.}%
    \renewcommand\color[2][]{}%
  }%
  \providecommand\includegraphics[2][]{%
    \GenericError{(gnuplot) \space\space\space\@spaces}{%
      Package graphicx or graphics not loaded%
    }{See the gnuplot documentation for explanation.%
    }{The gnuplot epslatex terminal needs graphicx.sty or graphics.sty.}%
    \renewcommand\includegraphics[2][]{}%
  }%
  \providecommand\rotatebox[2]{#2}%
  \@ifundefined{ifGPcolor}{%
    \newif\ifGPcolor
    \GPcolortrue
  }{}%
  \@ifundefined{ifGPblacktext}{%
    \newif\ifGPblacktext
    \GPblacktexttrue
  }{}%
  \let\gplgaddtomacro\g@addto@macro
  \gdef\gplbacktext{}%
  \gdef\gplfronttext{}%
  \makeatother
  \ifGPblacktext
    \def\colorrgb#1{}%
    \def\colorgray#1{}%
  \else
    \ifGPcolor
      \def\colorrgb#1{\color[rgb]{#1}}%
      \def\colorgray#1{\color[gray]{#1}}%
      \expandafter\def\csname LTw\endcsname{\color{white}}%
      \expandafter\def\csname LTb\endcsname{\color{black}}%
      \expandafter\def\csname LTa\endcsname{\color{black}}%
      \expandafter\def\csname LT0\endcsname{\color[rgb]{1,0,0}}%
      \expandafter\def\csname LT1\endcsname{\color[rgb]{0,1,0}}%
      \expandafter\def\csname LT2\endcsname{\color[rgb]{0,0,1}}%
      \expandafter\def\csname LT3\endcsname{\color[rgb]{1,0,1}}%
      \expandafter\def\csname LT4\endcsname{\color[rgb]{0,1,1}}%
      \expandafter\def\csname LT5\endcsname{\color[rgb]{1,1,0}}%
      \expandafter\def\csname LT6\endcsname{\color[rgb]{0,0,0}}%
      \expandafter\def\csname LT7\endcsname{\color[rgb]{1,0.3,0}}%
      \expandafter\def\csname LT8\endcsname{\color[rgb]{0.5,0.5,0.5}}%
    \else
      \def\colorrgb#1{\color{black}}%
      \def\colorgray#1{\color[gray]{#1}}%
      \expandafter\def\csname LTw\endcsname{\color{white}}%
      \expandafter\def\csname LTb\endcsname{\color{black}}%
      \expandafter\def\csname LTa\endcsname{\color{black}}%
      \expandafter\def\csname LT0\endcsname{\color{black}}%
      \expandafter\def\csname LT1\endcsname{\color{black}}%
      \expandafter\def\csname LT2\endcsname{\color{black}}%
      \expandafter\def\csname LT3\endcsname{\color{black}}%
      \expandafter\def\csname LT4\endcsname{\color{black}}%
      \expandafter\def\csname LT5\endcsname{\color{black}}%
      \expandafter\def\csname LT6\endcsname{\color{black}}%
      \expandafter\def\csname LT7\endcsname{\color{black}}%
      \expandafter\def\csname LT8\endcsname{\color{black}}%
    \fi
  \fi
    \setlength{\unitlength}{0.0500bp}%
    \ifx\gptboxheight\undefined%
      \newlength{\gptboxheight}%
      \newlength{\gptboxwidth}%
      \newsavebox{\gptboxtext}%
    \fi%
    \setlength{\fboxrule}{0.5pt}%
    \setlength{\fboxsep}{1pt}%
\begin{picture}(7200.00,5040.00)%
    \gplgaddtomacro\gplbacktext{%
      \csname LTb\endcsname
      \put(946,704){\makebox(0,0)[r]{\strut{}$10^{-6}$}}%
      \csname LTb\endcsname
      \put(946,1357){\makebox(0,0)[r]{\strut{}$10^{-5}$}}%
      \csname LTb\endcsname
      \put(946,2010){\makebox(0,0)[r]{\strut{}$10^{-4}$}}%
      \csname LTb\endcsname
      \put(946,2663){\makebox(0,0)[r]{\strut{}$10^{-3}$}}%
      \csname LTb\endcsname
      \put(946,3316){\makebox(0,0)[r]{\strut{}$10^{-2}$}}%
      \csname LTb\endcsname
      \put(946,3969){\makebox(0,0)[r]{\strut{}$10^{-1}$}}%
      \csname LTb\endcsname
      \put(946,4622){\makebox(0,0)[r]{\strut{}$10^{0}$}}%
      \csname LTb\endcsname
      \put(3941,484){\makebox(0,0){\strut{}$10$}}%
    }%
    \gplgaddtomacro\gplfronttext{%
      \csname LTb\endcsname
      \put(209,2761){\rotatebox{-270}{\makebox(0,0){\strut{}error}}}%
      \put(3940,154){\makebox(0,0){\strut{}$q/h$}}%
      \csname LTb\endcsname
      \put(4114,2417){\makebox(0,0)[r]{\strut{}Forest-Ruth~$(4,3)$~\eqref{eq:3-forest-ruth}}}%
      \csname LTb\endcsname
      \put(4114,2197){\makebox(0,0)[r]{\strut{}FR-Type~$(4,4)$~\eqref{eq:4-fr-type}}}%
      \csname LTb\endcsname
      \put(4114,1977){\makebox(0,0)[r]{\strut{}Non-Unitary~$(4,4)$~\eqref{eq:4-non-unitary1}}}%
      \csname LTb\endcsname
      \put(4114,1757){\makebox(0,0)[r]{\strut{}Suzuki~$(4,5)$~\eqref{eq:5-suzuki4}}}%
      \csname LTb\endcsname
      \put(4114,1537){\makebox(0,0)[r]{\strut{}Opt.~4th~order~$(4,5)$~\eqref{eq:5-opt-4th-ord}}}%
      \csname LTb\endcsname
      \put(4114,1317){\makebox(0,0)[r]{\strut{}Non-Unitary~$(4,5)$~\eqref{eq:5-non-unitary2}}}%
      \csname LTb\endcsname
      \put(4114,1097){\makebox(0,0)[r]{\strut{}Unif.~Non-Unitary~$(4,5)$~\eqref{eq:5-non-unitary-const}}}%
      \csname LTb\endcsname
      \put(4114,877){\makebox(0,0)[r]{\strut{}Blanes\&Moan~$(4,6)$~\eqref{eq:6-blanes4}}}%
    }%
    \gplbacktext
    \put(0,0){\includegraphics{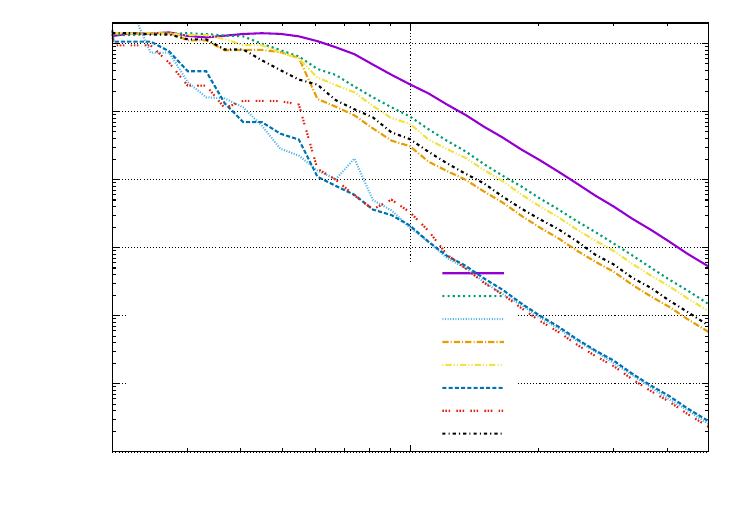}}%
    \gplfronttext
  \end{picture}%
\endgroup

%% file: simulation/benchmark/3-stage_fix-t_ord4.tex
\begingroup
  \inputencoding{latin1}%
  \makeatletter
  \providecommand\color[2][]{%
    \GenericError{(gnuplot) \space\space\space\@spaces}{%
      Package color not loaded in conjunction with
      terminal option `colourtext'%
    }{See the gnuplot documentation for explanation.%
    }{Either use 'blacktext' in gnuplot or load the package
      color.sty in LaTeX.}%
    \renewcommand\color[2][]{}%
  }%
  \providecommand\includegraphics[2][]{%
    \GenericError{(gnuplot) \space\space\space\@spaces}{%
      Package graphicx or graphics not loaded%
    }{See the gnuplot documentation for explanation.%
    }{The gnuplot epslatex terminal needs graphicx.sty or graphics.sty.}%
    \renewcommand\includegraphics[2][]{}%
  }%
  \providecommand\rotatebox[2]{#2}%
  \@ifundefined{ifGPcolor}{%
    \newif\ifGPcolor
    \GPcolortrue
  }{}%
  \@ifundefined{ifGPblacktext}{%
    \newif\ifGPblacktext
    \GPblacktexttrue
  }{}%
  \let\gplgaddtomacro\g@addto@macro
  \gdef\gplbacktext{}%
  \gdef\gplfronttext{}%
  \makeatother
  \ifGPblacktext
    \def\colorrgb#1{}%
    \def\colorgray#1{}%
  \else
    \ifGPcolor
      \def\colorrgb#1{\color[rgb]{#1}}%
      \def\colorgray#1{\color[gray]{#1}}%
      \expandafter\def\csname LTw\endcsname{\color{white}}%
      \expandafter\def\csname LTb\endcsname{\color{black}}%
      \expandafter\def\csname LTa\endcsname{\color{black}}%
      \expandafter\def\csname LT0\endcsname{\color[rgb]{1,0,0}}%
      \expandafter\def\csname LT1\endcsname{\color[rgb]{0,1,0}}%
      \expandafter\def\csname LT2\endcsname{\color[rgb]{0,0,1}}%
      \expandafter\def\csname LT3\endcsname{\color[rgb]{1,0,1}}%
      \expandafter\def\csname LT4\endcsname{\color[rgb]{0,1,1}}%
      \expandafter\def\csname LT5\endcsname{\color[rgb]{1,1,0}}%
      \expandafter\def\csname LT6\endcsname{\color[rgb]{0,0,0}}%
      \expandafter\def\csname LT7\endcsname{\color[rgb]{1,0.3,0}}%
      \expandafter\def\csname LT8\endcsname{\color[rgb]{0.5,0.5,0.5}}%
    \else
      \def\colorrgb#1{\color{black}}%
      \def\colorgray#1{\color[gray]{#1}}%
      \expandafter\def\csname LTw\endcsname{\color{white}}%
      \expandafter\def\csname LTb\endcsname{\color{black}}%
      \expandafter\def\csname LTa\endcsname{\color{black}}%
      \expandafter\def\csname LT0\endcsname{\color{black}}%
      \expandafter\def\csname LT1\endcsname{\color{black}}%
      \expandafter\def\csname LT2\endcsname{\color{black}}%
      \expandafter\def\csname LT3\endcsname{\color{black}}%
      \expandafter\def\csname LT4\endcsname{\color{black}}%
      \expandafter\def\csname LT5\endcsname{\color{black}}%
      \expandafter\def\csname LT6\endcsname{\color{black}}%
      \expandafter\def\csname LT7\endcsname{\color{black}}%
      \expandafter\def\csname LT8\endcsname{\color{black}}%
    \fi
  \fi
    \setlength{\unitlength}{0.0500bp}%
    \ifx\gptboxheight\undefined%
      \newlength{\gptboxheight}%
      \newlength{\gptboxwidth}%
      \newsavebox{\gptboxtext}%
    \fi%
    \setlength{\fboxrule}{0.5pt}%
    \setlength{\fboxsep}{1pt}%
\begin{picture}(7200.00,5040.00)%
    \gplgaddtomacro\gplbacktext{%
      \csname LTb\endcsname
      \put(946,704){\makebox(0,0)[r]{\strut{}$10^{-6}$}}%
      \csname LTb\endcsname
      \put(946,1357){\makebox(0,0)[r]{\strut{}$10^{-5}$}}%
      \csname LTb\endcsname
      \put(946,2010){\makebox(0,0)[r]{\strut{}$10^{-4}$}}%
      \csname LTb\endcsname
      \put(946,2663){\makebox(0,0)[r]{\strut{}$10^{-3}$}}%
      \csname LTb\endcsname
      \put(946,3316){\makebox(0,0)[r]{\strut{}$10^{-2}$}}%
      \csname LTb\endcsname
      \put(946,3969){\makebox(0,0)[r]{\strut{}$10^{-1}$}}%
      \csname LTb\endcsname
      \put(946,4622){\makebox(0,0)[r]{\strut{}$10^{0}$}}%
      \csname LTb\endcsname
      \put(3941,484){\makebox(0,0){\strut{}$10$}}%
    }%
    \gplgaddtomacro\gplfronttext{%
      \csname LTb\endcsname
      \put(209,2761){\rotatebox{-270}{\makebox(0,0){\strut{}error}}}%
      \put(3940,154){\makebox(0,0){\strut{}$q/h$}}%
      \csname LTb\endcsname
      \put(4114,2417){\makebox(0,0)[r]{\strut{}Forest-Ruth~$(4,3)$~\eqref{eq:3-forest-ruth}}}%
      \csname LTb\endcsname
      \put(4114,2197){\makebox(0,0)[r]{\strut{}FR-Type~$(4,4)$~\eqref{eq:4-fr-type}}}%
      \csname LTb\endcsname
      \put(4114,1977){\makebox(0,0)[r]{\strut{}Non-Unitary~$(4,4)$~\eqref{eq:4-non-unitary1}}}%
      \csname LTb\endcsname
      \put(4114,1757){\makebox(0,0)[r]{\strut{}Suzuki~$(4,5)$~\eqref{eq:5-suzuki4}}}%
      \csname LTb\endcsname
      \put(4114,1537){\makebox(0,0)[r]{\strut{}Opt.~4th~order~$(4,5)$~\eqref{eq:5-opt-4th-ord}}}%
      \csname LTb\endcsname
      \put(4114,1317){\makebox(0,0)[r]{\strut{}Non-Unitary~$(4,5)$~\eqref{eq:5-non-unitary2}}}%
      \csname LTb\endcsname
      \put(4114,1097){\makebox(0,0)[r]{\strut{}Unif.~Non-Unitary~$(4,5)$~\eqref{eq:5-non-unitary-const}}}%
      \csname LTb\endcsname
      \put(4114,877){\makebox(0,0)[r]{\strut{}Blanes\&Moan~$(4,6)$~\eqref{eq:6-blanes4}}}%
    }%
    \gplbacktext
    \put(0,0){\includegraphics{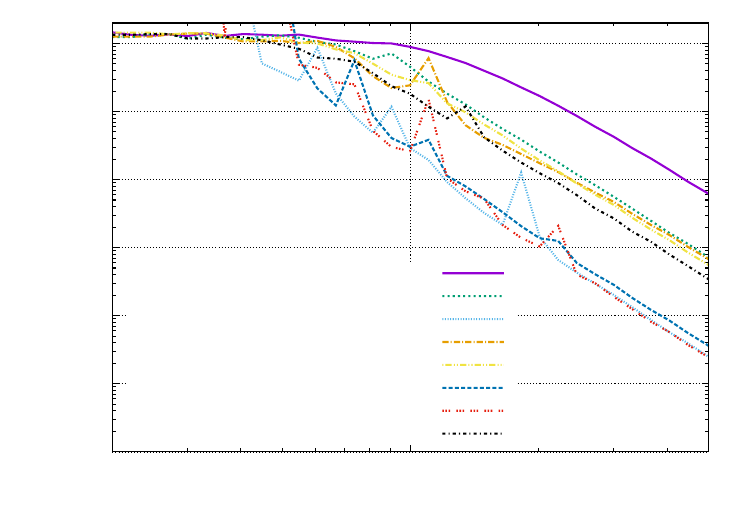}}%
    \gplfronttext
  \end{picture}%
\endgroup

%% file: simulation/benchmark/3L-stage_fix-t_ord4.tex
\begingroup
  \inputencoding{latin1}%
  \makeatletter
  \providecommand\color[2][]{%
    \GenericError{(gnuplot) \space\space\space\@spaces}{%
      Package color not loaded in conjunction with
      terminal option `colourtext'%
    }{See the gnuplot documentation for explanation.%
    }{Either use 'blacktext' in gnuplot or load the package
      color.sty in LaTeX.}%
    \renewcommand\color[2][]{}%
  }%
  \providecommand\includegraphics[2][]{%
    \GenericError{(gnuplot) \space\space\space\@spaces}{%
      Package graphicx or graphics not loaded%
    }{See the gnuplot documentation for explanation.%
    }{The gnuplot epslatex terminal needs graphicx.sty or graphics.sty.}%
    \renewcommand\includegraphics[2][]{}%
  }%
  \providecommand\rotatebox[2]{#2}%
  \@ifundefined{ifGPcolor}{%
    \newif\ifGPcolor
    \GPcolortrue
  }{}%
  \@ifundefined{ifGPblacktext}{%
    \newif\ifGPblacktext
    \GPblacktexttrue
  }{}%
  \let\gplgaddtomacro\g@addto@macro
  \gdef\gplbacktext{}%
  \gdef\gplfronttext{}%
  \makeatother
  \ifGPblacktext
    \def\colorrgb#1{}%
    \def\colorgray#1{}%
  \else
    \ifGPcolor
      \def\colorrgb#1{\color[rgb]{#1}}%
      \def\colorgray#1{\color[gray]{#1}}%
      \expandafter\def\csname LTw\endcsname{\color{white}}%
      \expandafter\def\csname LTb\endcsname{\color{black}}%
      \expandafter\def\csname LTa\endcsname{\color{black}}%
      \expandafter\def\csname LT0\endcsname{\color[rgb]{1,0,0}}%
      \expandafter\def\csname LT1\endcsname{\color[rgb]{0,1,0}}%
      \expandafter\def\csname LT2\endcsname{\color[rgb]{0,0,1}}%
      \expandafter\def\csname LT3\endcsname{\color[rgb]{1,0,1}}%
      \expandafter\def\csname LT4\endcsname{\color[rgb]{0,1,1}}%
      \expandafter\def\csname LT5\endcsname{\color[rgb]{1,1,0}}%
      \expandafter\def\csname LT6\endcsname{\color[rgb]{0,0,0}}%
      \expandafter\def\csname LT7\endcsname{\color[rgb]{1,0.3,0}}%
      \expandafter\def\csname LT8\endcsname{\color[rgb]{0.5,0.5,0.5}}%
    \else
      \def\colorrgb#1{\color{black}}%
      \def\colorgray#1{\color[gray]{#1}}%
      \expandafter\def\csname LTw\endcsname{\color{white}}%
      \expandafter\def\csname LTb\endcsname{\color{black}}%
      \expandafter\def\csname LTa\endcsname{\color{black}}%
      \expandafter\def\csname LT0\endcsname{\color{black}}%
      \expandafter\def\csname LT1\endcsname{\color{black}}%
      \expandafter\def\csname LT2\endcsname{\color{black}}%
      \expandafter\def\csname LT3\endcsname{\color{black}}%
      \expandafter\def\csname LT4\endcsname{\color{black}}%
      \expandafter\def\csname LT5\endcsname{\color{black}}%
      \expandafter\def\csname LT6\endcsname{\color{black}}%
      \expandafter\def\csname LT7\endcsname{\color{black}}%
      \expandafter\def\csname LT8\endcsname{\color{black}}%
    \fi
  \fi
    \setlength{\unitlength}{0.0500bp}%
    \ifx\gptboxheight\undefined%
      \newlength{\gptboxheight}%
      \newlength{\gptboxwidth}%
      \newsavebox{\gptboxtext}%
    \fi%
    \setlength{\fboxrule}{0.5pt}%
    \setlength{\fboxsep}{1pt}%
\begin{picture}(7200.00,5040.00)%
    \gplgaddtomacro\gplbacktext{%
      \csname LTb\endcsname
      \put(946,704){\makebox(0,0)[r]{\strut{}$10^{-6}$}}%
      \csname LTb\endcsname
      \put(946,1357){\makebox(0,0)[r]{\strut{}$10^{-5}$}}%
      \csname LTb\endcsname
      \put(946,2010){\makebox(0,0)[r]{\strut{}$10^{-4}$}}%
      \csname LTb\endcsname
      \put(946,2663){\makebox(0,0)[r]{\strut{}$10^{-3}$}}%
      \csname LTb\endcsname
      \put(946,3316){\makebox(0,0)[r]{\strut{}$10^{-2}$}}%
      \csname LTb\endcsname
      \put(946,3969){\makebox(0,0)[r]{\strut{}$10^{-1}$}}%
      \csname LTb\endcsname
      \put(946,4622){\makebox(0,0)[r]{\strut{}$10^{0}$}}%
      \csname LTb\endcsname
      \put(3941,484){\makebox(0,0){\strut{}$10$}}%
    }%
    \gplgaddtomacro\gplfronttext{%
      \csname LTb\endcsname
      \put(209,2761){\rotatebox{-270}{\makebox(0,0){\strut{}error}}}%
      \put(3940,154){\makebox(0,0){\strut{}$q/h$}}%
      \csname LTb\endcsname
      \put(4114,2417){\makebox(0,0)[r]{\strut{}Forest-Ruth~$(4,3)$~\eqref{eq:3-forest-ruth}}}%
      \csname LTb\endcsname
      \put(4114,2197){\makebox(0,0)[r]{\strut{}FR-Type~$(4,4)$~\eqref{eq:4-fr-type}}}%
      \csname LTb\endcsname
      \put(4114,1977){\makebox(0,0)[r]{\strut{}Non-Unitary~$(4,4)$~\eqref{eq:4-non-unitary1}}}%
      \csname LTb\endcsname
      \put(4114,1757){\makebox(0,0)[r]{\strut{}Suzuki~$(4,5)$~\eqref{eq:5-suzuki4}}}%
      \csname LTb\endcsname
      \put(4114,1537){\makebox(0,0)[r]{\strut{}Opt.~4th~order~$(4,5)$~\eqref{eq:5-opt-4th-ord}}}%
      \csname LTb\endcsname
      \put(4114,1317){\makebox(0,0)[r]{\strut{}Non-Unitary~$(4,5)$~\eqref{eq:5-non-unitary2}}}%
      \csname LTb\endcsname
      \put(4114,1097){\makebox(0,0)[r]{\strut{}Unif.~Non-Unitary~$(4,5)$~\eqref{eq:5-non-unitary-const}}}%
      \csname LTb\endcsname
      \put(4114,877){\makebox(0,0)[r]{\strut{}Blanes\&Moan~$(4,6)$~\eqref{eq:6-blanes4}}}%
    }%
    \gplbacktext
    \put(0,0){\includegraphics{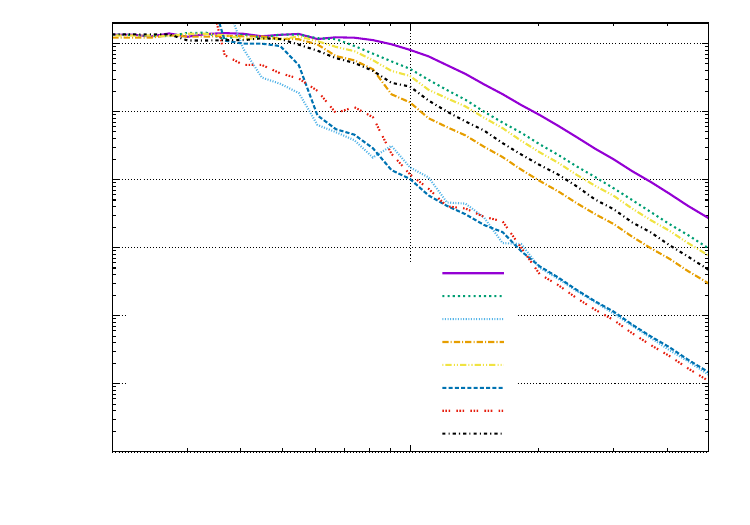}}%
    \gplfronttext
  \end{picture}%
\endgroup

%% file: simulation/benchmark/2-stage_fix-t_ordRest.tex
\begingroup
  \inputencoding{latin1}%
  \makeatletter
  \providecommand\color[2][]{%
    \GenericError{(gnuplot) \space\space\space\@spaces}{%
      Package color not loaded in conjunction with
      terminal option `colourtext'%
    }{See the gnuplot documentation for explanation.%
    }{Either use 'blacktext' in gnuplot or load the package
      color.sty in LaTeX.}%
    \renewcommand\color[2][]{}%
  }%
  \providecommand\includegraphics[2][]{%
    \GenericError{(gnuplot) \space\space\space\@spaces}{%
      Package graphicx or graphics not loaded%
    }{See the gnuplot documentation for explanation.%
    }{The gnuplot epslatex terminal needs graphicx.sty or graphics.sty.}%
    \renewcommand\includegraphics[2][]{}%
  }%
  \providecommand\rotatebox[2]{#2}%
  \@ifundefined{ifGPcolor}{%
    \newif\ifGPcolor
    \GPcolortrue
  }{}%
  \@ifundefined{ifGPblacktext}{%
    \newif\ifGPblacktext
    \GPblacktexttrue
  }{}%
  \let\gplgaddtomacro\g@addto@macro
  \gdef\gplbacktext{}%
  \gdef\gplfronttext{}%
  \makeatother
  \ifGPblacktext
    \def\colorrgb#1{}%
    \def\colorgray#1{}%
  \else
    \ifGPcolor
      \def\colorrgb#1{\color[rgb]{#1}}%
      \def\colorgray#1{\color[gray]{#1}}%
      \expandafter\def\csname LTw\endcsname{\color{white}}%
      \expandafter\def\csname LTb\endcsname{\color{black}}%
      \expandafter\def\csname LTa\endcsname{\color{black}}%
      \expandafter\def\csname LT0\endcsname{\color[rgb]{1,0,0}}%
      \expandafter\def\csname LT1\endcsname{\color[rgb]{0,1,0}}%
      \expandafter\def\csname LT2\endcsname{\color[rgb]{0,0,1}}%
      \expandafter\def\csname LT3\endcsname{\color[rgb]{1,0,1}}%
      \expandafter\def\csname LT4\endcsname{\color[rgb]{0,1,1}}%
      \expandafter\def\csname LT5\endcsname{\color[rgb]{1,1,0}}%
      \expandafter\def\csname LT6\endcsname{\color[rgb]{0,0,0}}%
      \expandafter\def\csname LT7\endcsname{\color[rgb]{1,0.3,0}}%
      \expandafter\def\csname LT8\endcsname{\color[rgb]{0.5,0.5,0.5}}%
    \else
      \def\colorrgb#1{\color{black}}%
      \def\colorgray#1{\color[gray]{#1}}%
      \expandafter\def\csname LTw\endcsname{\color{white}}%
      \expandafter\def\csname LTb\endcsname{\color{black}}%
      \expandafter\def\csname LTa\endcsname{\color{black}}%
      \expandafter\def\csname LT0\endcsname{\color{black}}%
      \expandafter\def\csname LT1\endcsname{\color{black}}%
      \expandafter\def\csname LT2\endcsname{\color{black}}%
      \expandafter\def\csname LT3\endcsname{\color{black}}%
      \expandafter\def\csname LT4\endcsname{\color{black}}%
      \expandafter\def\csname LT5\endcsname{\color{black}}%
      \expandafter\def\csname LT6\endcsname{\color{black}}%
      \expandafter\def\csname LT7\endcsname{\color{black}}%
      \expandafter\def\csname LT8\endcsname{\color{black}}%
    \fi
  \fi
    \setlength{\unitlength}{0.0500bp}%
    \ifx\gptboxheight\undefined%
      \newlength{\gptboxheight}%
      \newlength{\gptboxwidth}%
      \newsavebox{\gptboxtext}%
    \fi%
    \setlength{\fboxrule}{0.5pt}%
    \setlength{\fboxsep}{1pt}%
\begin{picture}(7200.00,5040.00)%
    \gplgaddtomacro\gplbacktext{%
      \csname LTb\endcsname
      \put(1078,704){\makebox(0,0)[r]{\strut{}$10^{-14}$}}%
      \csname LTb\endcsname
      \put(1078,1272){\makebox(0,0)[r]{\strut{}$10^{-12}$}}%
      \csname LTb\endcsname
      \put(1078,1841){\makebox(0,0)[r]{\strut{}$10^{-10}$}}%
      \csname LTb\endcsname
      \put(1078,2409){\makebox(0,0)[r]{\strut{}$10^{-8}$}}%
      \csname LTb\endcsname
      \put(1078,2978){\makebox(0,0)[r]{\strut{}$10^{-6}$}}%
      \csname LTb\endcsname
      \put(1078,3546){\makebox(0,0)[r]{\strut{}$10^{-4}$}}%
      \csname LTb\endcsname
      \put(1078,4115){\makebox(0,0)[r]{\strut{}$10^{-2}$}}%
      \csname LTb\endcsname
      \put(1078,4683){\makebox(0,0)[r]{\strut{}$10^{0}$}}%
      \csname LTb\endcsname
      \put(2906,484){\makebox(0,0){\strut{}$10$}}%
      \csname LTb\endcsname
      \put(5333,484){\makebox(0,0){\strut{}$100$}}%
    }%
    \gplgaddtomacro\gplfronttext{%
      \csname LTb\endcsname
      \put(209,2761){\rotatebox{-270}{\makebox(0,0){\strut{}error}}}%
      \put(4006,154){\makebox(0,0){\strut{}$q/h$}}%
      \csname LTb\endcsname
      \put(4246,2857){\makebox(0,0)[r]{\strut{}Verlet~$(2,1)$~\eqref{eq:1-leap-frog}}}%
      \csname LTb\endcsname
      \put(4246,2637){\makebox(0,0)[r]{\strut{}Omelyan~$(2,2)$~\eqref{eq:2-omelyan2}}}%
      \csname LTb\endcsname
      \put(4246,2417){\makebox(0,0)[r]{\strut{}Yoshida~$(6,7)$~\eqref{eq:7-yoshida}}}%
      \csname LTb\endcsname
      \put(4246,2197){\makebox(0,0)[r]{\strut{}Blanes\&Moan~$(6,10)$~\eqref{eq:10-blanes6}}}%
      \csname LTb\endcsname
      \put(4246,1977){\makebox(0,0)[r]{\strut{}Suzuki~$(6,25)$~(sec.~\ref{sec:suzuki6})}}%
      \csname LTb\endcsname
      \put(4246,1757){\makebox(0,0)[r]{\strut{}Morales~$(8,17)$~\eqref{eq:17-morales}}}%
      \csname LTb\endcsname
      \put(4246,1537){\makebox(0,0)[r]{\strut{}BM6+S~$(8,50)$~(sec.~\ref{sec:very_high_orders})}}%
      \csname LTb\endcsname
      \put(4246,1317){\makebox(0,0)[r]{\strut{}Suzuki~$(8,125)$~(sec.~\ref{sec:very_high_orders})}}%
      \csname LTb\endcsname
      \put(4246,1097){\makebox(0,0)[r]{\strut{}Taylor~(sec.~\ref{sec:taylor_expansion})}}%
    }%
    \gplbacktext
    \put(0,0){\includegraphics{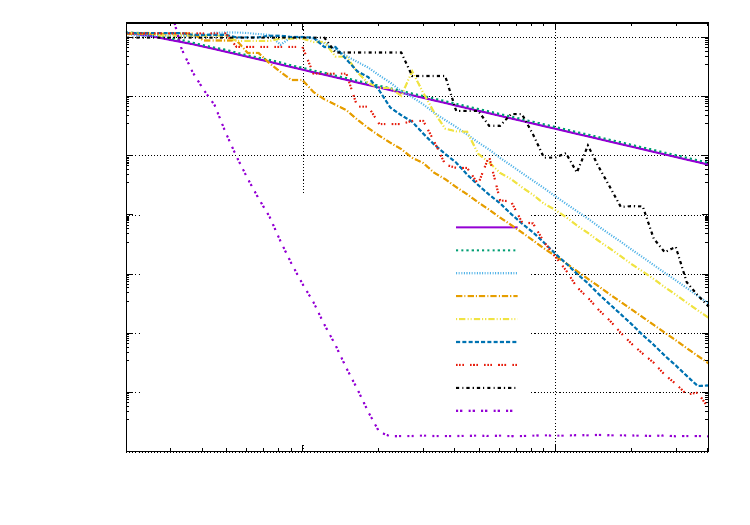}}%
    \gplfronttext
  \end{picture}%
\endgroup

%% file: simulation/benchmark/2L-stage_fix-t_ordRest.tex
\begingroup
  \inputencoding{latin1}%
  \makeatletter
  \providecommand\color[2][]{%
    \GenericError{(gnuplot) \space\space\space\@spaces}{%
      Package color not loaded in conjunction with
      terminal option `colourtext'%
    }{See the gnuplot documentation for explanation.%
    }{Either use 'blacktext' in gnuplot or load the package
      color.sty in LaTeX.}%
    \renewcommand\color[2][]{}%
  }%
  \providecommand\includegraphics[2][]{%
    \GenericError{(gnuplot) \space\space\space\@spaces}{%
      Package graphicx or graphics not loaded%
    }{See the gnuplot documentation for explanation.%
    }{The gnuplot epslatex terminal needs graphicx.sty or graphics.sty.}%
    \renewcommand\includegraphics[2][]{}%
  }%
  \providecommand\rotatebox[2]{#2}%
  \@ifundefined{ifGPcolor}{%
    \newif\ifGPcolor
    \GPcolortrue
  }{}%
  \@ifundefined{ifGPblacktext}{%
    \newif\ifGPblacktext
    \GPblacktexttrue
  }{}%
  \let\gplgaddtomacro\g@addto@macro
  \gdef\gplbacktext{}%
  \gdef\gplfronttext{}%
  \makeatother
  \ifGPblacktext
    \def\colorrgb#1{}%
    \def\colorgray#1{}%
  \else
    \ifGPcolor
      \def\colorrgb#1{\color[rgb]{#1}}%
      \def\colorgray#1{\color[gray]{#1}}%
      \expandafter\def\csname LTw\endcsname{\color{white}}%
      \expandafter\def\csname LTb\endcsname{\color{black}}%
      \expandafter\def\csname LTa\endcsname{\color{black}}%
      \expandafter\def\csname LT0\endcsname{\color[rgb]{1,0,0}}%
      \expandafter\def\csname LT1\endcsname{\color[rgb]{0,1,0}}%
      \expandafter\def\csname LT2\endcsname{\color[rgb]{0,0,1}}%
      \expandafter\def\csname LT3\endcsname{\color[rgb]{1,0,1}}%
      \expandafter\def\csname LT4\endcsname{\color[rgb]{0,1,1}}%
      \expandafter\def\csname LT5\endcsname{\color[rgb]{1,1,0}}%
      \expandafter\def\csname LT6\endcsname{\color[rgb]{0,0,0}}%
      \expandafter\def\csname LT7\endcsname{\color[rgb]{1,0.3,0}}%
      \expandafter\def\csname LT8\endcsname{\color[rgb]{0.5,0.5,0.5}}%
    \else
      \def\colorrgb#1{\color{black}}%
      \def\colorgray#1{\color[gray]{#1}}%
      \expandafter\def\csname LTw\endcsname{\color{white}}%
      \expandafter\def\csname LTb\endcsname{\color{black}}%
      \expandafter\def\csname LTa\endcsname{\color{black}}%
      \expandafter\def\csname LT0\endcsname{\color{black}}%
      \expandafter\def\csname LT1\endcsname{\color{black}}%
      \expandafter\def\csname LT2\endcsname{\color{black}}%
      \expandafter\def\csname LT3\endcsname{\color{black}}%
      \expandafter\def\csname LT4\endcsname{\color{black}}%
      \expandafter\def\csname LT5\endcsname{\color{black}}%
      \expandafter\def\csname LT6\endcsname{\color{black}}%
      \expandafter\def\csname LT7\endcsname{\color{black}}%
      \expandafter\def\csname LT8\endcsname{\color{black}}%
    \fi
  \fi
    \setlength{\unitlength}{0.0500bp}%
    \ifx\gptboxheight\undefined%
      \newlength{\gptboxheight}%
      \newlength{\gptboxwidth}%
      \newsavebox{\gptboxtext}%
    \fi%
    \setlength{\fboxrule}{0.5pt}%
    \setlength{\fboxsep}{1pt}%
\begin{picture}(7200.00,5040.00)%
    \gplgaddtomacro\gplbacktext{%
      \csname LTb\endcsname
      \put(1078,704){\makebox(0,0)[r]{\strut{}$10^{-14}$}}%
      \csname LTb\endcsname
      \put(1078,1272){\makebox(0,0)[r]{\strut{}$10^{-12}$}}%
      \csname LTb\endcsname
      \put(1078,1841){\makebox(0,0)[r]{\strut{}$10^{-10}$}}%
      \csname LTb\endcsname
      \put(1078,2409){\makebox(0,0)[r]{\strut{}$10^{-8}$}}%
      \csname LTb\endcsname
      \put(1078,2978){\makebox(0,0)[r]{\strut{}$10^{-6}$}}%
      \csname LTb\endcsname
      \put(1078,3546){\makebox(0,0)[r]{\strut{}$10^{-4}$}}%
      \csname LTb\endcsname
      \put(1078,4115){\makebox(0,0)[r]{\strut{}$10^{-2}$}}%
      \csname LTb\endcsname
      \put(1078,4683){\makebox(0,0)[r]{\strut{}$10^{0}$}}%
      \csname LTb\endcsname
      \put(2906,484){\makebox(0,0){\strut{}$10$}}%
      \csname LTb\endcsname
      \put(5333,484){\makebox(0,0){\strut{}$100$}}%
    }%
    \gplgaddtomacro\gplfronttext{%
      \csname LTb\endcsname
      \put(209,2761){\rotatebox{-270}{\makebox(0,0){\strut{}error}}}%
      \put(4006,154){\makebox(0,0){\strut{}$q/h$}}%
      \csname LTb\endcsname
      \put(4246,2857){\makebox(0,0)[r]{\strut{}Verlet~$(2,1)$~\eqref{eq:1-leap-frog}}}%
      \csname LTb\endcsname
      \put(4246,2637){\makebox(0,0)[r]{\strut{}Omelyan~$(2,2)$~\eqref{eq:2-omelyan2}}}%
      \csname LTb\endcsname
      \put(4246,2417){\makebox(0,0)[r]{\strut{}Yoshida~$(6,7)$~\eqref{eq:7-yoshida}}}%
      \csname LTb\endcsname
      \put(4246,2197){\makebox(0,0)[r]{\strut{}Blanes\&Moan~$(6,10)$~\eqref{eq:10-blanes6}}}%
      \csname LTb\endcsname
      \put(4246,1977){\makebox(0,0)[r]{\strut{}Suzuki~$(6,25)$~(sec.~\ref{sec:suzuki6})}}%
      \csname LTb\endcsname
      \put(4246,1757){\makebox(0,0)[r]{\strut{}Morales~$(8,17)$~\eqref{eq:17-morales}}}%
      \csname LTb\endcsname
      \put(4246,1537){\makebox(0,0)[r]{\strut{}BM6+S~$(8,50)$~(sec.~\ref{sec:very_high_orders})}}%
      \csname LTb\endcsname
      \put(4246,1317){\makebox(0,0)[r]{\strut{}Suzuki~$(8,125)$~(sec.~\ref{sec:very_high_orders})}}%
      \csname LTb\endcsname
      \put(4246,1097){\makebox(0,0)[r]{\strut{}Taylor~(sec.~\ref{sec:taylor_expansion})}}%
    }%
    \gplbacktext
    \put(0,0){\includegraphics{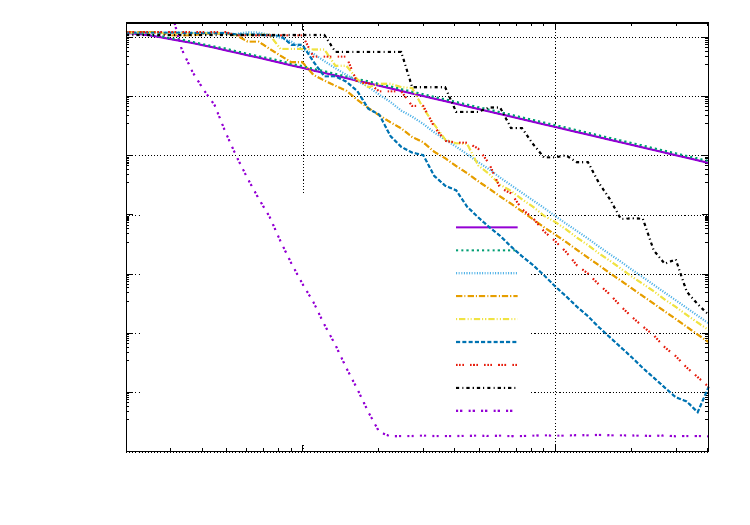}}%
    \gplfronttext
  \end{picture}%
\endgroup

%% file: simulation/benchmark/3-stage_fix-t_ordRest.tex
\begingroup
  \inputencoding{latin1}%
  \makeatletter
  \providecommand\color[2][]{%
    \GenericError{(gnuplot) \space\space\space\@spaces}{%
      Package color not loaded in conjunction with
      terminal option `colourtext'%
    }{See the gnuplot documentation for explanation.%
    }{Either use 'blacktext' in gnuplot or load the package
      color.sty in LaTeX.}%
    \renewcommand\color[2][]{}%
  }%
  \providecommand\includegraphics[2][]{%
    \GenericError{(gnuplot) \space\space\space\@spaces}{%
      Package graphicx or graphics not loaded%
    }{See the gnuplot documentation for explanation.%
    }{The gnuplot epslatex terminal needs graphicx.sty or graphics.sty.}%
    \renewcommand\includegraphics[2][]{}%
  }%
  \providecommand\rotatebox[2]{#2}%
  \@ifundefined{ifGPcolor}{%
    \newif\ifGPcolor
    \GPcolortrue
  }{}%
  \@ifundefined{ifGPblacktext}{%
    \newif\ifGPblacktext
    \GPblacktexttrue
  }{}%
  \let\gplgaddtomacro\g@addto@macro
  \gdef\gplbacktext{}%
  \gdef\gplfronttext{}%
  \makeatother
  \ifGPblacktext
    \def\colorrgb#1{}%
    \def\colorgray#1{}%
  \else
    \ifGPcolor
      \def\colorrgb#1{\color[rgb]{#1}}%
      \def\colorgray#1{\color[gray]{#1}}%
      \expandafter\def\csname LTw\endcsname{\color{white}}%
      \expandafter\def\csname LTb\endcsname{\color{black}}%
      \expandafter\def\csname LTa\endcsname{\color{black}}%
      \expandafter\def\csname LT0\endcsname{\color[rgb]{1,0,0}}%
      \expandafter\def\csname LT1\endcsname{\color[rgb]{0,1,0}}%
      \expandafter\def\csname LT2\endcsname{\color[rgb]{0,0,1}}%
      \expandafter\def\csname LT3\endcsname{\color[rgb]{1,0,1}}%
      \expandafter\def\csname LT4\endcsname{\color[rgb]{0,1,1}}%
      \expandafter\def\csname LT5\endcsname{\color[rgb]{1,1,0}}%
      \expandafter\def\csname LT6\endcsname{\color[rgb]{0,0,0}}%
      \expandafter\def\csname LT7\endcsname{\color[rgb]{1,0.3,0}}%
      \expandafter\def\csname LT8\endcsname{\color[rgb]{0.5,0.5,0.5}}%
    \else
      \def\colorrgb#1{\color{black}}%
      \def\colorgray#1{\color[gray]{#1}}%
      \expandafter\def\csname LTw\endcsname{\color{white}}%
      \expandafter\def\csname LTb\endcsname{\color{black}}%
      \expandafter\def\csname LTa\endcsname{\color{black}}%
      \expandafter\def\csname LT0\endcsname{\color{black}}%
      \expandafter\def\csname LT1\endcsname{\color{black}}%
      \expandafter\def\csname LT2\endcsname{\color{black}}%
      \expandafter\def\csname LT3\endcsname{\color{black}}%
      \expandafter\def\csname LT4\endcsname{\color{black}}%
      \expandafter\def\csname LT5\endcsname{\color{black}}%
      \expandafter\def\csname LT6\endcsname{\color{black}}%
      \expandafter\def\csname LT7\endcsname{\color{black}}%
      \expandafter\def\csname LT8\endcsname{\color{black}}%
    \fi
  \fi
    \setlength{\unitlength}{0.0500bp}%
    \ifx\gptboxheight\undefined%
      \newlength{\gptboxheight}%
      \newlength{\gptboxwidth}%
      \newsavebox{\gptboxtext}%
    \fi%
    \setlength{\fboxrule}{0.5pt}%
    \setlength{\fboxsep}{1pt}%
\begin{picture}(7200.00,5040.00)%
    \gplgaddtomacro\gplbacktext{%
      \csname LTb\endcsname
      \put(1078,704){\makebox(0,0)[r]{\strut{}$10^{-14}$}}%
      \csname LTb\endcsname
      \put(1078,1272){\makebox(0,0)[r]{\strut{}$10^{-12}$}}%
      \csname LTb\endcsname
      \put(1078,1841){\makebox(0,0)[r]{\strut{}$10^{-10}$}}%
      \csname LTb\endcsname
      \put(1078,2409){\makebox(0,0)[r]{\strut{}$10^{-8}$}}%
      \csname LTb\endcsname
      \put(1078,2978){\makebox(0,0)[r]{\strut{}$10^{-6}$}}%
      \csname LTb\endcsname
      \put(1078,3546){\makebox(0,0)[r]{\strut{}$10^{-4}$}}%
      \csname LTb\endcsname
      \put(1078,4115){\makebox(0,0)[r]{\strut{}$10^{-2}$}}%
      \csname LTb\endcsname
      \put(1078,4683){\makebox(0,0)[r]{\strut{}$10^{0}$}}%
      \csname LTb\endcsname
      \put(2906,484){\makebox(0,0){\strut{}$10$}}%
      \csname LTb\endcsname
      \put(5333,484){\makebox(0,0){\strut{}$100$}}%
    }%
    \gplgaddtomacro\gplfronttext{%
      \csname LTb\endcsname
      \put(209,2761){\rotatebox{-270}{\makebox(0,0){\strut{}error}}}%
      \put(4006,154){\makebox(0,0){\strut{}$q/h$}}%
      \csname LTb\endcsname
      \put(4246,2857){\makebox(0,0)[r]{\strut{}Verlet~$(2,1)$~\eqref{eq:1-leap-frog}}}%
      \csname LTb\endcsname
      \put(4246,2637){\makebox(0,0)[r]{\strut{}Omelyan~$(2,2)$~\eqref{eq:2-omelyan2}}}%
      \csname LTb\endcsname
      \put(4246,2417){\makebox(0,0)[r]{\strut{}Yoshida~$(6,7)$~\eqref{eq:7-yoshida}}}%
      \csname LTb\endcsname
      \put(4246,2197){\makebox(0,0)[r]{\strut{}Blanes\&Moan~$(6,10)$~\eqref{eq:10-blanes6}}}%
      \csname LTb\endcsname
      \put(4246,1977){\makebox(0,0)[r]{\strut{}Suzuki~$(6,25)$~(sec.~\ref{sec:suzuki6})}}%
      \csname LTb\endcsname
      \put(4246,1757){\makebox(0,0)[r]{\strut{}Morales~$(8,17)$~\eqref{eq:17-morales}}}%
      \csname LTb\endcsname
      \put(4246,1537){\makebox(0,0)[r]{\strut{}BM6+S~$(8,50)$~(sec.~\ref{sec:very_high_orders})}}%
      \csname LTb\endcsname
      \put(4246,1317){\makebox(0,0)[r]{\strut{}Suzuki~$(8,125)$~(sec.~\ref{sec:very_high_orders})}}%
      \csname LTb\endcsname
      \put(4246,1097){\makebox(0,0)[r]{\strut{}Taylor~(sec.~\ref{sec:taylor_expansion})}}%
    }%
    \gplbacktext
    \put(0,0){\includegraphics{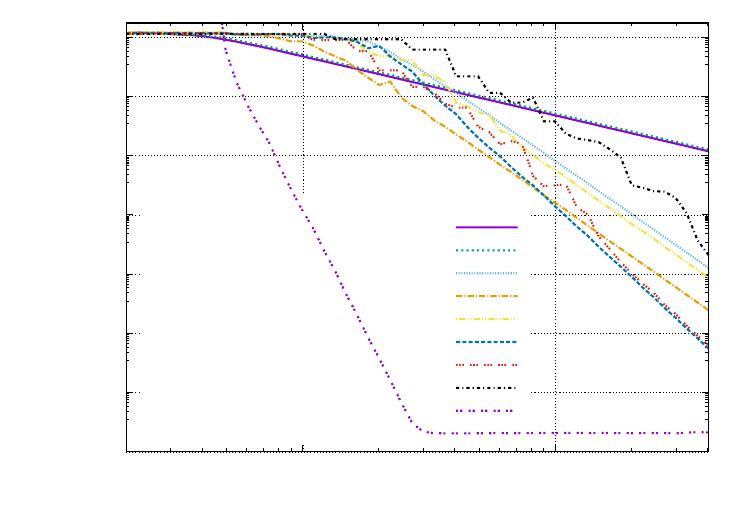}}%
    \gplfronttext
  \end{picture}%
\endgroup

%% file: simulation/benchmark/3L-stage_fix-t_ordRest.tex
\begingroup
  \inputencoding{latin1}%
  \makeatletter
  \providecommand\color[2][]{%
    \GenericError{(gnuplot) \space\space\space\@spaces}{%
      Package color not loaded in conjunction with
      terminal option `colourtext'%
    }{See the gnuplot documentation for explanation.%
    }{Either use 'blacktext' in gnuplot or load the package
      color.sty in LaTeX.}%
    \renewcommand\color[2][]{}%
  }%
  \providecommand\includegraphics[2][]{%
    \GenericError{(gnuplot) \space\space\space\@spaces}{%
      Package graphicx or graphics not loaded%
    }{See the gnuplot documentation for explanation.%
    }{The gnuplot epslatex terminal needs graphicx.sty or graphics.sty.}%
    \renewcommand\includegraphics[2][]{}%
  }%
  \providecommand\rotatebox[2]{#2}%
  \@ifundefined{ifGPcolor}{%
    \newif\ifGPcolor
    \GPcolortrue
  }{}%
  \@ifundefined{ifGPblacktext}{%
    \newif\ifGPblacktext
    \GPblacktexttrue
  }{}%
  \let\gplgaddtomacro\g@addto@macro
  \gdef\gplbacktext{}%
  \gdef\gplfronttext{}%
  \makeatother
  \ifGPblacktext
    \def\colorrgb#1{}%
    \def\colorgray#1{}%
  \else
    \ifGPcolor
      \def\colorrgb#1{\color[rgb]{#1}}%
      \def\colorgray#1{\color[gray]{#1}}%
      \expandafter\def\csname LTw\endcsname{\color{white}}%
      \expandafter\def\csname LTb\endcsname{\color{black}}%
      \expandafter\def\csname LTa\endcsname{\color{black}}%
      \expandafter\def\csname LT0\endcsname{\color[rgb]{1,0,0}}%
      \expandafter\def\csname LT1\endcsname{\color[rgb]{0,1,0}}%
      \expandafter\def\csname LT2\endcsname{\color[rgb]{0,0,1}}%
      \expandafter\def\csname LT3\endcsname{\color[rgb]{1,0,1}}%
      \expandafter\def\csname LT4\endcsname{\color[rgb]{0,1,1}}%
      \expandafter\def\csname LT5\endcsname{\color[rgb]{1,1,0}}%
      \expandafter\def\csname LT6\endcsname{\color[rgb]{0,0,0}}%
      \expandafter\def\csname LT7\endcsname{\color[rgb]{1,0.3,0}}%
      \expandafter\def\csname LT8\endcsname{\color[rgb]{0.5,0.5,0.5}}%
    \else
      \def\colorrgb#1{\color{black}}%
      \def\colorgray#1{\color[gray]{#1}}%
      \expandafter\def\csname LTw\endcsname{\color{white}}%
      \expandafter\def\csname LTb\endcsname{\color{black}}%
      \expandafter\def\csname LTa\endcsname{\color{black}}%
      \expandafter\def\csname LT0\endcsname{\color{black}}%
      \expandafter\def\csname LT1\endcsname{\color{black}}%
      \expandafter\def\csname LT2\endcsname{\color{black}}%
      \expandafter\def\csname LT3\endcsname{\color{black}}%
      \expandafter\def\csname LT4\endcsname{\color{black}}%
      \expandafter\def\csname LT5\endcsname{\color{black}}%
      \expandafter\def\csname LT6\endcsname{\color{black}}%
      \expandafter\def\csname LT7\endcsname{\color{black}}%
      \expandafter\def\csname LT8\endcsname{\color{black}}%
    \fi
  \fi
    \setlength{\unitlength}{0.0500bp}%
    \ifx\gptboxheight\undefined%
      \newlength{\gptboxheight}%
      \newlength{\gptboxwidth}%
      \newsavebox{\gptboxtext}%
    \fi%
    \setlength{\fboxrule}{0.5pt}%
    \setlength{\fboxsep}{1pt}%
\begin{picture}(7200.00,5040.00)%
    \gplgaddtomacro\gplbacktext{%
      \csname LTb\endcsname
      \put(1078,704){\makebox(0,0)[r]{\strut{}$10^{-14}$}}%
      \csname LTb\endcsname
      \put(1078,1272){\makebox(0,0)[r]{\strut{}$10^{-12}$}}%
      \csname LTb\endcsname
      \put(1078,1841){\makebox(0,0)[r]{\strut{}$10^{-10}$}}%
      \csname LTb\endcsname
      \put(1078,2409){\makebox(0,0)[r]{\strut{}$10^{-8}$}}%
      \csname LTb\endcsname
      \put(1078,2978){\makebox(0,0)[r]{\strut{}$10^{-6}$}}%
      \csname LTb\endcsname
      \put(1078,3546){\makebox(0,0)[r]{\strut{}$10^{-4}$}}%
      \csname LTb\endcsname
      \put(1078,4115){\makebox(0,0)[r]{\strut{}$10^{-2}$}}%
      \csname LTb\endcsname
      \put(1078,4683){\makebox(0,0)[r]{\strut{}$10^{0}$}}%
      \csname LTb\endcsname
      \put(2906,484){\makebox(0,0){\strut{}$10$}}%
      \csname LTb\endcsname
      \put(5333,484){\makebox(0,0){\strut{}$100$}}%
    }%
    \gplgaddtomacro\gplfronttext{%
      \csname LTb\endcsname
      \put(209,2761){\rotatebox{-270}{\makebox(0,0){\strut{}error}}}%
      \put(4006,154){\makebox(0,0){\strut{}$q/h$}}%
      \csname LTb\endcsname
      \put(4246,2857){\makebox(0,0)[r]{\strut{}Verlet~$(2,1)$~\eqref{eq:1-leap-frog}}}%
      \csname LTb\endcsname
      \put(4246,2637){\makebox(0,0)[r]{\strut{}Omelyan~$(2,2)$~\eqref{eq:2-omelyan2}}}%
      \csname LTb\endcsname
      \put(4246,2417){\makebox(0,0)[r]{\strut{}Yoshida~$(6,7)$~\eqref{eq:7-yoshida}}}%
      \csname LTb\endcsname
      \put(4246,2197){\makebox(0,0)[r]{\strut{}Blanes\&Moan~$(6,10)$~\eqref{eq:10-blanes6}}}%
      \csname LTb\endcsname
      \put(4246,1977){\makebox(0,0)[r]{\strut{}Suzuki~$(6,25)$~(sec.~\ref{sec:suzuki6})}}%
      \csname LTb\endcsname
      \put(4246,1757){\makebox(0,0)[r]{\strut{}Morales~$(8,17)$~\eqref{eq:17-morales}}}%
      \csname LTb\endcsname
      \put(4246,1537){\makebox(0,0)[r]{\strut{}BM6+S~$(8,50)$~(sec.~\ref{sec:very_high_orders})}}%
      \csname LTb\endcsname
      \put(4246,1317){\makebox(0,0)[r]{\strut{}Suzuki~$(8,125)$~(sec.~\ref{sec:very_high_orders})}}%
      \csname LTb\endcsname
      \put(4246,1097){\makebox(0,0)[r]{\strut{}Taylor~(sec.~\ref{sec:taylor_expansion})}}%
    }%
    \gplbacktext
    \put(0,0){\includegraphics{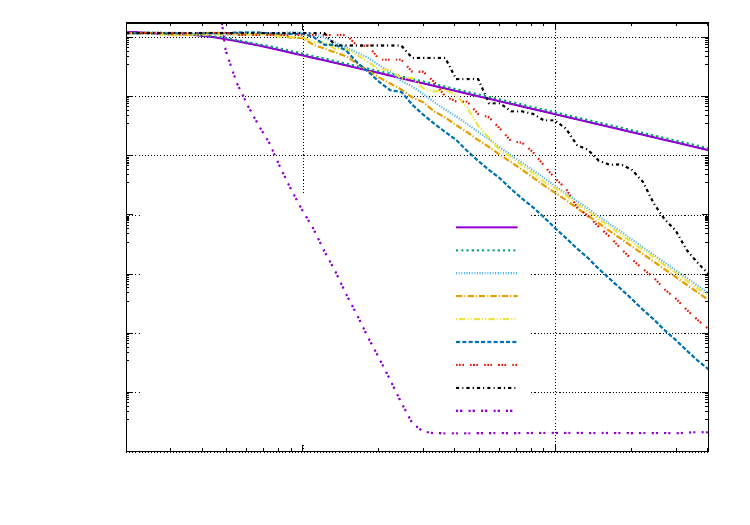}}%
    \gplfronttext
  \end{picture}%
\endgroup

%% file: simulation/benchmark/2-stage_fix-dt_ord4.tex
\begingroup
  \inputencoding{latin1}%
  \makeatletter
  \providecommand\color[2][]{%
    \GenericError{(gnuplot) \space\space\space\@spaces}{%
      Package color not loaded in conjunction with
      terminal option `colourtext'%
    }{See the gnuplot documentation for explanation.%
    }{Either use 'blacktext' in gnuplot or load the package
      color.sty in LaTeX.}%
    \renewcommand\color[2][]{}%
  }%
  \providecommand\includegraphics[2][]{%
    \GenericError{(gnuplot) \space\space\space\@spaces}{%
      Package graphicx or graphics not loaded%
    }{See the gnuplot documentation for explanation.%
    }{The gnuplot epslatex terminal needs graphicx.sty or graphics.sty.}%
    \renewcommand\includegraphics[2][]{}%
  }%
  \providecommand\rotatebox[2]{#2}%
  \@ifundefined{ifGPcolor}{%
    \newif\ifGPcolor
    \GPcolortrue
  }{}%
  \@ifundefined{ifGPblacktext}{%
    \newif\ifGPblacktext
    \GPblacktexttrue
  }{}%
  \let\gplgaddtomacro\g@addto@macro
  \gdef\gplbacktext{}%
  \gdef\gplfronttext{}%
  \makeatother
  \ifGPblacktext
    \def\colorrgb#1{}%
    \def\colorgray#1{}%
  \else
    \ifGPcolor
      \def\colorrgb#1{\color[rgb]{#1}}%
      \def\colorgray#1{\color[gray]{#1}}%
      \expandafter\def\csname LTw\endcsname{\color{white}}%
      \expandafter\def\csname LTb\endcsname{\color{black}}%
      \expandafter\def\csname LTa\endcsname{\color{black}}%
      \expandafter\def\csname LT0\endcsname{\color[rgb]{1,0,0}}%
      \expandafter\def\csname LT1\endcsname{\color[rgb]{0,1,0}}%
      \expandafter\def\csname LT2\endcsname{\color[rgb]{0,0,1}}%
      \expandafter\def\csname LT3\endcsname{\color[rgb]{1,0,1}}%
      \expandafter\def\csname LT4\endcsname{\color[rgb]{0,1,1}}%
      \expandafter\def\csname LT5\endcsname{\color[rgb]{1,1,0}}%
      \expandafter\def\csname LT6\endcsname{\color[rgb]{0,0,0}}%
      \expandafter\def\csname LT7\endcsname{\color[rgb]{1,0.3,0}}%
      \expandafter\def\csname LT8\endcsname{\color[rgb]{0.5,0.5,0.5}}%
    \else
      \def\colorrgb#1{\color{black}}%
      \def\colorgray#1{\color[gray]{#1}}%
      \expandafter\def\csname LTw\endcsname{\color{white}}%
      \expandafter\def\csname LTb\endcsname{\color{black}}%
      \expandafter\def\csname LTa\endcsname{\color{black}}%
      \expandafter\def\csname LT0\endcsname{\color{black}}%
      \expandafter\def\csname LT1\endcsname{\color{black}}%
      \expandafter\def\csname LT2\endcsname{\color{black}}%
      \expandafter\def\csname LT3\endcsname{\color{black}}%
      \expandafter\def\csname LT4\endcsname{\color{black}}%
      \expandafter\def\csname LT5\endcsname{\color{black}}%
      \expandafter\def\csname LT6\endcsname{\color{black}}%
      \expandafter\def\csname LT7\endcsname{\color{black}}%
      \expandafter\def\csname LT8\endcsname{\color{black}}%
    \fi
  \fi
    \setlength{\unitlength}{0.0500bp}%
    \ifx\gptboxheight\undefined%
      \newlength{\gptboxheight}%
      \newlength{\gptboxwidth}%
      \newsavebox{\gptboxtext}%
    \fi%
    \setlength{\fboxrule}{0.5pt}%
    \setlength{\fboxsep}{1pt}%
\begin{picture}(7200.00,5040.00)%
    \gplgaddtomacro\gplbacktext{%
      \csname LTb\endcsname
      \put(946,1704){\makebox(0,0)[r]{\strut{}$10^{-7}$}}%
      \csname LTb\endcsname
      \put(946,3136){\makebox(0,0)[r]{\strut{}$10^{-6}$}}%
      \csname LTb\endcsname
      \put(946,4567){\makebox(0,0)[r]{\strut{}$10^{-5}$}}%
      \csname LTb\endcsname
      \put(1078,484){\makebox(0,0){\strut{}$0$}}%
      \csname LTb\endcsname
      \put(2223,484){\makebox(0,0){\strut{}$2$}}%
      \csname LTb\endcsname
      \put(3368,484){\makebox(0,0){\strut{}$4$}}%
      \csname LTb\endcsname
      \put(4513,484){\makebox(0,0){\strut{}$6$}}%
      \csname LTb\endcsname
      \put(5658,484){\makebox(0,0){\strut{}$8$}}%
      \csname LTb\endcsname
      \put(6803,484){\makebox(0,0){\strut{}$10$}}%
    }%
    \gplgaddtomacro\gplfronttext{%
      \csname LTb\endcsname
      \put(209,2761){\rotatebox{-270}{\makebox(0,0){\strut{}$\mathrm{error}/t$}}}%
      \put(3940,154){\makebox(0,0){\strut{}$t$}}%
      \csname LTb\endcsname
      \put(5816,4646){\makebox(0,0)[r]{\strut{}Forest-Ruth~$(4,3)$~\eqref{eq:3-forest-ruth}}}%
      \csname LTb\endcsname
      \put(5816,4426){\makebox(0,0)[r]{\strut{}FR-Type~$(4,4)$~\eqref{eq:4-fr-type}}}%
      \csname LTb\endcsname
      \put(5816,4206){\makebox(0,0)[r]{\strut{}Non-Unitary~$(4,4)$~\eqref{eq:4-non-unitary1}}}%
      \csname LTb\endcsname
      \put(5816,3986){\makebox(0,0)[r]{\strut{}Suzuki~$(4,5)$~\eqref{eq:5-suzuki4}}}%
      \csname LTb\endcsname
      \put(5816,3766){\makebox(0,0)[r]{\strut{}Opt.~4th~order~$(4,5)$~\eqref{eq:5-opt-4th-ord}}}%
      \csname LTb\endcsname
      \put(5816,3546){\makebox(0,0)[r]{\strut{}Non-Unitary~$(4,5)$~\eqref{eq:5-non-unitary2}}}%
      \csname LTb\endcsname
      \put(5816,3326){\makebox(0,0)[r]{\strut{}Unif.~Non-Unitary~$(4,5)$~\eqref{eq:5-non-unitary-const}}}%
      \csname LTb\endcsname
      \put(5816,3106){\makebox(0,0)[r]{\strut{}Blanes\&Moan~$(4,6)$~\eqref{eq:6-blanes4}}}%
    }%
    \gplbacktext
    \put(0,0){\includegraphics{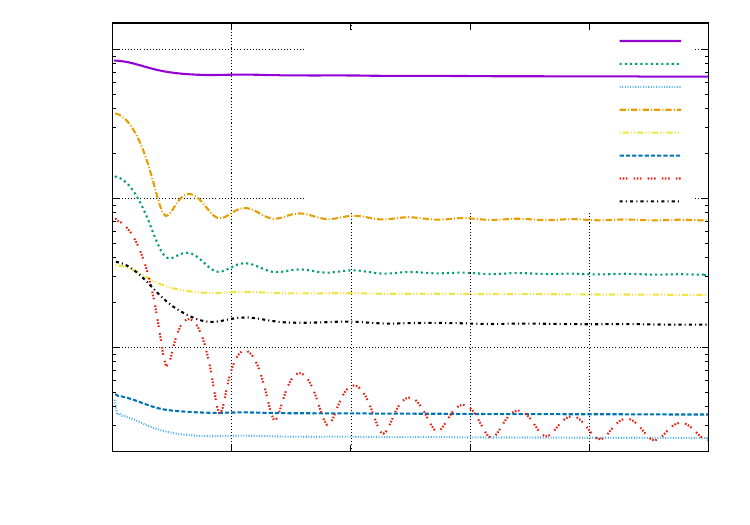}}%
    \gplfronttext
  \end{picture}%
\endgroup

%% file: simulation/benchmark/3L-stage_fix-dt_ord4.tex
\begingroup
  \inputencoding{latin1}%
  \makeatletter
  \providecommand\color[2][]{%
    \GenericError{(gnuplot) \space\space\space\@spaces}{%
      Package color not loaded in conjunction with
      terminal option `colourtext'%
    }{See the gnuplot documentation for explanation.%
    }{Either use 'blacktext' in gnuplot or load the package
      color.sty in LaTeX.}%
    \renewcommand\color[2][]{}%
  }%
  \providecommand\includegraphics[2][]{%
    \GenericError{(gnuplot) \space\space\space\@spaces}{%
      Package graphicx or graphics not loaded%
    }{See the gnuplot documentation for explanation.%
    }{The gnuplot epslatex terminal needs graphicx.sty or graphics.sty.}%
    \renewcommand\includegraphics[2][]{}%
  }%
  \providecommand\rotatebox[2]{#2}%
  \@ifundefined{ifGPcolor}{%
    \newif\ifGPcolor
    \GPcolortrue
  }{}%
  \@ifundefined{ifGPblacktext}{%
    \newif\ifGPblacktext
    \GPblacktexttrue
  }{}%
  \let\gplgaddtomacro\g@addto@macro
  \gdef\gplbacktext{}%
  \gdef\gplfronttext{}%
  \makeatother
  \ifGPblacktext
    \def\colorrgb#1{}%
    \def\colorgray#1{}%
  \else
    \ifGPcolor
      \def\colorrgb#1{\color[rgb]{#1}}%
      \def\colorgray#1{\color[gray]{#1}}%
      \expandafter\def\csname LTw\endcsname{\color{white}}%
      \expandafter\def\csname LTb\endcsname{\color{black}}%
      \expandafter\def\csname LTa\endcsname{\color{black}}%
      \expandafter\def\csname LT0\endcsname{\color[rgb]{1,0,0}}%
      \expandafter\def\csname LT1\endcsname{\color[rgb]{0,1,0}}%
      \expandafter\def\csname LT2\endcsname{\color[rgb]{0,0,1}}%
      \expandafter\def\csname LT3\endcsname{\color[rgb]{1,0,1}}%
      \expandafter\def\csname LT4\endcsname{\color[rgb]{0,1,1}}%
      \expandafter\def\csname LT5\endcsname{\color[rgb]{1,1,0}}%
      \expandafter\def\csname LT6\endcsname{\color[rgb]{0,0,0}}%
      \expandafter\def\csname LT7\endcsname{\color[rgb]{1,0.3,0}}%
      \expandafter\def\csname LT8\endcsname{\color[rgb]{0.5,0.5,0.5}}%
    \else
      \def\colorrgb#1{\color{black}}%
      \def\colorgray#1{\color[gray]{#1}}%
      \expandafter\def\csname LTw\endcsname{\color{white}}%
      \expandafter\def\csname LTb\endcsname{\color{black}}%
      \expandafter\def\csname LTa\endcsname{\color{black}}%
      \expandafter\def\csname LT0\endcsname{\color{black}}%
      \expandafter\def\csname LT1\endcsname{\color{black}}%
      \expandafter\def\csname LT2\endcsname{\color{black}}%
      \expandafter\def\csname LT3\endcsname{\color{black}}%
      \expandafter\def\csname LT4\endcsname{\color{black}}%
      \expandafter\def\csname LT5\endcsname{\color{black}}%
      \expandafter\def\csname LT6\endcsname{\color{black}}%
      \expandafter\def\csname LT7\endcsname{\color{black}}%
      \expandafter\def\csname LT8\endcsname{\color{black}}%
    \fi
  \fi
    \setlength{\unitlength}{0.0500bp}%
    \ifx\gptboxheight\undefined%
      \newlength{\gptboxheight}%
      \newlength{\gptboxwidth}%
      \newsavebox{\gptboxtext}%
    \fi%
    \setlength{\fboxrule}{0.5pt}%
    \setlength{\fboxsep}{1pt}%
\begin{picture}(7200.00,5040.00)%
    \gplgaddtomacro\gplbacktext{%
      \csname LTb\endcsname
      \put(946,1017){\makebox(0,0)[r]{\strut{}$10^{-7}$}}%
      \csname LTb\endcsname
      \put(946,2425){\makebox(0,0)[r]{\strut{}$10^{-6}$}}%
      \csname LTb\endcsname
      \put(946,3834){\makebox(0,0)[r]{\strut{}$10^{-5}$}}%
      \csname LTb\endcsname
      \put(1078,484){\makebox(0,0){\strut{}$0$}}%
      \csname LTb\endcsname
      \put(2223,484){\makebox(0,0){\strut{}$2$}}%
      \csname LTb\endcsname
      \put(3368,484){\makebox(0,0){\strut{}$4$}}%
      \csname LTb\endcsname
      \put(4513,484){\makebox(0,0){\strut{}$6$}}%
      \csname LTb\endcsname
      \put(5658,484){\makebox(0,0){\strut{}$8$}}%
      \csname LTb\endcsname
      \put(6803,484){\makebox(0,0){\strut{}$10$}}%
    }%
    \gplgaddtomacro\gplfronttext{%
      \csname LTb\endcsname
      \put(209,2761){\rotatebox{-270}{\makebox(0,0){\strut{}$\mathrm{error}/t$}}}%
      \put(3940,154){\makebox(0,0){\strut{}$t$}}%
      \csname LTb\endcsname
      \put(5816,2637){\makebox(0,0)[r]{\strut{}Forest-Ruth~$(4,3)$~\eqref{eq:3-forest-ruth}}}%
      \csname LTb\endcsname
      \put(5816,2417){\makebox(0,0)[r]{\strut{}FR-Type~$(4,4)$~\eqref{eq:4-fr-type}}}%
      \csname LTb\endcsname
      \put(5816,2197){\makebox(0,0)[r]{\strut{}Non-Unitary~$(4,4)$~\eqref{eq:4-non-unitary1}}}%
      \csname LTb\endcsname
      \put(5816,1977){\makebox(0,0)[r]{\strut{}Suzuki~$(4,5)$~\eqref{eq:5-suzuki4}}}%
      \csname LTb\endcsname
      \put(5816,1757){\makebox(0,0)[r]{\strut{}Opt.~4th~order~$(4,5)$~\eqref{eq:5-opt-4th-ord}}}%
      \csname LTb\endcsname
      \put(5816,1537){\makebox(0,0)[r]{\strut{}Non-Unitary~$(4,5)$~\eqref{eq:5-non-unitary2}}}%
      \csname LTb\endcsname
      \put(5816,1317){\makebox(0,0)[r]{\strut{}Unif.~Non-Unitary~$(4,5)$~\eqref{eq:5-non-unitary-const}}}%
      \csname LTb\endcsname
      \put(5816,1097){\makebox(0,0)[r]{\strut{}Blanes\&Moan~$(4,6)$~\eqref{eq:6-blanes4}}}%
    }%
    \gplbacktext
    \put(0,0){\includegraphics{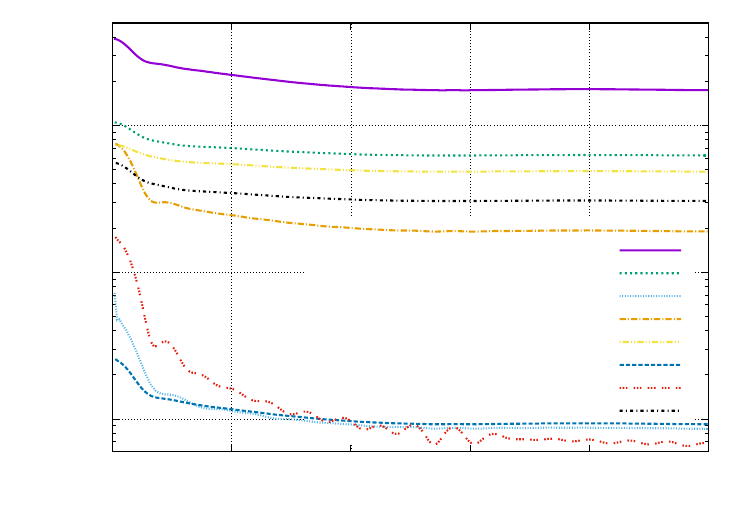}}%
    \gplfronttext
  \end{picture}%
\endgroup

%% file: simulation/benchmark/2-stage_fix-dt_ordRest.tex
\begingroup
  \inputencoding{latin1}%
  \makeatletter
  \providecommand\color[2][]{%
    \GenericError{(gnuplot) \space\space\space\@spaces}{%
      Package color not loaded in conjunction with
      terminal option `colourtext'%
    }{See the gnuplot documentation for explanation.%
    }{Either use 'blacktext' in gnuplot or load the package
      color.sty in LaTeX.}%
    \renewcommand\color[2][]{}%
  }%
  \providecommand\includegraphics[2][]{%
    \GenericError{(gnuplot) \space\space\space\@spaces}{%
      Package graphicx or graphics not loaded%
    }{See the gnuplot documentation for explanation.%
    }{The gnuplot epslatex terminal needs graphicx.sty or graphics.sty.}%
    \renewcommand\includegraphics[2][]{}%
  }%
  \providecommand\rotatebox[2]{#2}%
  \@ifundefined{ifGPcolor}{%
    \newif\ifGPcolor
    \GPcolortrue
  }{}%
  \@ifundefined{ifGPblacktext}{%
    \newif\ifGPblacktext
    \GPblacktexttrue
  }{}%
  \let\gplgaddtomacro\g@addto@macro
  \gdef\gplbacktext{}%
  \gdef\gplfronttext{}%
  \makeatother
  \ifGPblacktext
    \def\colorrgb#1{}%
    \def\colorgray#1{}%
  \else
    \ifGPcolor
      \def\colorrgb#1{\color[rgb]{#1}}%
      \def\colorgray#1{\color[gray]{#1}}%
      \expandafter\def\csname LTw\endcsname{\color{white}}%
      \expandafter\def\csname LTb\endcsname{\color{black}}%
      \expandafter\def\csname LTa\endcsname{\color{black}}%
      \expandafter\def\csname LT0\endcsname{\color[rgb]{1,0,0}}%
      \expandafter\def\csname LT1\endcsname{\color[rgb]{0,1,0}}%
      \expandafter\def\csname LT2\endcsname{\color[rgb]{0,0,1}}%
      \expandafter\def\csname LT3\endcsname{\color[rgb]{1,0,1}}%
      \expandafter\def\csname LT4\endcsname{\color[rgb]{0,1,1}}%
      \expandafter\def\csname LT5\endcsname{\color[rgb]{1,1,0}}%
      \expandafter\def\csname LT6\endcsname{\color[rgb]{0,0,0}}%
      \expandafter\def\csname LT7\endcsname{\color[rgb]{1,0.3,0}}%
      \expandafter\def\csname LT8\endcsname{\color[rgb]{0.5,0.5,0.5}}%
    \else
      \def\colorrgb#1{\color{black}}%
      \def\colorgray#1{\color[gray]{#1}}%
      \expandafter\def\csname LTw\endcsname{\color{white}}%
      \expandafter\def\csname LTb\endcsname{\color{black}}%
      \expandafter\def\csname LTa\endcsname{\color{black}}%
      \expandafter\def\csname LT0\endcsname{\color{black}}%
      \expandafter\def\csname LT1\endcsname{\color{black}}%
      \expandafter\def\csname LT2\endcsname{\color{black}}%
      \expandafter\def\csname LT3\endcsname{\color{black}}%
      \expandafter\def\csname LT4\endcsname{\color{black}}%
      \expandafter\def\csname LT5\endcsname{\color{black}}%
      \expandafter\def\csname LT6\endcsname{\color{black}}%
      \expandafter\def\csname LT7\endcsname{\color{black}}%
      \expandafter\def\csname LT8\endcsname{\color{black}}%
    \fi
  \fi
    \setlength{\unitlength}{0.0500bp}%
    \ifx\gptboxheight\undefined%
      \newlength{\gptboxheight}%
      \newlength{\gptboxwidth}%
      \newsavebox{\gptboxtext}%
    \fi%
    \setlength{\fboxrule}{0.5pt}%
    \setlength{\fboxsep}{1pt}%
\begin{picture}(7200.00,5040.00)%
    \gplgaddtomacro\gplbacktext{%
      \csname LTb\endcsname
      \put(1078,1047){\makebox(0,0)[r]{\strut{}$10^{-14}$}}%
      \csname LTb\endcsname
      \put(1078,1733){\makebox(0,0)[r]{\strut{}$10^{-12}$}}%
      \csname LTb\endcsname
      \put(1078,2419){\makebox(0,0)[r]{\strut{}$10^{-10}$}}%
      \csname LTb\endcsname
      \put(1078,3104){\makebox(0,0)[r]{\strut{}$10^{-8}$}}%
      \csname LTb\endcsname
      \put(1078,3790){\makebox(0,0)[r]{\strut{}$10^{-6}$}}%
      \csname LTb\endcsname
      \put(1078,4476){\makebox(0,0)[r]{\strut{}$10^{-4}$}}%
      \csname LTb\endcsname
      \put(1210,484){\makebox(0,0){\strut{}$0$}}%
      \csname LTb\endcsname
      \put(2329,484){\makebox(0,0){\strut{}$2$}}%
      \csname LTb\endcsname
      \put(3447,484){\makebox(0,0){\strut{}$4$}}%
      \csname LTb\endcsname
      \put(4566,484){\makebox(0,0){\strut{}$6$}}%
      \csname LTb\endcsname
      \put(5684,484){\makebox(0,0){\strut{}$8$}}%
      \csname LTb\endcsname
      \put(6803,484){\makebox(0,0){\strut{}$10$}}%
    }%
    \gplgaddtomacro\gplfronttext{%
      \csname LTb\endcsname
      \put(209,2761){\rotatebox{-270}{\makebox(0,0){\strut{}$\mathrm{error}/t$}}}%
      \put(4006,154){\makebox(0,0){\strut{}$t$}}%
      \csname LTb\endcsname
      \put(5816,2857){\makebox(0,0)[r]{\strut{}Verlet~$(2,1)$~\eqref{eq:1-leap-frog}}}%
      \csname LTb\endcsname
      \put(5816,2637){\makebox(0,0)[r]{\strut{}Omelyan~$(2,2)$~\eqref{eq:2-omelyan2}}}%
      \csname LTb\endcsname
      \put(5816,2417){\makebox(0,0)[r]{\strut{}Yoshida~$(6,7)$~\eqref{eq:7-yoshida}}}%
      \csname LTb\endcsname
      \put(5816,2197){\makebox(0,0)[r]{\strut{}Blanes\&Moan~$(6,10)$~\eqref{eq:10-blanes6}}}%
      \csname LTb\endcsname
      \put(5816,1977){\makebox(0,0)[r]{\strut{}Suzuki~$(6,25)$~(sec.~\ref{sec:suzuki6})}}%
      \csname LTb\endcsname
      \put(5816,1757){\makebox(0,0)[r]{\strut{}Morales~$(8,17)$~\eqref{eq:17-morales}}}%
      \csname LTb\endcsname
      \put(5816,1537){\makebox(0,0)[r]{\strut{}BM6+S~$(8,50)$~(sec.~\ref{sec:very_high_orders})}}%
      \csname LTb\endcsname
      \put(5816,1317){\makebox(0,0)[r]{\strut{}Suzuki~$(8,125)$~(sec.~\ref{sec:very_high_orders})}}%
      \csname LTb\endcsname
      \put(5816,1097){\makebox(0,0)[r]{\strut{}Taylor~(sec.~\ref{sec:taylor_expansion})}}%
    }%
    \gplbacktext
    \put(0,0){\includegraphics{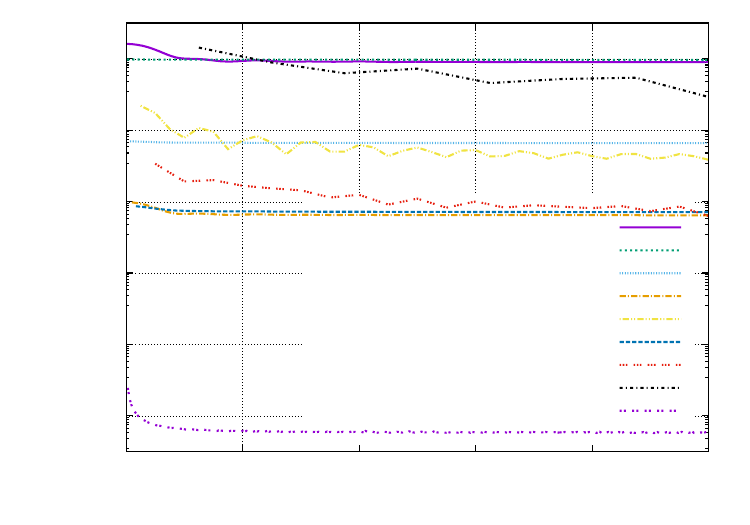}}%
    \gplfronttext
  \end{picture}%
\endgroup

%% file: simulation/benchmark/3L-stage_fix-dt_ordRest.tex
\begingroup
  \inputencoding{latin1}%
  \makeatletter
  \providecommand\color[2][]{%
    \GenericError{(gnuplot) \space\space\space\@spaces}{%
      Package color not loaded in conjunction with
      terminal option `colourtext'%
    }{See the gnuplot documentation for explanation.%
    }{Either use 'blacktext' in gnuplot or load the package
      color.sty in LaTeX.}%
    \renewcommand\color[2][]{}%
  }%
  \providecommand\includegraphics[2][]{%
    \GenericError{(gnuplot) \space\space\space\@spaces}{%
      Package graphicx or graphics not loaded%
    }{See the gnuplot documentation for explanation.%
    }{The gnuplot epslatex terminal needs graphicx.sty or graphics.sty.}%
    \renewcommand\includegraphics[2][]{}%
  }%
  \providecommand\rotatebox[2]{#2}%
  \@ifundefined{ifGPcolor}{%
    \newif\ifGPcolor
    \GPcolortrue
  }{}%
  \@ifundefined{ifGPblacktext}{%
    \newif\ifGPblacktext
    \GPblacktexttrue
  }{}%
  \let\gplgaddtomacro\g@addto@macro
  \gdef\gplbacktext{}%
  \gdef\gplfronttext{}%
  \makeatother
  \ifGPblacktext
    \def\colorrgb#1{}%
    \def\colorgray#1{}%
  \else
    \ifGPcolor
      \def\colorrgb#1{\color[rgb]{#1}}%
      \def\colorgray#1{\color[gray]{#1}}%
      \expandafter\def\csname LTw\endcsname{\color{white}}%
      \expandafter\def\csname LTb\endcsname{\color{black}}%
      \expandafter\def\csname LTa\endcsname{\color{black}}%
      \expandafter\def\csname LT0\endcsname{\color[rgb]{1,0,0}}%
      \expandafter\def\csname LT1\endcsname{\color[rgb]{0,1,0}}%
      \expandafter\def\csname LT2\endcsname{\color[rgb]{0,0,1}}%
      \expandafter\def\csname LT3\endcsname{\color[rgb]{1,0,1}}%
      \expandafter\def\csname LT4\endcsname{\color[rgb]{0,1,1}}%
      \expandafter\def\csname LT5\endcsname{\color[rgb]{1,1,0}}%
      \expandafter\def\csname LT6\endcsname{\color[rgb]{0,0,0}}%
      \expandafter\def\csname LT7\endcsname{\color[rgb]{1,0.3,0}}%
      \expandafter\def\csname LT8\endcsname{\color[rgb]{0.5,0.5,0.5}}%
    \else
      \def\colorrgb#1{\color{black}}%
      \def\colorgray#1{\color[gray]{#1}}%
      \expandafter\def\csname LTw\endcsname{\color{white}}%
      \expandafter\def\csname LTb\endcsname{\color{black}}%
      \expandafter\def\csname LTa\endcsname{\color{black}}%
      \expandafter\def\csname LT0\endcsname{\color{black}}%
      \expandafter\def\csname LT1\endcsname{\color{black}}%
      \expandafter\def\csname LT2\endcsname{\color{black}}%
      \expandafter\def\csname LT3\endcsname{\color{black}}%
      \expandafter\def\csname LT4\endcsname{\color{black}}%
      \expandafter\def\csname LT5\endcsname{\color{black}}%
      \expandafter\def\csname LT6\endcsname{\color{black}}%
      \expandafter\def\csname LT7\endcsname{\color{black}}%
      \expandafter\def\csname LT8\endcsname{\color{black}}%
    \fi
  \fi
    \setlength{\unitlength}{0.0500bp}%
    \ifx\gptboxheight\undefined%
      \newlength{\gptboxheight}%
      \newlength{\gptboxwidth}%
      \newsavebox{\gptboxtext}%
    \fi%
    \setlength{\fboxrule}{0.5pt}%
    \setlength{\fboxsep}{1pt}%
\begin{picture}(7200.00,5040.00)%
    \gplgaddtomacro\gplbacktext{%
      \csname LTb\endcsname
      \put(1078,1047){\makebox(0,0)[r]{\strut{}$10^{-14}$}}%
      \csname LTb\endcsname
      \put(1078,1733){\makebox(0,0)[r]{\strut{}$10^{-12}$}}%
      \csname LTb\endcsname
      \put(1078,2419){\makebox(0,0)[r]{\strut{}$10^{-10}$}}%
      \csname LTb\endcsname
      \put(1078,3104){\makebox(0,0)[r]{\strut{}$10^{-8}$}}%
      \csname LTb\endcsname
      \put(1078,3790){\makebox(0,0)[r]{\strut{}$10^{-6}$}}%
      \csname LTb\endcsname
      \put(1078,4476){\makebox(0,0)[r]{\strut{}$10^{-4}$}}%
      \csname LTb\endcsname
      \put(1210,484){\makebox(0,0){\strut{}$0$}}%
      \csname LTb\endcsname
      \put(2329,484){\makebox(0,0){\strut{}$2$}}%
      \csname LTb\endcsname
      \put(3447,484){\makebox(0,0){\strut{}$4$}}%
      \csname LTb\endcsname
      \put(4566,484){\makebox(0,0){\strut{}$6$}}%
      \csname LTb\endcsname
      \put(5684,484){\makebox(0,0){\strut{}$8$}}%
      \csname LTb\endcsname
      \put(6803,484){\makebox(0,0){\strut{}$10$}}%
    }%
    \gplgaddtomacro\gplfronttext{%
      \csname LTb\endcsname
      \put(209,2761){\rotatebox{-270}{\makebox(0,0){\strut{}$\mathrm{error}/t$}}}%
      \put(4006,154){\makebox(0,0){\strut{}$t$}}%
      \csname LTb\endcsname
      \put(5816,2857){\makebox(0,0)[r]{\strut{}Verlet~$(2,1)$~\eqref{eq:1-leap-frog}}}%
      \csname LTb\endcsname
      \put(5816,2637){\makebox(0,0)[r]{\strut{}Omelyan~$(2,2)$~\eqref{eq:2-omelyan2}}}%
      \csname LTb\endcsname
      \put(5816,2417){\makebox(0,0)[r]{\strut{}Yoshida~$(6,7)$~\eqref{eq:7-yoshida}}}%
      \csname LTb\endcsname
      \put(5816,2197){\makebox(0,0)[r]{\strut{}Blanes\&Moan~$(6,10)$~\eqref{eq:10-blanes6}}}%
      \csname LTb\endcsname
      \put(5816,1977){\makebox(0,0)[r]{\strut{}Suzuki~$(6,25)$~(sec.~\ref{sec:suzuki6})}}%
      \csname LTb\endcsname
      \put(5816,1757){\makebox(0,0)[r]{\strut{}Morales~$(8,17)$~\eqref{eq:17-morales}}}%
      \csname LTb\endcsname
      \put(5816,1537){\makebox(0,0)[r]{\strut{}BM6+S~$(8,50)$~(sec.~\ref{sec:very_high_orders})}}%
      \csname LTb\endcsname
      \put(5816,1317){\makebox(0,0)[r]{\strut{}Suzuki~$(8,125)$~(sec.~\ref{sec:very_high_orders})}}%
      \csname LTb\endcsname
      \put(5816,1097){\makebox(0,0)[r]{\strut{}Taylor~(sec.~\ref{sec:taylor_expansion})}}%
    }%
    \gplbacktext
    \put(0,0){\includegraphics{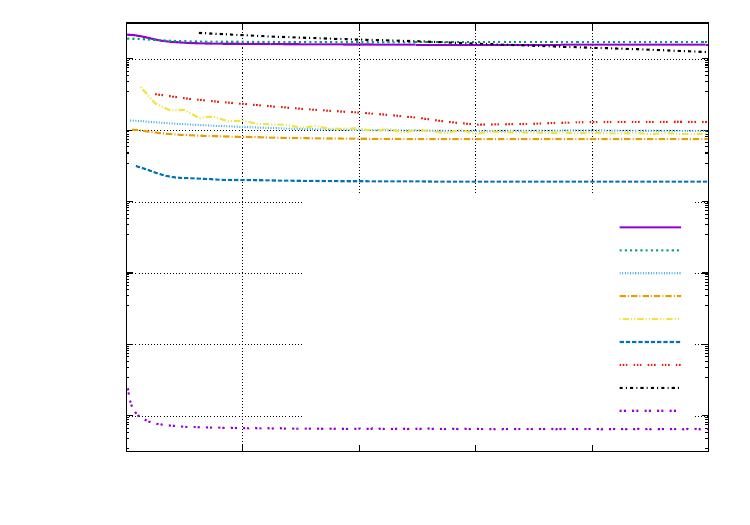}}%
    \gplfronttext
  \end{picture}%
\endgroup

%% file: simulation/benchmark/2-stage_fix-t_opt.tex
\begingroup
  \inputencoding{latin1}%
  \makeatletter
  \providecommand\color[2][]{%
    \GenericError{(gnuplot) \space\space\space\@spaces}{%
      Package color not loaded in conjunction with
      terminal option `colourtext'%
    }{See the gnuplot documentation for explanation.%
    }{Either use 'blacktext' in gnuplot or load the package
      color.sty in LaTeX.}%
    \renewcommand\color[2][]{}%
  }%
  \providecommand\includegraphics[2][]{%
    \GenericError{(gnuplot) \space\space\space\@spaces}{%
      Package graphicx or graphics not loaded%
    }{See the gnuplot documentation for explanation.%
    }{The gnuplot epslatex terminal needs graphicx.sty or graphics.sty.}%
    \renewcommand\includegraphics[2][]{}%
  }%
  \providecommand\rotatebox[2]{#2}%
  \@ifundefined{ifGPcolor}{%
    \newif\ifGPcolor
    \GPcolortrue
  }{}%
  \@ifundefined{ifGPblacktext}{%
    \newif\ifGPblacktext
    \GPblacktexttrue
  }{}%
  \let\gplgaddtomacro\g@addto@macro
  \gdef\gplbacktext{}%
  \gdef\gplfronttext{}%
  \makeatother
  \ifGPblacktext
    \def\colorrgb#1{}%
    \def\colorgray#1{}%
  \else
    \ifGPcolor
      \def\colorrgb#1{\color[rgb]{#1}}%
      \def\colorgray#1{\color[gray]{#1}}%
      \expandafter\def\csname LTw\endcsname{\color{white}}%
      \expandafter\def\csname LTb\endcsname{\color{black}}%
      \expandafter\def\csname LTa\endcsname{\color{black}}%
      \expandafter\def\csname LT0\endcsname{\color[rgb]{1,0,0}}%
      \expandafter\def\csname LT1\endcsname{\color[rgb]{0,1,0}}%
      \expandafter\def\csname LT2\endcsname{\color[rgb]{0,0,1}}%
      \expandafter\def\csname LT3\endcsname{\color[rgb]{1,0,1}}%
      \expandafter\def\csname LT4\endcsname{\color[rgb]{0,1,1}}%
      \expandafter\def\csname LT5\endcsname{\color[rgb]{1,1,0}}%
      \expandafter\def\csname LT6\endcsname{\color[rgb]{0,0,0}}%
      \expandafter\def\csname LT7\endcsname{\color[rgb]{1,0.3,0}}%
      \expandafter\def\csname LT8\endcsname{\color[rgb]{0.5,0.5,0.5}}%
    \else
      \def\colorrgb#1{\color{black}}%
      \def\colorgray#1{\color[gray]{#1}}%
      \expandafter\def\csname LTw\endcsname{\color{white}}%
      \expandafter\def\csname LTb\endcsname{\color{black}}%
      \expandafter\def\csname LTa\endcsname{\color{black}}%
      \expandafter\def\csname LT0\endcsname{\color{black}}%
      \expandafter\def\csname LT1\endcsname{\color{black}}%
      \expandafter\def\csname LT2\endcsname{\color{black}}%
      \expandafter\def\csname LT3\endcsname{\color{black}}%
      \expandafter\def\csname LT4\endcsname{\color{black}}%
      \expandafter\def\csname LT5\endcsname{\color{black}}%
      \expandafter\def\csname LT6\endcsname{\color{black}}%
      \expandafter\def\csname LT7\endcsname{\color{black}}%
      \expandafter\def\csname LT8\endcsname{\color{black}}%
    \fi
  \fi
    \setlength{\unitlength}{0.0500bp}%
    \ifx\gptboxheight\undefined%
      \newlength{\gptboxheight}%
      \newlength{\gptboxwidth}%
      \newsavebox{\gptboxtext}%
    \fi%
    \setlength{\fboxrule}{0.5pt}%
    \setlength{\fboxsep}{1pt}%
\begin{picture}(7200.00,5040.00)%
    \gplgaddtomacro\gplbacktext{%
      \csname LTb\endcsname
      \put(1078,704){\makebox(0,0)[r]{\strut{}$10^{-14}$}}%
      \csname LTb\endcsname
      \put(1078,1272){\makebox(0,0)[r]{\strut{}$10^{-12}$}}%
      \csname LTb\endcsname
      \put(1078,1841){\makebox(0,0)[r]{\strut{}$10^{-10}$}}%
      \csname LTb\endcsname
      \put(1078,2409){\makebox(0,0)[r]{\strut{}$10^{-8}$}}%
      \csname LTb\endcsname
      \put(1078,2978){\makebox(0,0)[r]{\strut{}$10^{-6}$}}%
      \csname LTb\endcsname
      \put(1078,3546){\makebox(0,0)[r]{\strut{}$10^{-4}$}}%
      \csname LTb\endcsname
      \put(1078,4115){\makebox(0,0)[r]{\strut{}$10^{-2}$}}%
      \csname LTb\endcsname
      \put(1078,4683){\makebox(0,0)[r]{\strut{}$10^{0}$}}%
      \csname LTb\endcsname
      \put(2906,484){\makebox(0,0){\strut{}$10$}}%
      \csname LTb\endcsname
      \put(5333,484){\makebox(0,0){\strut{}$100$}}%
    }%
    \gplgaddtomacro\gplfronttext{%
      \csname LTb\endcsname
      \put(209,2761){\rotatebox{-270}{\makebox(0,0){\strut{}error}}}%
      \put(4006,154){\makebox(0,0){\strut{}$q/h$}}%
      \csname LTb\endcsname
      \put(3982,2417){\makebox(0,0)[r]{\strut{}Verlet~$(2,1)$~\eqref{eq:1-leap-frog}}}%
      \csname LTb\endcsname
      \put(3982,2197){\makebox(0,0)[r]{\strut{}Non-Unitary~$(4,4)$~\eqref{eq:4-non-unitary1}}}%
      \csname LTb\endcsname
      \put(3982,1977){\makebox(0,0)[r]{\strut{}Suzuki~$(4,5)$~\eqref{eq:5-suzuki4}}}%
      \csname LTb\endcsname
      \put(3982,1757){\makebox(0,0)[r]{\strut{}Blanes\&Moan~$(4,6)$~\eqref{eq:6-blanes4}}}%
      \csname LTb\endcsname
      \put(3982,1537){\makebox(0,0)[r]{\strut{}Blanes\&Moan~$(6,10)$~\eqref{eq:10-blanes6}}}%
      \csname LTb\endcsname
      \put(3982,1317){\makebox(0,0)[r]{\strut{}Morales~$(8,17)$~\eqref{eq:17-morales}}}%
      \csname LTb\endcsname
      \put(3982,1097){\makebox(0,0)[r]{\strut{}Taylor~(sec.~\ref{sec:taylor_expansion})}}%
    }%
    \gplbacktext
    \put(0,0){\includegraphics{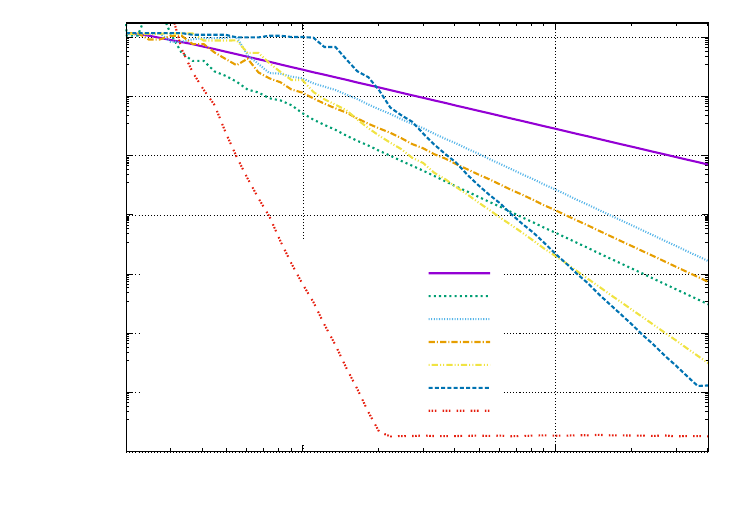}}%
    \gplfronttext
  \end{picture}%
\endgroup

%% file: simulation/benchmark/3L-stage_fix-t_opt.tex
\begingroup
  \inputencoding{latin1}%
  \makeatletter
  \providecommand\color[2][]{%
    \GenericError{(gnuplot) \space\space\space\@spaces}{%
      Package color not loaded in conjunction with
      terminal option `colourtext'%
    }{See the gnuplot documentation for explanation.%
    }{Either use 'blacktext' in gnuplot or load the package
      color.sty in LaTeX.}%
    \renewcommand\color[2][]{}%
  }%
  \providecommand\includegraphics[2][]{%
    \GenericError{(gnuplot) \space\space\space\@spaces}{%
      Package graphicx or graphics not loaded%
    }{See the gnuplot documentation for explanation.%
    }{The gnuplot epslatex terminal needs graphicx.sty or graphics.sty.}%
    \renewcommand\includegraphics[2][]{}%
  }%
  \providecommand\rotatebox[2]{#2}%
  \@ifundefined{ifGPcolor}{%
    \newif\ifGPcolor
    \GPcolortrue
  }{}%
  \@ifundefined{ifGPblacktext}{%
    \newif\ifGPblacktext
    \GPblacktexttrue
  }{}%
  \let\gplgaddtomacro\g@addto@macro
  \gdef\gplbacktext{}%
  \gdef\gplfronttext{}%
  \makeatother
  \ifGPblacktext
    \def\colorrgb#1{}%
    \def\colorgray#1{}%
  \else
    \ifGPcolor
      \def\colorrgb#1{\color[rgb]{#1}}%
      \def\colorgray#1{\color[gray]{#1}}%
      \expandafter\def\csname LTw\endcsname{\color{white}}%
      \expandafter\def\csname LTb\endcsname{\color{black}}%
      \expandafter\def\csname LTa\endcsname{\color{black}}%
      \expandafter\def\csname LT0\endcsname{\color[rgb]{1,0,0}}%
      \expandafter\def\csname LT1\endcsname{\color[rgb]{0,1,0}}%
      \expandafter\def\csname LT2\endcsname{\color[rgb]{0,0,1}}%
      \expandafter\def\csname LT3\endcsname{\color[rgb]{1,0,1}}%
      \expandafter\def\csname LT4\endcsname{\color[rgb]{0,1,1}}%
      \expandafter\def\csname LT5\endcsname{\color[rgb]{1,1,0}}%
      \expandafter\def\csname LT6\endcsname{\color[rgb]{0,0,0}}%
      \expandafter\def\csname LT7\endcsname{\color[rgb]{1,0.3,0}}%
      \expandafter\def\csname LT8\endcsname{\color[rgb]{0.5,0.5,0.5}}%
    \else
      \def\colorrgb#1{\color{black}}%
      \def\colorgray#1{\color[gray]{#1}}%
      \expandafter\def\csname LTw\endcsname{\color{white}}%
      \expandafter\def\csname LTb\endcsname{\color{black}}%
      \expandafter\def\csname LTa\endcsname{\color{black}}%
      \expandafter\def\csname LT0\endcsname{\color{black}}%
      \expandafter\def\csname LT1\endcsname{\color{black}}%
      \expandafter\def\csname LT2\endcsname{\color{black}}%
      \expandafter\def\csname LT3\endcsname{\color{black}}%
      \expandafter\def\csname LT4\endcsname{\color{black}}%
      \expandafter\def\csname LT5\endcsname{\color{black}}%
      \expandafter\def\csname LT6\endcsname{\color{black}}%
      \expandafter\def\csname LT7\endcsname{\color{black}}%
      \expandafter\def\csname LT8\endcsname{\color{black}}%
    \fi
  \fi
    \setlength{\unitlength}{0.0500bp}%
    \ifx\gptboxheight\undefined%
      \newlength{\gptboxheight}%
      \newlength{\gptboxwidth}%
      \newsavebox{\gptboxtext}%
    \fi%
    \setlength{\fboxrule}{0.5pt}%
    \setlength{\fboxsep}{1pt}%
\begin{picture}(7200.00,5040.00)%
    \gplgaddtomacro\gplbacktext{%
      \csname LTb\endcsname
      \put(1078,704){\makebox(0,0)[r]{\strut{}$10^{-14}$}}%
      \csname LTb\endcsname
      \put(1078,1272){\makebox(0,0)[r]{\strut{}$10^{-12}$}}%
      \csname LTb\endcsname
      \put(1078,1841){\makebox(0,0)[r]{\strut{}$10^{-10}$}}%
      \csname LTb\endcsname
      \put(1078,2409){\makebox(0,0)[r]{\strut{}$10^{-8}$}}%
      \csname LTb\endcsname
      \put(1078,2978){\makebox(0,0)[r]{\strut{}$10^{-6}$}}%
      \csname LTb\endcsname
      \put(1078,3546){\makebox(0,0)[r]{\strut{}$10^{-4}$}}%
      \csname LTb\endcsname
      \put(1078,4115){\makebox(0,0)[r]{\strut{}$10^{-2}$}}%
      \csname LTb\endcsname
      \put(1078,4683){\makebox(0,0)[r]{\strut{}$10^{0}$}}%
      \csname LTb\endcsname
      \put(2906,484){\makebox(0,0){\strut{}$10$}}%
      \csname LTb\endcsname
      \put(5333,484){\makebox(0,0){\strut{}$100$}}%
    }%
    \gplgaddtomacro\gplfronttext{%
      \csname LTb\endcsname
      \put(209,2761){\rotatebox{-270}{\makebox(0,0){\strut{}error}}}%
      \put(4006,154){\makebox(0,0){\strut{}$q/h$}}%
      \csname LTb\endcsname
      \put(3982,2417){\makebox(0,0)[r]{\strut{}Verlet~$(2,1)$~\eqref{eq:1-leap-frog}}}%
      \csname LTb\endcsname
      \put(3982,2197){\makebox(0,0)[r]{\strut{}Non-Unitary~$(4,4)$~\eqref{eq:4-non-unitary1}}}%
      \csname LTb\endcsname
      \put(3982,1977){\makebox(0,0)[r]{\strut{}Suzuki~$(4,5)$~\eqref{eq:5-suzuki4}}}%
      \csname LTb\endcsname
      \put(3982,1757){\makebox(0,0)[r]{\strut{}Blanes\&Moan~$(4,6)$~\eqref{eq:6-blanes4}}}%
      \csname LTb\endcsname
      \put(3982,1537){\makebox(0,0)[r]{\strut{}Blanes\&Moan~$(6,10)$~\eqref{eq:10-blanes6}}}%
      \csname LTb\endcsname
      \put(3982,1317){\makebox(0,0)[r]{\strut{}Morales~$(8,17)$~\eqref{eq:17-morales}}}%
      \csname LTb\endcsname
      \put(3982,1097){\makebox(0,0)[r]{\strut{}Taylor~(sec.~\ref{sec:taylor_expansion})}}%
    }%
    \gplbacktext
    \put(0,0){\includegraphics{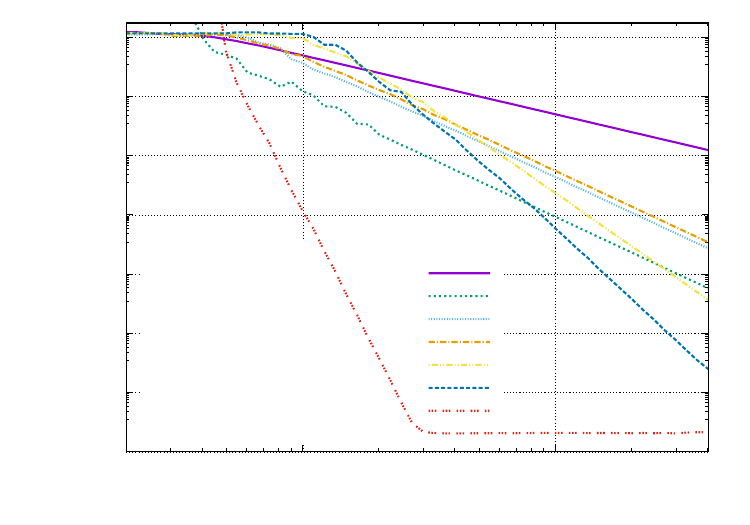}}%
    \gplfronttext
  \end{picture}%
\endgroup

%% file: inkscape/flow-chart.pdf_tex
\begingroup%
  \makeatletter%
  \providecommand\color[2][]{%
    \errmessage{(Inkscape) Color is used for the text in Inkscape, but the package 'color.sty' is not loaded}%
    \renewcommand\color[2][]{}%
  }%
  \providecommand\transparent[1]{%
    \errmessage{(Inkscape) Transparency is used (non-zero) for the text in Inkscape, but the package 'transparent.sty' is not loaded}%
    \renewcommand\transparent[1]{}%
  }%
  \providecommand\rotatebox[2]{#2}%
  \newcommand*\fsize{\dimexpr\f@size pt\relax}%
  \newcommand*\lineheight[1]{\fontsize{\fsize}{#1\fsize}\selectfont}%
  \ifx\svgwidth\undefined%
    \setlength{\unitlength}{567.41320195bp}%
    \ifx\svgscale\undefined%
      \relax%
    \else%
      \setlength{\unitlength}{\unitlength * \real{\svgscale}}%
    \fi%
  \else%
    \setlength{\unitlength}{\svgwidth}%
  \fi%
  \global\let\svgwidth\undefined%
  \global\let\svgscale\undefined%
  \makeatother%
  \begin{picture}(1,1.02559072)%
    \lineheight{1}%
    \setlength\tabcolsep{0pt}%
    \put(0.3107717,0.95858672){\color[rgb]{0,0,0}\makebox(0,0)[lt]{\lineheight{1.25}\smash{\begin{tabular}[t]{l}Do you care about\\performance?\end{tabular}}}}%
    \put(0,0){\includegraphics[width=\unitlength,page=1]{flow-chart.pdf}}%
    \put(0.21442205,0.95191067){\color[rgb]{0,0,0}\makebox(0,0)[lt]{\lineheight{1.25}\smash{\begin{tabular}[t]{l}somewhat\end{tabular}}}}%
    \put(0.02239363,0.97313983){\color[rgb]{0,0,0}\makebox(0,0)[lt]{\lineheight{1.25}\smash{\begin{tabular}[t]{l}Use Blanes\&Moan,\\order $n=4$~\eqref{eq:6-blanes4}\\(Not always optimal,\\but always good)\end{tabular}}}}%
    \put(0,0){\includegraphics[width=\unitlength,page=2]{flow-chart.pdf}}%
    \put(0.48029176,0.95057007){\color[rgb]{0,0,0}\makebox(0,0)[lt]{\lineheight{1.25}\smash{\begin{tabular}[t]{l}no\end{tabular}}}}%
    \put(0.56283271,0.97191735){\color[rgb]{0,0,0}\makebox(0,0)[lt]{\lineheight{1.25}\smash{\begin{tabular}[t]{l}Use Verlet,\\order $n=2$~\eqref{eq:1-leap-frog}\\(Quick and easy\\to code)\end{tabular}}}}%
    \put(0,0){\includegraphics[width=\unitlength,page=3]{flow-chart.pdf}}%
    \put(0.37857281,0.87775885){\color[rgb]{0,0,0}\makebox(0,0)[lt]{\lineheight{1.25}\smash{\begin{tabular}[t]{l}yes\end{tabular}}}}%
    \put(0.29197805,0.82439503){\color[rgb]{0,0,0}\makebox(0,0)[lt]{\lineheight{1.25}\smash{\begin{tabular}[t]{l}Do you need exact\\unitarity (see sec.~\ref{sec:non-unitary1})?\end{tabular}}}}%
    \put(0,0){\includegraphics[width=\unitlength,page=4]{flow-chart.pdf}}%
    \put(0.12777427,0.69226475){\color[rgb]{0,0,0}\makebox(0,0)[lt]{\lineheight{1.25}\smash{\begin{tabular}[t]{l}Do you need high precision\\(relative error $<10^{-4}$)?\end{tabular}}}}%
    \put(0.40534995,0.69226475){\color[rgb]{0,0,0}\makebox(0,0)[lt]{\lineheight{1.25}\smash{\begin{tabular}[t]{l}Is the hamiltonian $H$ bounded\\and can it be applied instead of $\eto{H}$?\end{tabular}}}}%
    \put(0,0){\includegraphics[width=\unitlength,page=5]{flow-chart.pdf}}%
    \put(0.30899046,0.74820526){\color[rgb]{0,0,0}\makebox(0,0)[lt]{\lineheight{1.25}\smash{\begin{tabular}[t]{l}yes\end{tabular}}}}%
    \put(0.47553558,0.74820526){\color[rgb]{0,0,0}\makebox(0,0)[lt]{\lineheight{1.25}\smash{\begin{tabular}[t]{l}no\end{tabular}}}}%
    \put(0.02768079,0.5528108){\color[rgb]{0,0,0}\makebox(0,0)[lt]{\lineheight{1.25}\smash{\begin{tabular}[t]{l}Use Blanes\&Moan,\\order $n=6$~\eqref{eq:10-blanes6}\\or higher (sec.~\ref{sec:very_high_orders})\\(High precision\\made affordable)\end{tabular}}}}%
    \put(0,0){\includegraphics[width=\unitlength,page=6]{flow-chart.pdf}}%
    \put(0.15039238,0.62179235){\color[rgb]{0,0,0}\makebox(0,0)[lt]{\lineheight{1.25}\smash{\begin{tabular}[t]{l}yes\end{tabular}}}}%
    \put(0.56094437,0.55744239){\color[rgb]{0,0,0}\makebox(0,0)[lt]{\lineheight{1.25}\smash{\begin{tabular}[t]{l}Use non-unitary\\scheme~\eqref{eq:4-non-unitary1}, order $n=4$\\or higher (sec.~\ref{sec:very_high_orders})\\(As cheap as it gets,\\yet precise)\end{tabular}}}}%
    \put(0.29639666,0.55876416){\color[rgb]{0,0,0}\makebox(0,0)[lt]{\lineheight{1.25}\smash{\begin{tabular}[t]{l}Use Taylor expansion,\\sec.~\ref{sec:taylor_expansion}\\(Only the exact solution\\is more precise)\end{tabular}}}}%
    \put(0,0){\includegraphics[width=\unitlength,page=7]{flow-chart.pdf}}%
    \put(0.44292632,0.62466473){\color[rgb]{0,0,0}\makebox(0,0)[lt]{\lineheight{1.25}\smash{\begin{tabular}[t]{l}yes\end{tabular}}}}%
    \put(0.60682683,0.62466473){\color[rgb]{0,0,0}\makebox(0,0)[lt]{\lineheight{1.25}\smash{\begin{tabular}[t]{l}no\end{tabular}}}}%
    \put(0.13136204,0.37314716){\color[rgb]{0,0,0}\makebox(0,0)[lt]{\lineheight{1.25}\smash{\begin{tabular}[t]{l}Can you split the hamiltonian $H$ into\\exactly two operators $H=A+B$ (see eq.~\eqref{eq:2-stage-decomposition})?\end{tabular}}}}%
    \put(0,0){\includegraphics[width=\unitlength,page=8]{flow-chart.pdf}}%
    \put(0.23748273,0.61409046){\color[rgb]{0,0,0}\makebox(0,0)[lt]{\lineheight{1.25}\smash{\begin{tabular}[t]{l}no\end{tabular}}}}%
    \put(0.50288714,0.25861597){\color[rgb]{0,0,0}\makebox(0,0)[lt]{\lineheight{1.25}\smash{\begin{tabular}[t]{l}Use Suzuki,\\order $n=4$~\eqref{eq:5-suzuki4}\\(It's well established\\for a reason)\end{tabular}}}}%
    \put(0.115677,0.25327507){\color[rgb]{0,0,0}\makebox(0,0)[lt]{\lineheight{1.25}\smash{\begin{tabular}[t]{l}Is one of the operators\\dominant (e.g.~$A\ll B$)?\end{tabular}}}}%
    \put(0.04654139,0.1162056){\color[rgb]{0,0,0}\makebox(0,0)[lt]{\lineheight{1.25}\smash{\begin{tabular}[t]{l}Use Omelyan's small $A$\\scheme~\eqref{eq:4-small-b}, order $n=4$\\(Make most out of\\your prior knowledge)\end{tabular}}}}%
    \put(0,0){\includegraphics[width=\unitlength,page=9]{flow-chart.pdf}}%
    \put(0.37970352,0.11426801){\color[rgb]{0,0,0}\makebox(0,0)[lt]{\lineheight{1.25}\smash{\begin{tabular}[t]{l}Use Blanes\&Moan,\\order $n=4$~\eqref{eq:6-blanes4}\\(It's always good,\\but here it's optimal)\end{tabular}}}}%
    \put(0,0){\includegraphics[width=\unitlength,page=10]{flow-chart.pdf}}%
    \put(0.21715276,0.30668337){\color[rgb]{0,0,0}\makebox(0,0)[lt]{\lineheight{1.25}\smash{\begin{tabular}[t]{l}yes\end{tabular}}}}%
    \put(0.46950694,0.31202419){\color[rgb]{0,0,0}\makebox(0,0)[lt]{\lineheight{1.25}\smash{\begin{tabular}[t]{l}no\end{tabular}}}}%
    \put(0.14571917,0.19652878){\color[rgb]{0,0,0}\makebox(0,0)[lt]{\lineheight{1.25}\smash{\begin{tabular}[t]{l}yes\end{tabular}}}}%
    \put(0.34666791,0.18918512){\color[rgb]{0,0,0}\makebox(0,0)[lt]{\lineheight{1.25}\smash{\begin{tabular}[t]{l}no\end{tabular}}}}%
  \end{picture}%
\endgroup%